\documentclass[a4paper,11pt]{article}
\pdfoutput=1 
\usepackage{jheppub} 
\usepackage[T1]{fontenc} 
\usepackage{subfig}
\usepackage{float}
\usepackage{xcolor}
\usepackage{adjustbox}
\usepackage{blindtext}
\usepackage{multirow}
\usepackage{verbatim}
\usepackage{bm}
\usepackage{mathrsfs}
\usepackage{amsfonts}
\usepackage{tikz-feynman}
\usepackage{amsmath}
\usepackage{slashed}
\usepackage[normalem]{ulem}
\usepackage{cleveref}
\usepackage{feynmf}
\usepackage{multicol}
\usepackage{booktabs}
\usepackage{contour}
\usepackage{url}

\definecolor{Gray}{gray}{0.9}
\def\lra#1{\overset{\text{\scriptsize$\leftrightarrow$}}{#1}}

\newcommand{\beq}{\begin{equation}}
\newcommand{\eeq}{\end{equation}}
\newcommand{\bea}{\begin{eqnarray}}
\newcommand{\eea}{\end{eqnarray}}

\renewcommand{\phi}{\ensuremath{\varphi}}
\newcommand{\sss}{\scriptscriptstyle}

\newcommand{\MA}{\mathcal{A}}

\newcommand{\OO}{\mathcal{O}}

\newcommand{\QQ}{\ensuremath{\mathcal{Q}}}
\newcommand{\Op}[1]{\OO_{\sss #1}}
\newcommand{\Opp}[2]{\OO_{\sss #1}^{\sss #2}}

\newcommand{\Qp}[1]{\QQ_{\sss #1}}
\newcommand{\Qpp}[2]{\QQ_{\sss #1}^{\sss #2}}
\newcommand{\Qppd}[2]{^{\dagger}\QQ_{\sss #1}^{\sss #2}}

\newcommand{\lb}{\left(}
\newcommand{\rb}{\right)}

\newcommand{\OctTr}{$\Opp{Qq}{3,8}$}
\newcommand{\OctSi}{$\Opp{Qq}{1,8}$}
\newcommand{\QuOct}{$\Opp{Qu}{8}$}
\newcommand{\tqOct}{$\Opp{tq}{8}$}
\newcommand{\QdOct}{$\Opp{Qd}{8}$}
\newcommand{\tuOct}{$\Opp{tu}{8}$}
\newcommand{\tdOct}{$\Opp{td}{8}$}
\newcommand{\TriSi}{$\Opp{Qq}{3,1}$}
\newcommand{\SiSi}{$\Opp{Qq}{1,1}$}
\newcommand{\QuSi}{$\Opp{Qu}{1}$}
\newcommand{\tqSi}{$\Opp{tq}{1}$}
\newcommand{\QdSi}{$\Opp{Qd}{1}$}
\newcommand{\tuSi}{$\Opp{tu}{1}$}
\newcommand{\tdSi}{$\Opp{td}{1}$}
\newcommand{\QQSi}{$\Opp{QQ}{1}$}
\newcommand{\QQOct}{$\Opp{QQ}{8}$}
\newcommand{\ttSi}{$\Opp{tt}{1}$}
\newcommand{\QtSi}{$\Opp{Qt}{1}$}
\newcommand{\QtOct}{$\Opp{Qt}{8}$}
\newcommand{\ctZ}{$\Op{tZ}$}
\newcommand{\ctW}{$\Op{tW}$}
\newcommand{\ctG}{$\Op{tG}$}
\newcommand{\ctp}{$\Op{t\phi}$}
\newcommand{\cpt}{$\Op{\phi t}$}
\newcommand{\cpQM}{$\Opp{\phi Q}{(-)}$}
\newcommand{\cpG}{$\Op{\phi G}$}
\newcommand{\cG}{$\Op{G}$}
\newcommand{\tttt}{$t\bar{t}t\bar{t}$\xspace}
\DeclareRobustCommand{\rchi}{{\mathpalette\irchi\relax}}
\newcommand{\irchi}[2]{\raisebox{\depth}{$#1\chi$}}

\newcommand{\mybox}[1]{  
	\iftrue 
	\vspace{0.5cm} \noindent\fbox{%
		\parbox{\textwidth}{%
			#1
	}} 
	\vspace{0.5cm}
	\fi}

\subheader{\normalsize
	\vspace*{-3.7em}
	\begin{flushright}
		CP3-22-37\\
	\end{flushright}
}

\title{\boldmath Complete SMEFT predictions for four top quark production at hadron colliders}

\author[a]{Rafael Aoude,} 
\author[a,b]{Hesham El Faham,}
\author[a,c]{Fabio Maltoni,}
\author[d]{Eleni Vryonidou}
\affiliation[a]{Centre for Cosmology, Particle Physics and Phenomenology (CP3),\\ Universit\'e Catholique de Louvain,\\ Chemin du Cyclotron, B-1348 Louvain la Neuve, Belgium}
\affiliation[b]{Inter-University Institute for High Energies (IIHE), Vrije Universiteit Brussel, \\Pleinlaan 2, 1050 Brussels, Belgium}
\affiliation[c]{Dipartimento di Fisica e Astronomia, Universit\`a di Bologna and INFN, Sezione di Bologna,\\ Via Irnerio 46, 40126 Bologna, Italy}
\affiliation[d]{Department of Physics and Astronomy, University of Manchester, \\
Oxford Road, Manchester M13 9PL, United Kingdom}

\emailAdd{rafael.aoude@uclouvain.be}
\emailAdd{hesham.el.faham@vub.be}
\emailAdd{fabio.maltoni@uclouvain.be}
\emailAdd{eleni.vryonidou@manchester.ac.uk}

\abstract{We study four top quark production at hadron colliders in the Standard Model Effective Field Theory (SMEFT). 
We perform an analysis at the tree-level, including all possible QCD- and EW-coupling orders and relevant dimension-six operators.  
We find several cases where formally subleading terms give rise to significant contributions, potentially providing sensitivity to a broad class of operators.  
Inclusive and differential predictions are presented for the LHC and a future $pp$ circular collider operating at $100$ TeV. 
We estimate the sensitivity of different operators and perform a simplified chi-square fit to set limits on SMEFT Wilson coefficients. In so doing, we assess the importance of including subleading terms and differential information in constraining new physics contributions. 
Finally, we compute the SMEFT predictions for the double insertion of dimension-six operators and scrutinise the possible enhancements to the sensitivity induced by a specific class of higher order terms in the EFT series.}

\keywords{\textit{Top quark, QCD, SMEFT, Monte-Carlo, Collider physics}}

\begin{document} 

\maketitle

\section{Introduction} \label{sec:intro}
In the last decade, the Large Hadron Collider (LHC) experiment at CERN has tested our understanding of fundamental interactions up to several TeV's of energy.
The unexpected success of the Standard Model (SM), on the one hand, and its inherent incompleteness as a theory of nature, on the other hand, have led to vigorous efforts by the experimental and theoretical communities to study where new physics could lurk.  
In this endeavour, a unique role is played by precision physics, where accurate theoretical predictions of SM processes are compared with experimental measurements searching for deviations. 
The upcoming third run of the LHC - characterised by a 4.5\% increase in collision centre of mass-energy, from 13 TeV to 13.6 TeV and a two-fold increase of luminosity- will provide a new handle on rare phenomena (for a review of the latest experimental measurements of \tttt and future prospects at the LHC, see Ref.~\cite{Blekman:2022jag}). 
Among the rarest and most spectacular processes at the LHC is the production of two pairs of top-antitop quarks, i.e., four top quark production. 
This process is characterised by a tiny SM cross-section at 13 TeV, of about 12 fb~\cite{Frederix:2017wme}, i.e., around five orders of magnitude lower than that of $t\bar{t}$ production. However, despite the tiny rates, \tttt signatures are distinctive, leading to a wealthy and energetic final state, which is challenging to mimic through other processes. 
Therefore, the very high-$Q^2$ and low backgrounds offer a unique opportunity for probing new physics~\cite{Lillie:2007hd,Pomarol:2008bh,Kumar:2009vs,Deandrea:2014raa,Berger:2011af,Aguilar-Saavedra:2011mam,Beck:2015cga,BhupalDev:2014bir,Acharya:2009gb,Gregoire:2011ka,Degrande:2010kt,Cao:2021qqt,Darme:2021gtt,Banelli:2020iau,Englert:2019zmt}.

Given its promise, precise predictions of four-top production have become necessary.
Next-to-leading order (NLO) corrections in QCD have been computed first in Ref.~\cite{Bevilacqua:2012em} and then also become available in event generators~\cite{Alwall:2014hca,Maltoni:2015ena,Jezo:2021smh}.
The complete NLO predictions, including all possible QCD and electroweak (EW) orders, have been calculated in Ref.~\cite{Frederix:2017wme}.  
They revealed a peculiar and unexpected interplay between EW and QCD contributions, with significant contributions with opposite signs arising from formally subleading terms.  
Significant subleading contributions have also been observed in the case of new physics contributions to four-top production~\cite{Darme:2021gtt,Degrande:2020evl}. 
Moreover, it has been suggested in the study of Ref.~\cite{Cao:2016wib} that four-top production may be a valuable probe of the top-Yukawa coupling, $y_t$, at the tree-level.
 
In the SM, representative diagrams of the pure-QCD $\mathscr{O}(\alpha_s^{2})$ four-top production are shown in \cref{fig:4tops_LO_diags}, occurring through $gg$ and $q\bar{q}$ initial states. 
QCD-induced diagrams typically provide the leading contribution. 
However, formally subleading diagrams with insertions of EW couplings, i.e. $\mathscr{O}(\alpha_{s}\alpha_{\mathrm{w}})$, can also be significant.
Examples of the latter are shown in \cref{fig:4tops_LO_diags_scattering}, where top quarks scatter through the exchange of a Higgs boson or $Z/\gamma^*$. 
\begin{figure}[h!]
    \centering
    \scalebox{0.9}{\begin{tikzpicture}
\begin{feynman}[small]
\vertex (a0) {};
\vertex[right = of a0] (a1) {};
\vertex[right = of a1] (a2) {$\bar{t}$};
\vertex[below = of a0] (c0) {$g$};
\vertex[right = of c0] (c1);
\vertex[right = of c1] (c2) {};
\vertex[right = of c2] (c3) {};
\vertex[right = of c3] (c4) {$\bar{t}$};
\vertex[right = of c4] (c5) {};
\vertex[below = of c0] (b0) {};
\vertex[right = of b0] (b1) {};
\vertex[right = of b1] (b2);
\vertex[right = of b2] (b3);
\vertex[right = of b3] (b4) {};
\vertex[right = of b4] (b5) {};
\vertex[below = of b0] (d0) {$g$};
\vertex[right = of d0] (d1);
\vertex[right = of d1] (d2) {};
\vertex[right = of d2] (d3) {};
\vertex[right = of d3] (d4) {$t$};
\vertex[right = of d4] (d5) {};
\vertex[below = of d0] (e0) {};
\vertex[right = of e0] (e1) {};
\vertex[right = of e1] (e2) {$t$};
\diagram*{
(c0) -- [gluon] (c1),
(d0) -- [gluon] (d1),
(c1) -- [fermion] (b2),
(c1) -- [anti fermion] (a2),
(d1) -- [anti fermion] (b2),
(d1) -- [fermion] (e2),
(b2) -- [gluon] (b3),
(b3) -- [anti fermion] (c4),
(b3) -- [fermion] (d4),
};
\end{feynman}
\end{tikzpicture}
\begin{tikzpicture}
\begin{feynman}[small]
\vertex (a0) {};
\vertex[right = of a0] (a1) {};
\vertex[right = of a1] (a2) {};
\vertex[right = of a2] (a3) {};
\vertex[right = of a3] (a4) {$\bar{t}$};
\vertex[below = of a0] (c0) {$q$};
\vertex[right = of c0] (c1) {};
\vertex[right = of c1] (c2);
\vertex[right = of c2] (c3);
\vertex[right = of c3] (c4) {$t$};
\vertex[right = of c4] (c5) {};
\vertex[below = of c0] (b0) {};
\vertex[right = of b0] (b1) {};
\vertex[right = of b1] (b2);
\vertex[right = of b2] (b3);
\vertex[right = of b3] (b4) {};
\vertex[right = of b4] (b5) {};
\vertex[below = of b0] (d0) {$\bar{q}$};
\vertex[right = of d0] (d1){};
\vertex[right = of d1] (d2);
\vertex[right = of d2] (d3);
\vertex[right = of d3] (d4) {$\bar{t}$};
\vertex[right = of d4] (d5) {};
\vertex[below = of d0] (e0) {};
\vertex[right = of e0] (e1) {};
\vertex[right = of e1] (e2) {};
\vertex[right = of e2] (e3) {};
\vertex[right = of e3] (e4) {$t$};
\diagram*{
(c0) -- [fermion] (c2),
(d0) -- [anti fermion] (d2),
(c2) -- [fermion] (d2),
(c2) -- [gluon] (c3),
(d2) -- [gluon] (d3),
(c3) -- [anti fermion] (a4),
(c3) -- [fermion] (c4),
(d3) -- [anti fermion] (d4),
(d3) -- [fermion] (e4),
};
\end{feynman}
\end{tikzpicture}}
    \caption{Representative leading order tree-level Feynman diagrams of $\mathscr{O}(\alpha_{s}^{2})$ for the $gg$-initiated (\emph{left}) and the $q\bar{q}$-initiated (\emph{right}) SM four-top production at the LHC.}
    \label{fig:4tops_LO_diags}
\end{figure}
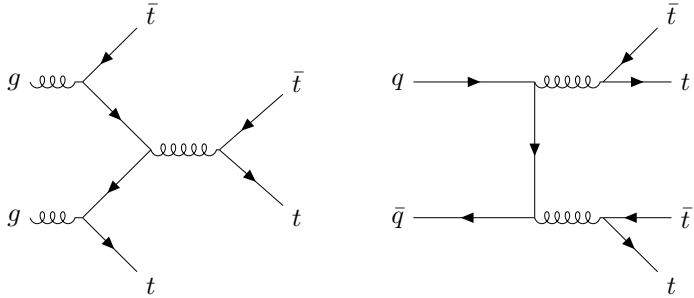
\begin{figure}[h!]
    \centering
    \scalebox{0.9}{\begin{tikzpicture}
\begin{feynman}[small]
\vertex (a0) {};
\vertex[right = of a0] (a1) {};
\vertex[right = of a1] (a2) {};
\vertex[right = of a2] (a3) {};
\vertex[right = of a3] (a4) {$\bar{t}$};
\vertex[below = of a0] (c0) {$g$};
\vertex[right = of c0] (c1) {};
\vertex[right = of c1] (c2);
\vertex[right = of c2] (c3) ;
\vertex[right = of c3] (c4) {$t$};
\vertex[right = of c4] (c5) {};
\vertex[below = of c0] (b0) {};
\vertex[right = of b0] (b1) {};
\vertex[right = of b1] (b2);
\vertex[right = of b2] (b3);
\vertex[right = of b3] (b4) {};
\vertex[right = of b4] (b5) {};
\vertex[below = of b0] (d0) {$g$};
\vertex[right = of d0] (d1){};
\vertex[right = of d1] (d2);
\vertex[right = of d2] (d3);
\vertex[right = of d3] (d4) {$\bar{t}$};
\vertex[right = of d4] (d5) {};
\vertex[below = of d0] (e0) {};
\vertex[right = of e0] (e1) {};
\vertex[right = of e1] (e2) {};
\vertex[right = of e2] (e3) {};
\vertex[right = of e3] (e4) {$t$};
\diagram*{
(c0) -- [gluon] (c2),
(d0) -- [gluon] (d2),
(c2) -- [anti fermion] (a4),
(d2) -- [fermion] (e4),
(c2) -- [fermion] (c3),
(d2) -- [anti fermion] (d3),
(c3) -- [scalar, edge label = $H$] (d3),
(c3) -- [fermion] (c4),
(d3) -- [anti fermion] (d4),
};
\end{feynman}
\end{tikzpicture}
\begin{tikzpicture}
\begin{feynman}[small]
\vertex (a0) {};
\vertex[right = of a0] (a1) {};
\vertex[right = of a1] (a2) {};
\vertex[right = of a2] (a3) {};
\vertex[right = of a3] (a4) {$\bar{t}$};
\vertex[below = of a0] (c0) {$g$};
\vertex[right = of c0] (c1) {};
\vertex[right = of c1] (c2);
\vertex[right = of c2] (c3) ;
\vertex[right = of c3] (c4) {$t$};
\vertex[right = of c4] (c5) {};
\vertex[below = of c0] (b0) {};
\vertex[right = of b0] (b1) {};
\vertex[right = of b1] (b2);
\vertex[right = of b2] (b3);
\vertex[right = of b3] (b4) {};
\vertex[right = of b4] (b5) {};
\vertex[below = of b0] (d0) {$g$};
\vertex[right = of d0] (d1){};
\vertex[right = of d1] (d2);
\vertex[right = of d2] (d3);
\vertex[right = of d3] (d4) {$\bar{t}$};
\vertex[right = of d4] (d5) {};
\vertex[below = of d0] (e0) {};
\vertex[right = of e0] (e1) {};
\vertex[right = of e1] (e2) {};
\vertex[right = of e2] (e3) {};
\vertex[right = of e3] (e4) {$t$};
\diagram*{
(c0) -- [gluon] (c2),
(d0) -- [gluon] (d2),
(c2) -- [anti fermion] (a4),
(d2) -- [fermion] (e4),
(c2) -- [fermion] (c3),
(d2) -- [anti fermion] (d3),
(c3) -- [boson, edge label = $Z$/$\gamma^{*}$] (d3),
(c3) -- [fermion] (c4),
(d3) -- [anti fermion] (d4),
};
\end{feynman}
\end{tikzpicture}}
    \caption{Representative diagrams of $\mathscr{O}(\alpha_{s}\alpha_{\mathrm{w}})$ for the SM four-top production at the LHC. 
    The diagrams show the EW $tt \to tt$ scattering involving the exchange of a Higgs boson (\emph{left}) or a $Z$-boson/virtual photon (\emph{right}).}
    \label{fig:4tops_LO_diags_scattering}
\end{figure}
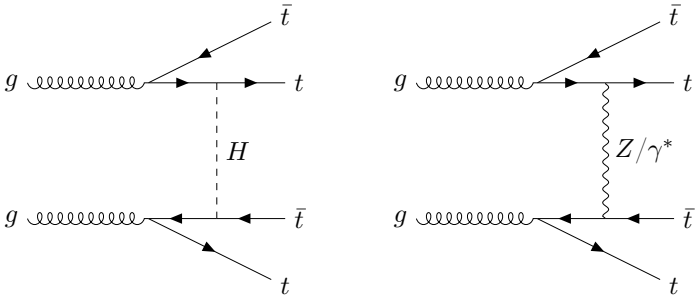
These diagrams can contribute through their interference with QCD ones or through their squares. 
In the SM at $\sqrt{s}=13$ TeV, it turns out that contributions from the interference of this class of diagrams with the leading QCD amplitude and from their squares are significant, reaching more than a third of the leading tree-level QCD  contributions. 
Nevertheless, these two contributions come with opposite signs, and there is a significant cancellation between them. 
This can be seen explicitly in table 7 of Ref.~\cite{Frederix:2017wme}.  
These large cancellations at the tree-level also motivate a high-order computation, including QCD and EW corrections, as discussed in Ref.~\cite{Frederix:2017wme}. 
At one-loop order, additional cancellations occur between different terms in the $\alpha_{s}$ and $\alpha_{\mathrm{w}}$ expansion, but in general, higher-order corrections are dominated by $\alpha_{s}$ corrections to the leading QCD term. 
It is worth noting that the size of various terms varies significantly depending on the choice of the renormalisation scale. 
Though it is clear from table 7 of Ref.~\cite{Frederix:2017wme} that several contributions are larger than what one would expect from the ratio of $\alpha_{\mathrm{w}}/\alpha_{s}$.
In summary, four-top production does not necessarily submit to the ``naive'' $\alpha_{s}$ and $\alpha_{\mathrm{w}}$ power counting. 
Therefore, a study of four-top production should be completed by including all QCD and EW-induced terms in the computations. This consideration constitutes the primary motivation behind the work presented in this paper.

The peculiar behaviour of the cross-section as a double series in $\alpha_s$ and $\alpha_\mathrm{w}$ for the SM process certainly motivates a detailed study in the case of including new physics effects, particularly in the SMEFT framework. 
The effective field theory approach assumes new physics to reside at a high scale $\Lambda$~\cite{Weinberg:1978kz,Buchmuller:1985jz,Leung:1984ni}. 
In the SMEFT, the SM Lagrangian is augmented with higher-dimensions operators built out of the SM fields and respecting the SM gauge symmetries, describing short-distance interactions generated by new physics at high scales. 
The pattern of such deformation depends on the details of the phenomena in the ultra-violet (UV) region. 
Being unknown, one assumes all possible operators to be there and studies their effects on low-energy observables. 
The beauty of this framework is that it allows perturbative calculations to be performed consistently order by order in the $1/\Lambda$ expansion. 
Such a powerful approach provides a consistent and calculable framework in which the potential deviations from the SM predictions can be encapsulated and predicted in type and pattern. 
Studies in the context of the SMEFT at the LHC are an ongoing effort in all SM sectors; the electroweak, the Higgs, the flavour, and the top sectors. 
Global fits combining a broad set of publicly available data have appeared~\cite{Brivio:2019ius,Hartland:2019bjb,Ethier:2021bye,Ellis:2020unq,Buckley:2015lku,Miralles:2021dyw}, indicating which directions (operators) in the fits can be constrained and whether complementary information or new strategies would be helpful. 
In the context of an EFT, four-top production is exciting as it is the simplest process where top quark self-interactions could be probed at the tree-level.  
In the SMEFT language, such interactions are described by a set of dimension-six operators of the form $\bar \psi \psi \bar \psi \psi$  operators with four top quark fields (left- or right-handed). 
Other processes at colliders do not directly constrain these operators at the tree-level. 
Therefore four-top production is naively expected to be the first place to see their effects\footnote{~Proposals for constraining four-top operators through loop effects have appeared in Refs.~\cite{Degrande:2020evl,Banelli:2020iau,Alasfar:2022zyr,Dawson:2022bxd}.}. 

This work considers all possible contributions of the SM and the SMEFT, including all the dimension-six CP-even SMEFT operators that enter at the tree-level. 
Part of our motivation is that in the SMEFT, the EFT interference with the SM is expected to provide the leading cross-section contribution. 
As the interference projects the kinematic and the colour structure of the SM amplitudes, its size can change significantly from one operator to another. 
It is also expected to vary depending on which contributions are included in the SM, as different operators have different colour and chirality structures. 
As mentioned previously, we retain all possible tree-level contributions at different orders in QCD and EW couplings in our computations. 
Specifically, we split the EW-induced contributions into the gauge and top-Yukawa ones and determine the inclusive and differential predictions for the LHC and FCC-hh. 
We organise our predictions as an expansion in $\alpha_{s}$, where each term is expanded in the weak parameters, highlighting the potential significance of the formally subleading terms.    

To assess the reach of constraining the SMEFT Wilson coefficients (WCs) at future colliders, we perform (\emph{i}) a signal-strength-based projection study for each operator at different collider energies obtaining theoretical limits on the corresponding WCs; (\emph{ii})
simplified chi-square ($\rchi^{2}$) fits at different collider energies on selected sets of operators. 
In the latter case, we also include differential information and assess its constraining power compared to using only inclusive measurements.

Finally, we scrutinise the claim of Ref.~\cite{Zhang:2017mls} on the enhanced EFT sensitivity of four-top production to 2-heavy-2-light four-fermion operators due to the contributions from double insertion. 
In  Ref.~\cite{Zhang:2017mls}, it was argued that though formally equivalent to single dimension-eight insertions, in some UV models, double insertion could provide the dominant terms and four-top production could compete with much more abundant top quark pair production processes through enhancements scaling as $\sim (cE^{2}/\Lambda^{2})^{4}$. 

This paper is structured as follows: in \cref{sec:smeft_framework}, we describe the theoretical tools for four-top production within the SMEFT framework, presenting the operators' definitions and the cross-section expansion in QCD and EW couplings. 
The inclusive and differential predictions are presented in \cref{sec:inclusive} and \cref{sec:differential}, respectively. 
The signal-strength projection study and the $\rchi^{2}$ fits are presented in \cref{sec:projections} and \cref{sec:toy}, respectively.
We discuss the results from the cross-section computation considering double EFT insertions in \cref{sec:double_insertions}. 
The work is summarised and concluded in \cref{sec:conclusions}.
\section{SMEFT framework to four top quark production}
\label{sec:smeft_framework}
\subsection{Operators definitions}
We compute the SMEFT contributions to four-top production using a specific flavour assumption which singles out the top quark interactions,
\begin{align}%
    U(3)_l\times U(3)_e \times U(2)_q\times U(2)_u\times U(3)_d \equiv U(2)^{2}\times U(3)^{3},
    \label{eq:flav_symmetry}
\end{align}%
where the subscripts refer to the five-fermion representations of the SM. 
This minimal relaxation of the $U(3)^5$ group gives rise to top quark chirality-flipping interactions, such as the dipole interactions and ones which modify the top-Yukawa coupling. 
We use the notation and operator conventions of Refs.~\cite{Aguilar-Saavedra:2018ksv,Degrande:2020evl} and study all operators in three classes of dimension-six SMEFT: four-fermion (4F), two-fermion (2F), and purely-bosonic (0F) operators, consistent with our flavour symmetry assumption.
We do not consider the two-fermion light quark operators as we expect them to be better constrained in other production processes.

\paragraph{Four-fermion operators}
Following the conventions and notation of Ref.~\cite{Aguilar-Saavedra:2018ksv}, the four-fermion operators are defined as follows:
\begin{multicols}{2}%
\noindent
\begin{align}%
    \Qpp{qq}{1(ijkl)}&=(\bar{q_i}\gamma^{\mu}q_{j})(\bar{q_k}\gamma_{\mu}q_{l}),\notag\\
    \Qpp{qu}{1(ijkl)}&=(\bar{q_i}\gamma^{\mu}q_{j})(\bar{u_k}\gamma_{\mu}u_{l}),\notag\\
    \Qpp{qd}{1(ijkl)}&=(\bar{q_i}\gamma^{\mu}q_{j})(\bar{d_k}\gamma_{\mu}d_{l}),\notag\\
    \Qpp{ud}{1(ijkl)}&=(\bar{u_i}\gamma^{\mu}u_{j})(\bar{d_k}\gamma_{\mu}d_{l}),\notag\\
    \Qppd{quqd}{1(ijkl)}&=(\bar{q_i}u_{j})\epsilon(\bar{q_k}d_{l}),\notag\\\notag
\end{align}%
\noindent
\begin{align}%
    \Qpp{qq}{3(ijkl)}&=(\bar{q_i}\gamma^{\mu}\tau^{I}q_{j})(\bar{q_k}\gamma_{\mu}\tau^{I}q_{l}),\notag\\
    \Qpp{qu}{8(ijkl)}&=(\bar{q_i}\gamma^{\mu}T^{A}q_{j})(\bar{u_k}\gamma_{\mu}T^{A}u_{l}),\notag\\
    \Qpp{qd}{8(ijkl)}&=(\bar{q_i}\gamma^{\mu}T^{A}q_{j})(\bar{d_k}\gamma_{\mu}T^{A}d_{l}),\notag\\
    \Qpp{ud}{8(ijkl)}&=(\bar{u_i}\gamma^{\mu}T^{A}u_{j})(\bar{d_k}\gamma_{\mu}T^{A}d_{l}),\notag\\
    \Qppd{quqd}{8(ijkl)}&=(\bar{q_i}T^{A}u_{j})\epsilon(\bar{q_k}T^{A}d_{l}),\notag\\
    \Qpp{uu}{(ijkl)}&=(\bar{u_i}\gamma^{\mu}u_{j})(\bar{u_k}\gamma_{\mu}u_{l}),
    \label{eq:dim64f_op_warsaw}
\end{align}%
\end{multicols}%
\noindent
where the notation $\mathcal{Q}$ indicates operators are given in the original Warsaw basis~\cite{Grzadkowski:2010es}.
In this work however, we use operators aligned with the \texttt{SMEFTatNLO}~\cite{Degrande:2020evl} conventions, hereafter referred to as the ``top-basis'' and denoted by $\mathcal{O}$. 
The difference lies in the slight modification of the four-fermion operators rendering them more suitable for top quark physics, as well as normalising the operators $\Op{tG}$ and $\Op{G}$ through the inclusion of an extra $g_s$ factor in their definitions. 
The consequences of the latter normalisation are later discussed when presenting the inclusive predictions in \cref{sec:inclusive}.
The translations of all four-fermion operators from the Warsaw basis to the top-basis are given in \cref{tab:dim64f_smeftatnlo_basis} of \cref{sec:app_4f_def_constraints}.
We also present the recent constraints on their corresponding coefficients in \cref{tab:dim64f_wc_bounds} of the same appendix based on the global analysis of Ref.~\cite{Ethier:2021bye}.

\paragraph{Two-fermion and purely-bosonic operators}
The set of potentially relevant two-fermion operators to four-top production are defined as follows:
\begin{multicols}{2}%
\noindent
\begin{align}%
    \Qp{tB} 
    &=i\big(\bar{Q}\tau^{\mu\nu}\,t\big)\,\tilde{\phi}\,B_{\mu\nu} +\text{h.c.},\notag\\
    \Op{tW} 
    &= i\big(\bar{Q}\tau^{\mu\nu}\,\tau_{\sss I}\,t\big)\,\tilde{\phi}\,W^I_{\mu\nu} + \text{h.c.}, \notag\\
    \Op{tG} 
    &= ig{\sss S}\,\big(\bar{Q}\tau^{\mu\nu}\,T_{\sss A}\,t\big)\,\tilde{\phi}\,G^A_{\mu\nu}  + \text{h.c.},\notag
\end{align}%
\noindent
\begin{align}%
    \Op{t\phi} 
    &= \big(\phi^\dagger\phi-v^{2}/2\big)\bar{Q}t\tilde{\phi} + \text{h.c.} ,\notag\\
    \Op{\phi t} 
    &= i\big(\phi^\dagger\,\lra{D}_\mu\,\,\phi\big)\big(\bar{t}\,\gamma^\mu\,t\big),\notag\\
    \Qpp{\phi Q}{(1)} 
    &= i\big(\phi^\dagger\lra{D}_\mu\,\phi\big)\big(\bar{Q}\,\gamma^\mu\,Q\big),\notag\\
    \Qpp{\phi Q}{(3)} 
    &= i\big(\phi^\dagger\lra{D}_\mu\,\tau_I\,\phi\big)\big(\bar{Q}\,\gamma^\mu\,\tau^I\,Q\big).
    \label{eq:dim62f_op_smeftatnlo}
\end{align}%
\end{multicols}%
\noindent
For convenience and following the conventions of \cite{Degrande:2020evl,Aguilar-Saavedra:2018ksv} we consider the following linear combinations of Warsaw operators' coefficients: 
\begin{equation}%
    c_{tW}=C_{tW}, \hspace{0.5ex} c_{tZ}=-\sin{\theta_{\mathrm{w}}}C_{tB}+\cos{\theta_{\mathrm{w}}}C_{tW},
\end{equation}%
where we kept the notation $c_i$ for coefficients of operators written in the top-basis while $C_i$ denotes coefficients in Warsaw basis.
Similarly, we use the linear combination $c^{(-)}_{\varphi Q} = C^{(1)}_{\varphi Q} - C^{(3)}_{\varphi Q}$ instead of the singlet piece (this combination is notated as $\texttt{cpQM}$ in \texttt{SMEFTatNLO} while the triplet $c_{\varphi Q}^{(3)}$ as \texttt{cpQ3}).
The coefficient $c_{\varphi Q}^{(3)}$ is irrelevant to four-top production since it modifies the $tWb$ vertex.

On the other hand, the relevant purely-bosonic operators in the top-basis are defined as follows:
\begin{align}%
    \Op{\phi G} &= 
    \bigg(\phi^\dagger\phi -\frac{v^{2}}{2}\bigg)G_{A}^{\mu\nu}G^A_{\mu\nu} ,\qquad\qquad
    \Op{G} = 
    g_{s}f_{ABC}G^A_{\mu\nu}G^{B,\nu\rho}G^{C,\mu}_{\rho}.
    \label{eq:dim60f_op_smeftatnlo}
\end{align}%
The latter is constrained by studies including multi-jet production~\cite{Krauss:2016ely,Hirschi:2018etq}.
Bounds on the \cpG{} coefficient, as well as all two-fermion coefficients in \cref{eq:dim62f_op_smeftatnlo}, are given in \cref{tab:contributing_operators_wc_bounds} of \cref{sec:app_4f_def_constraints}.
Operators that modify the Higgs couplings to the gauge bosons and those that enter via fields' redefinition have negligible contributions in four-top production. 
Therefore, we omit them in what follows as we expect them to be constrained much better in other processes.

Having defined the potential operators relevant to four-top production, we now move to analyse the SM and EFT amplitudes in orders of QCD and EW couplings and subsequently examine different terms contributing to the cross-section. 

\subsection{Leading order coupling expansion}
In the presence of SMEFT operators, a generic scattering amplitude expanded in the $1/\Lambda$ parameter can be written as follows:
\begin{align}%
    \mathcal{A} = \mathcal{A}_{\rm SM} + \frac{1}{\Lambda^2} \mathcal{A}_{\rm (d6)}
    +\frac{1}{\Lambda^4}\big( \mathcal{A}_{\rm (d6)^2}
    + \mathcal{A}_{\rm (d8)}\big),
\end{align}%
leading to the decomposition of the partonic differential cross-section up to $\mathscr{O}(\Lambda^{-4})$,
\begin{align}%
    d\sigma = d\sigma_{\rm SM} +\frac{1}{\Lambda^2} d\sigma_{\rm int} + 
    \frac{1}{\Lambda^4}
    \big(
    d\sigma_{\rm quad} +  d\sigma_{\rm dbl} + d\sigma_{\rm d8}
    \big).
\end{align}%
The leading SMEFT contribution, $d\sigma_{\rm int}$, arises as the linear interference between $\mathcal{A}_{\rm SM}$ and $\mathcal{A}_{\rm (d6)}$, while the $d\sigma_{\rm quad}$ and $d\sigma_{\rm dbl}$ are the squared single-insertion (also known as the quadratic), and double-insertion contributions, respectively. 
All the $\mathscr{O}(\Lambda^{-4})$ contributions can be schematically written in terms of the amplitudes,
\begin{align}%
    d\sigma_{\rm quad} \sim |\mathcal{A}_{\rm (d6)}|^2, \qquad 
    d\sigma_{\rm dbl} \sim |\mathcal{A}_{\rm SM}\,\mathcal{A}_{\rm (d6)^2}|,
    \qquad
    d\sigma_{\rm d8} \sim |\mathcal{A}_{\rm SM}\,\mathcal{A}_{\rm (d8)}|,
\end{align}%
where the latter is the contribution arising from amplitudes with a single insertion of a dimension-eight operator interfering with the SM ones.
The construction of the SMEFT dimension-eight basis was explored in~\cite{Murphy:2020rsh,Li:2020gnx}, however, the systematic treatment of those operators is beyond the scope of this work.
In this work, we study all the contributions arising from dimension-six operators, including the particular case for which we examine the double-insertion ones, $d\sigma_{\rm dbl}$, in \cref{sec:double_insertions}.

The SM differential cross-section can be expanded in orders of the QCD and EW couplings, 
\begin{align}%
    d\sigma_{\rm SM} = \sum_{n,m} \alpha_s^n\,\alpha_{\rm w}^m \, d\sigma^{(n,m)}_{\rm SM}= 
    \sum_{i,j,k} \alpha_s^i\,\alpha^j\,\alpha_t^k \, d\sigma^{(i,j,k)}_{\rm SM},
\end{align}%
with $\alpha_{\rm w}$ collectively representing $\alpha$ and $\alpha_t$, and
\begin{equation}%
    \alpha_s = \frac{g_s^2}{4\pi}, \qquad   \alpha = \frac{e^2}{4\pi}, \qquad   \alpha_t = \frac{y_t^2}{4\pi}.
\end{equation}%
The four-top production occurs via the $gg-$ and $q\bar{q}$-initiated channels. 
Each of the amplitudes is a six-point diagram and thus has four couplings, i.e. $(i+j+k)=2$, and so we expand the $gg-$ and $q\bar{q}$-initiated SM amplitudes in terms of the QCD and EW couplings as follows:
\begin{subequations}%
\begin{align}%
    \MA^{(i,j,k)}_{{\rm SM},gg} 
    &=\alpha_s^2\,\MA^{(2,0,0)}_{{\rm SM},gg} 
    +\alpha_s
    \left(\,\alpha\,\MA^{(1,1,0)}_{{\rm SM},gg} 
    +\alpha_t\,\MA^{(1,0,1)}_{{\rm SM},gg}\right), \label{eq:sm_gg_amp}
    \\
    \MA^{(i,j,k)}_{{\rm SM},qq} 
    &=\alpha_s^2\,\MA^{(2,0,0)}_{{\rm SM},qq} 
    +\alpha_s\,\left(
    \alpha\,\MA^{(1,1,0)}_{{\rm SM},qq}
    +\alpha_t\,\MA^{(1,0,1)}_{{\rm SM},qq}\right)
    \label{eq:sm_qq_amp} \\
    &+\left(\alpha^2\,\MA^{(0,2,0)}_{{\rm SM},qq} 
    +\alpha^{3/2}\,\alpha_t^{1/2}\,\MA^{(0,3/2,1/2)}_{{\rm SM},qq}
    +\alpha\,\alpha_t\,\MA^{(0,1,1)}_{{\rm SM},qq}\right), \notag
\end{align}%
\end{subequations}%
with the term containing half-integer couplings, i.e. $\mathscr{O}(\alpha^{3/2}\,\alpha_t^{1/2})$, arising from diagrams containing a Higgs boson coupling to a top quark via one top-Yukawa vertex and a coupling to two EW bosons via one EW vertex. 
Each of the two $W$ bosons couples with a fermion line. 

\begin{figure}[h!]
    \centering
    \scalebox{0.9}{\begin{tikzpicture}
\begin{feynman}[small]
\vertex (a0) {};
\vertex[right = of a0] (a1) {};
\vertex[right = of a1] (a2) {$\bar{t}$};
\vertex[below = of a0] (c0) {$g$};
\vertex[right = of c0] (c1);
\vertex[right = of c1] (c2) {};
\vertex[right = of c2] (c3) {$\bar{t}$};
\vertex[right = of c3] (c4) {};
\vertex[right = of c4] (c5) {};
\vertex[below = of c0] (b0) {};
\vertex[right = of b0] (b1) {};
\vertex [blob] (b2) at (2,-2) {};
\vertex[right = of b2] (b3);
\vertex[right = of b3] (b4) {};
\vertex[right = of b4] (b5) {};
\vertex[below = of b0] (d0) {$g$};
\vertex[right = of d0] (d1);
\vertex[right = of d1] (d2) {};
\vertex[right = of d2] (d3) {$t$};
\vertex[right = of d3] (d4) {};
\vertex[right = of d4] (d5) {};
\vertex[below = of d0] (e0) {};
\vertex[right = of e0] (e1) {};
\vertex[right = of e1] (e2) {$t$};
\diagram*{
(c0) -- [gluon] (c1),
(d0) -- [gluon] (d1),
(c1) -- [fermion] (b2),
(c1) -- [anti fermion] (a2),
(d1) -- [anti fermion] (b2),
(d1) -- [fermion] (e2),
(b2) -- [anti fermion] (c3),
(b2) -- [fermion] (d3),
};
\end{feynman}
\end{tikzpicture}
\begin{tikzpicture}
\begin{feynman}[small]
\vertex (a0) {};
\vertex[right = of a0] (a1) {};
\vertex[right = of a1] (a2) {};
\vertex[right = of a2] (a3) {$\bar{t}$};
\vertex[right = of a3] (a4) {};
\vertex[below = of a0] (c0) {$q$};
\vertex[right = of c0] (c1) {};
\vertex [blob] (c2) at (2,-1){};
\vertex[right = of c2] (c3) {$t$};
\vertex[right = of c3] (c4) {};
\vertex[right = of c4] (c5) {};
\vertex[below = of c0] (b0) {};
\vertex[right = of b0] (b1) {};
\vertex[right = of b1] (b2);
\vertex[right = of b2] (b3);
\vertex[right = of b3] (b4) {};
\vertex[right = of b4] (b5) {};
\vertex[below = of b0] (d0) {$\bar{q}$};
\vertex[right = of d0] (d1){};
\vertex[right = of d1] (d2);
\vertex[right = of d2] (d3);
\vertex[right = of d3] (d4) {$\bar{t}$};
\vertex[right = of d4] (d5) {};
\vertex[below = of d0] (e0) {};
\vertex[right = of e0] (e1) {};
\vertex[right = of e1] (e2) {};
\vertex[right = of e2] (e3) {};
\vertex[right = of e3] (e4) {$t$};
\diagram*{
(c0) -- [fermion] (c2),
(d0) -- [anti fermion] (d2),
(c2) -- [fermion] (d2),
(d2) -- [gluon] (d3),
(c2) -- [anti fermion] (a3),
(c2) -- [fermion] (c3),
(d3) -- [anti fermion] (d4),
(d3) -- [fermion] (e4),
};
\end{feynman}
\end{tikzpicture}}
    \caption{Representative diagrams for four-top production with blobs representing the one dimension-six EFT insertion, in the $gg$-initiated production mode (\emph{left}) and in the $q\bar{q}$-initiated production mode (\emph{right}).}
    \label{fig:diags_EFT_4F}
\end{figure}
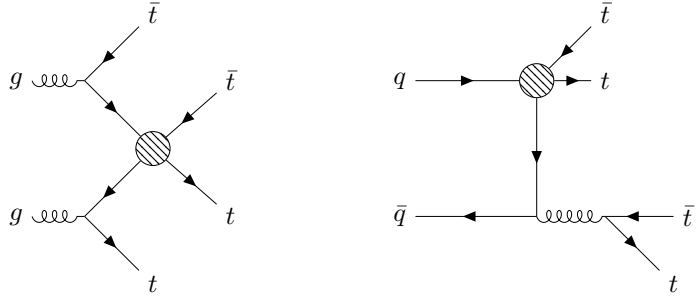

Moving to the EFT case, as an example, we show the cross-section expansion for one class of operators; the four-fermion ones. 
The insertions of a single dimension-six four-fermion operator in the amplitudes of both production channels are depicted in \cref{fig:diags_EFT_4F}.
The EFT linear interference cross-section can be decomposed in the same way as the SM,
\begin{align}%
    d\sigma_{\rm int} = \sum_{n,m} \alpha_s^n\,\alpha_{\rm w}^m \, d\sigma^{(n,m)}_{\rm int}= 
    \sum_{i,j,k} \alpha_s^i\,\alpha^j\,\alpha_{t}^k \, d\sigma^{(i,j,k)}_{\rm int},
\end{align}%
where each of the expanded partial cross-section is a sum of contributions from different WCs,
\begin{equation}%
    d\sigma_{\rm int}^{(n,m)} = c_{[r]} d\sigma^{(n,m)}_{{\rm int}\,[r]}.
\end{equation}%
The index $[r]$ runs over all possible dimension-six operators. 
For the discussion, we write the EFT linear interference cross-section in terms of the SM and EFT amplitudes,
\begin{align}%
    d\sigma_{\rm int}  &= d\sigma_{{\rm int},gg}  + d\sigma_{{\rm int},qq} \sim
    2\mathfrak{R}\lb\MA_{{\rm SM},gg}\, \MA_{{\rm EFT},gg}^\dagger\rb  + 
    2\mathfrak{R}\lb\MA_{{\rm SM},qq}\, \MA_{{\rm EFT},qq}^\dagger\rb,
\end{align}%
(we have omitted the PDF dependence for simplicity) with the four-fermion EFT amplitudes reading
\begin{subequations}%
\begin{align}%
    \MA^{(i,j,k)}_{{\rm EFT},gg,[\texttt{4F}]}
    &= \alpha_s \MA^{(1,0,0)}_{{\rm EFT},gg \,[\texttt{4F}]},  \label{eq:eft_gg_dim64f_amp}
    \\
    \MA^{(i,j,k)}_{{\rm EFT},qq\,[\texttt{4F}]}  
    &=\alpha_s \MA^{(1,0,0)}_{{\rm EFT},qq\,[\texttt{4F}]}   
    +\alpha \MA^{(0,1,0)}_{{\rm EFT},qq\,[\texttt{4F}]} 
    +\alpha_{t} \MA^{(0,0,1)}_{{\rm EFT},qq\,[\texttt{4F}]}.
    \label{eq:eft_qq_dim64f_amp}
\end{align}%
\end{subequations}%
We can then write the $gg$ and the $q\bar{q}$-initiated interference cross-section contributions induced by the four-fermion operators in terms of all the QCD and EW coupling orders,
\begin{subequations}
\begin{align}
    d\sigma_{{\rm int}, gg, \texttt{[4F]}}  
    &=\alpha_s^3\,d\sigma^{(3,0,0)}_{{\rm int},gg} 
    +\alpha_s^2\left(\,\alpha\,d\sigma^{(2,1,0)}_{{\rm int},gg} +\alpha_t\,d\sigma^{(2,0,1)}_{{\rm int},gg}\right). 
    \label{eq:gg_xsec_4f}
    \\
    d\sigma_{{\rm int}, qq, \texttt{[4F]}}  
    &=\alpha_s^3 \,d\sigma^{(3,0,0)}_{{\rm int}, qq} \notag\\
    &+\alpha_s^2\left(\alpha\,d\sigma^{(2,1,0)}_{{\rm int}, qq}
    +\alpha_t\,d\sigma^{(2,0,1)}_{{\rm int}, qq}\right)  
    \notag\\
    &+\alpha_s\left(\alpha^2\,d\sigma^{(1,2,0)}_{{\rm int}, qq}   
    +\alpha^{3/2}\,\alpha_{t}^{1/2} \,d\sigma^{(1,3/2,1/2)}_{{\rm int}, qq}
    +\alpha\alpha_t \,d\sigma^{(1,1,1)}_{{\rm int}, qq} 
    +\alpha_t^2 \,d\sigma^{(1,0,2)}_{{\rm int}, qq}\right) \notag \\
    &+(\alpha^3) \,d\sigma^{(0,3,0)}_{{\rm int}, qq}   
    +(\alpha^{5/2}\,\alpha_t^{1/2}) \,d\sigma^{(0,5/2,1/2)}_{{\rm int}, qq}  \notag\\
    &+(\alpha^2\,\alpha_t) \,d\sigma^{(0,2,1)}_{{\rm int}, qq}
    +(\alpha^{3/2}\,\alpha_t^{3/2}) \,d\sigma^{(0,3/2,3/2)}_{{\rm int}, qq}
    +(\alpha\,\alpha_t^2) \,d\sigma^{(0,1,2)}_{{\rm int}, qq}.
    \label{eq:qq_xsec_4f} 
\end{align}
\end{subequations}
\section{Hierarchy of inclusive predictions}
\label{sec:inclusive}
This section presents the numerical results from the complete LO SMEFT predictions of the \tttt production process at $\sqrt{s}=13$ TeV for the LHC, and at $\sqrt{s}=100$ TeV for future circular $pp$ colliders.
The computations were performed via \texttt{MadGraph5\_aMC@NLO}~\cite{Alwall:2014hca,Frederix:2018nkq} with the use of the \texttt{SMEFTatNLO} model~\cite{Degrande:2020evl}, and with the mass of the top quark, $m_{t}$, set to 172 GeV. 
Since \texttt{MadGraph5\_aMC@NLO} does not evolve the operator coefficients, and as recommended, the factorisation ($\mu_F$), renormalization ($\mu_R$) scales are fixed to 340 GeV$\sim(4m_{t})/2$\footnote{~The EFT renormalisation scale (\texttt{mueft}) parameter of the \texttt{SMEFTatNLO} in \texttt{MadGraph5\_aMC@NLO} is not relevant unless the running of the EFT coefficients is included.
We do not consider the running of the EFT coefficients in this work.}.
The proton PDFs and their uncertainties are evaluated employing reference sets and error replicas from the \texttt{NNPDF3.1 NLO} global fit~\cite{NNPDF:2017mvq} in the five flavour scheme (5FS), in which the bottom quark is taken to be massless.
Unless otherwise explicitly mentioned, no parton-level cuts are imposed. 
Before discussing the SMEFT results, we first show the decomposition of the four-top SM cross-sections in \cref{tab:sm_decomposition}.
\begin{table}[h!]
    \resizebox{\textwidth}{!}{
\renewcommand{\arraystretch}{1.1}
\centering
\begin{tabular}{ccccccc|c}
\hline
$\sqrt{s}$ &
$\mathscr{O}(\alpha_{s}^{4})$ & 
$\mathscr{O}(\alpha_{s}^{3}\alpha)$ &
$\mathscr{O}(\alpha_{s}^{3}\alpha_{t})$ & 
$\sum_{n,m}\mathscr{O}(\alpha_{s}^{2}\alpha^{n}\alpha^{m}_{t})$ & 
$\sum_{n,m}\mathscr{O}(\alpha_{s}\alpha^{n}\alpha^{m}_{t})$ &      
$\sum_{n,m}\mathscr{O}(\alpha^{n}\alpha^{m}_{t})$ &
Inclusive
\\\hline
13 TeV           
& 6.15                                                                      
& -1.44                                                                     
& -0.58                                                                     
& 2.33                                                                         
& $\times$        
& $\times$
& 6.46
\\
100 TeV          
&2570                                                           
&-313                                                              
&-197                                                               
&753                                                              
&$\times$                                                              
&$\times$
&2812
\\\hline
\end{tabular}}

    \caption{Entries in each column correspond to the LO \tttt SM cross-section [fb] in line of $\alpha_s$ order, at the LHC ($\sqrt{s}=13$) and FCC-hh ($\sqrt{s}=100$).
    For contributions at $\mathscr{O}(\alpha_{s}^N)$ with $N=0-2$, we sum all possible weak and Yukawa coupling combinations and present the total cross-section at a given order in $\alpha_{s}$. 
    The notation ``$\times$'' denotes negligible contributions. 
    The column titled ``Inclusive'' shows the total cross-section.}
    \label{tab:sm_decomposition}
\end{table}

The strengths of the linear interference of all the dimension-six SMEFT operators belonging to the four-fermion, two-fermion, and purely-bosonic classes are presented. 
The interference strength is the interference cross-section in fb with the corresponding WC individually set to unity, and the scale of new physics $\Lambda$ is fixed to 1(3) TeV, for the $\sqrt{s}=13(100)$ TeV scenario.
In presenting the results, the four-fermion operators are categorised into two sub-classes; contact terms involving four heavy quarks (4-heavy) and contact terms involving two heavy and two light quarks (2-heavy-2-light). 
Respectively, those insertions are depicted by the blobs shown in the \emph{left} and \emph{right} diagrams of \cref{fig:diags_EFT_4F}. 
On the other hand, since not all the two-fermion and purely-bosonic operators are relevant to \tttt production, only the \emph{contributing} operators from these classes are presented;
\begin{equation}%
    \{\mathcal{O}_{t\phi},\mathcal{O}_{tZ},\mathcal{O}_{tW},\mathcal{O}_{tG},\mathcal{O}_{\phi Q}^{(-)},\mathcal{O}_{\phi t},\mathcal{O}_{G},\mathcal{O}_{\phi G}\}.
    \label{eq:relevant_operators}
\end{equation}%

\begin{figure}[h!]
    \centering
    \includegraphics[width=0.5\textwidth]{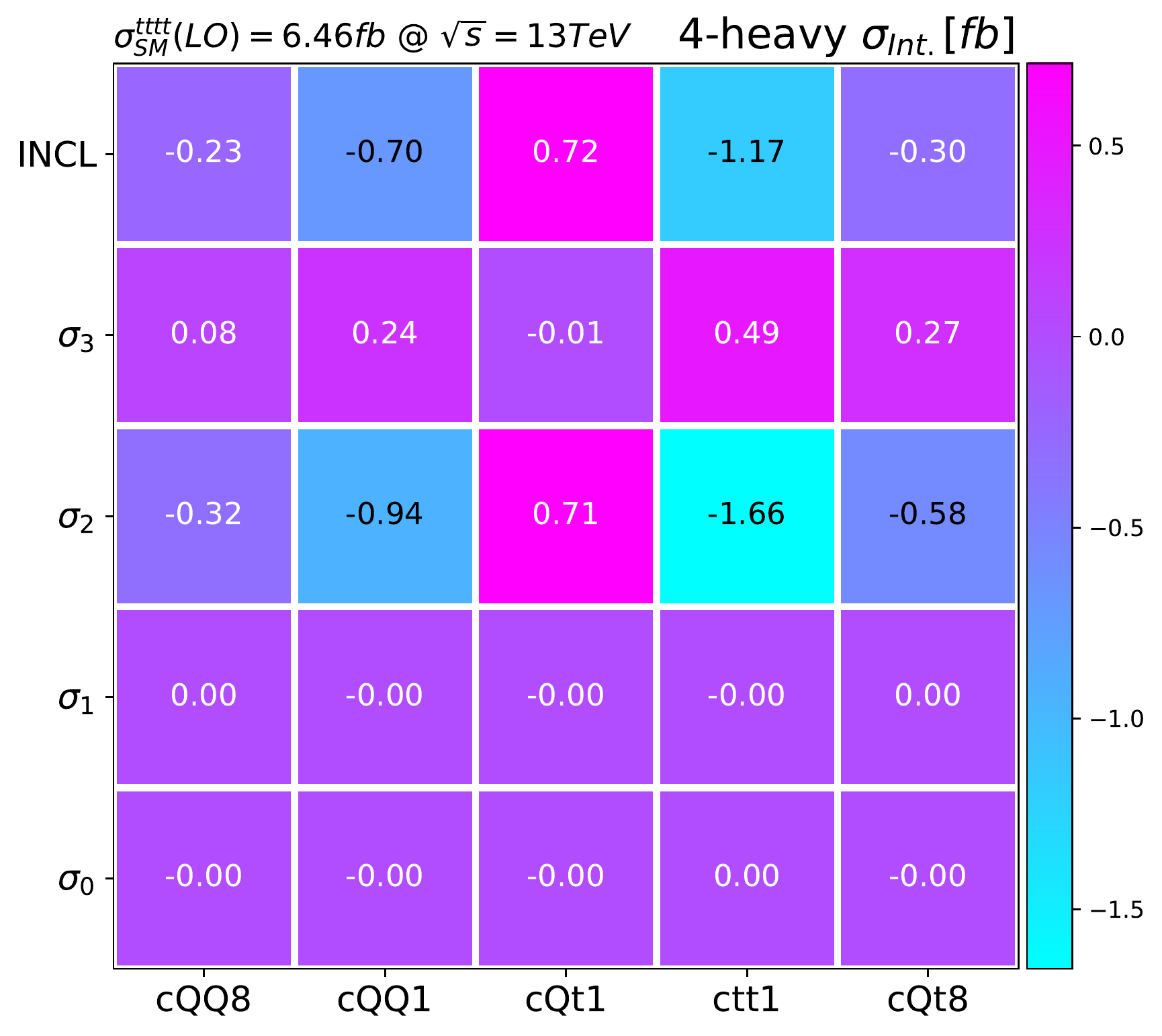}
    \caption{\label{fig:dim64f_4heavy_map_13TeV}
    Depiction of the interference strength of the 4-heavy operators.
    The columns denote the WCs in their UFO notation.
    The top row shows the total inclusive predictions, i.e. summing all the QCD and EW-induced contributions. 
    Each of the subsequent rows indicates the summation of all terms at a given order of $\alpha_{s}$ (an example is given in the main text).
    The predictions are obtained at $c=1$, $\sqrt{s}=13$ TeV, and $\Lambda=1$ TeV. 
    The SM LO cross-section at $\sqrt{s}=13$ TeV is presented for reference.}
\end{figure}

\begin{figure}[h!]
    \centering
    \includegraphics[width=\textwidth]{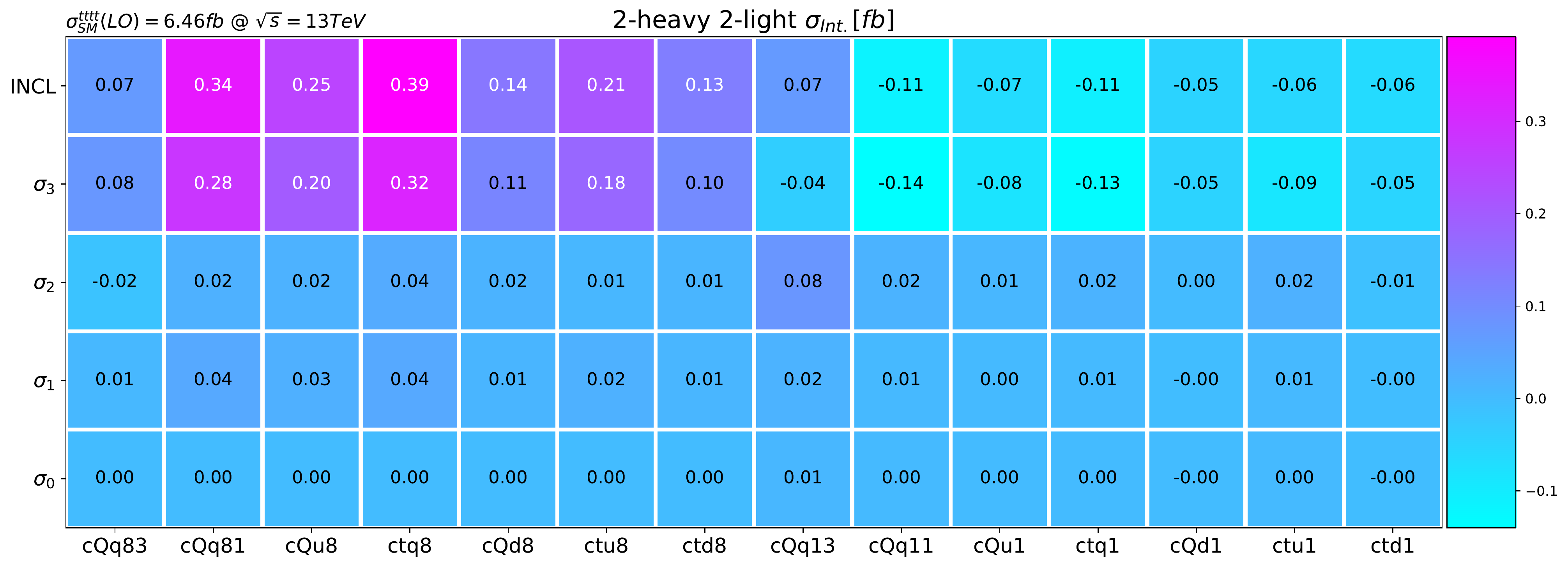}
    \caption{\label{fig:dim64f_2heavy_map_13TeV} 
    Same as \cref{fig:dim64f_4heavy_map_13TeV} but for the 2-heavy 2-light operators}
\end{figure}

\begin{figure}[h!]
    \centering
    \includegraphics[width=0.6\textwidth]{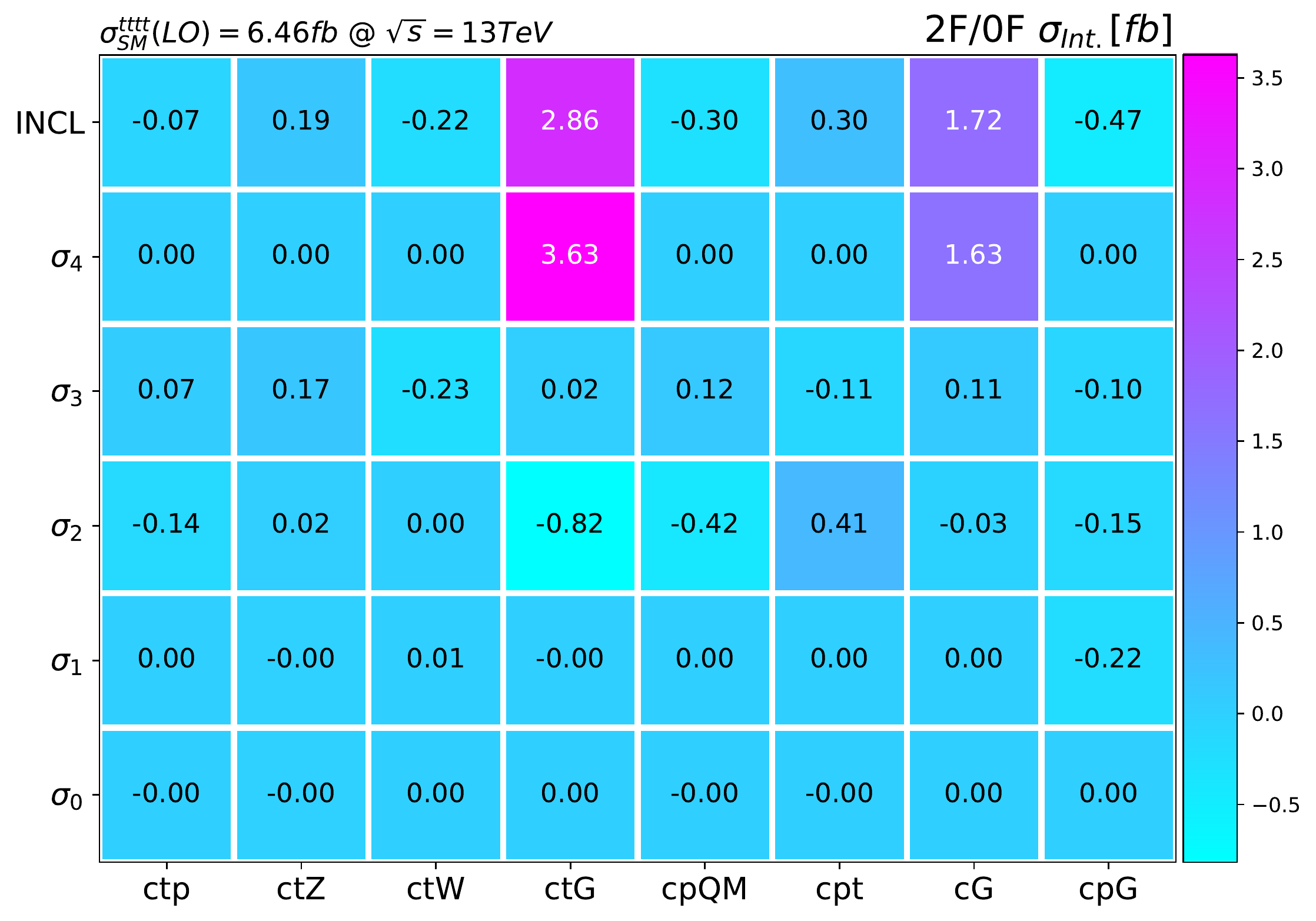}
    \caption{\label{fig:dim62f_dim60f_relevant_map_13TeV} 
    Same as \cref{fig:dim64f_4heavy_map_13TeV} but for the contributing operators (see \cref{eq:relevant_operators}).}
\end{figure}

\begin{figure}[h!]
    \centering
    \includegraphics[width=0.5\textwidth]{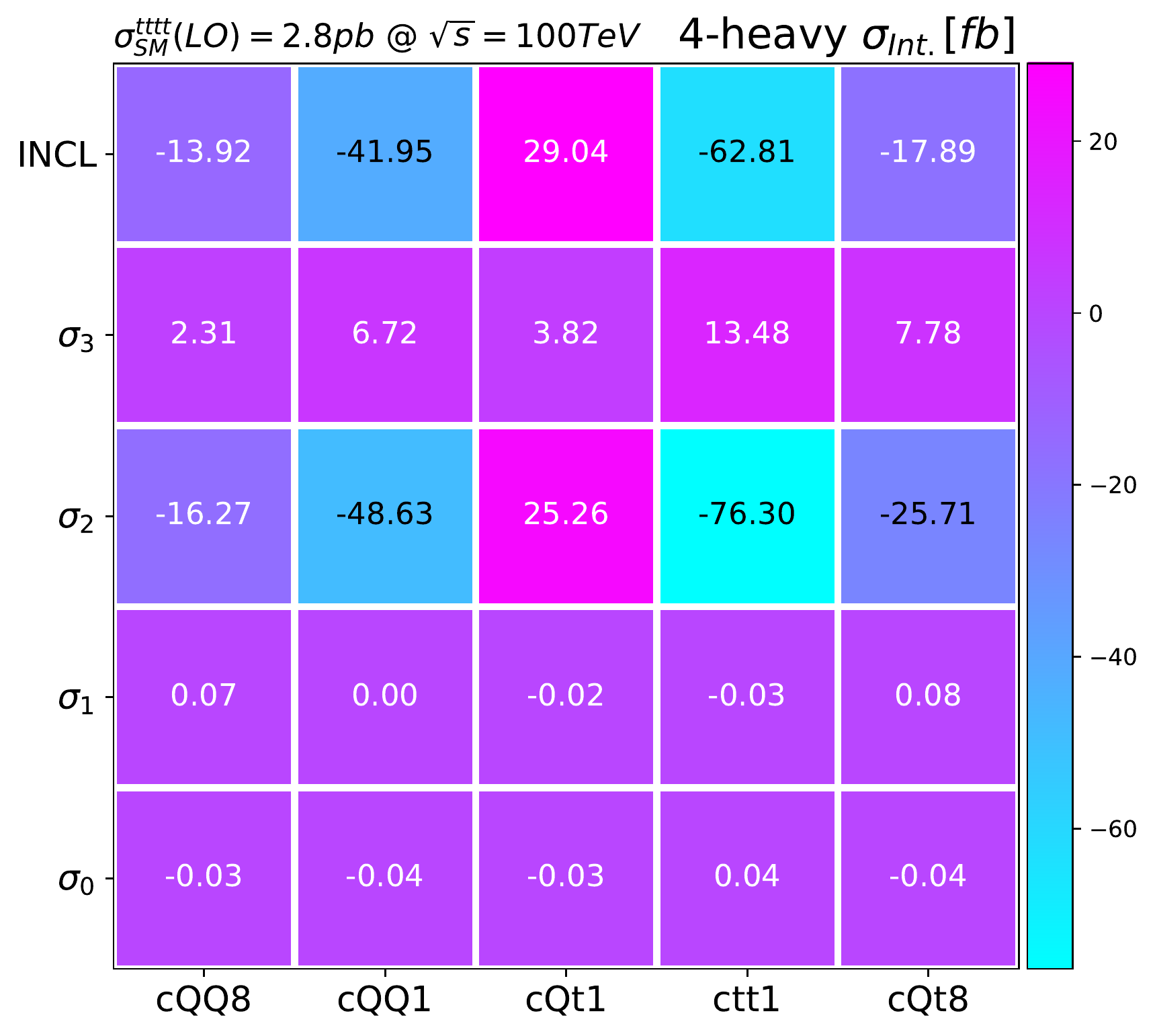}
    \caption{\label{fig:dim64f_4heavy_map_100TeV} Same as \cref{fig:dim64f_4heavy_map_13TeV} but for the FCC-hh scenario. 
    The predictions are obtained at $c=1$, $\sqrt{s}=100$ TeV, and $\Lambda=3$ TeV. 
    The SM LO cross-section at $\sqrt{s}=100$ TeV is presented for reference.}
\end{figure}

\begin{figure}[h!]
    \centering
    \includegraphics[width=\textwidth]{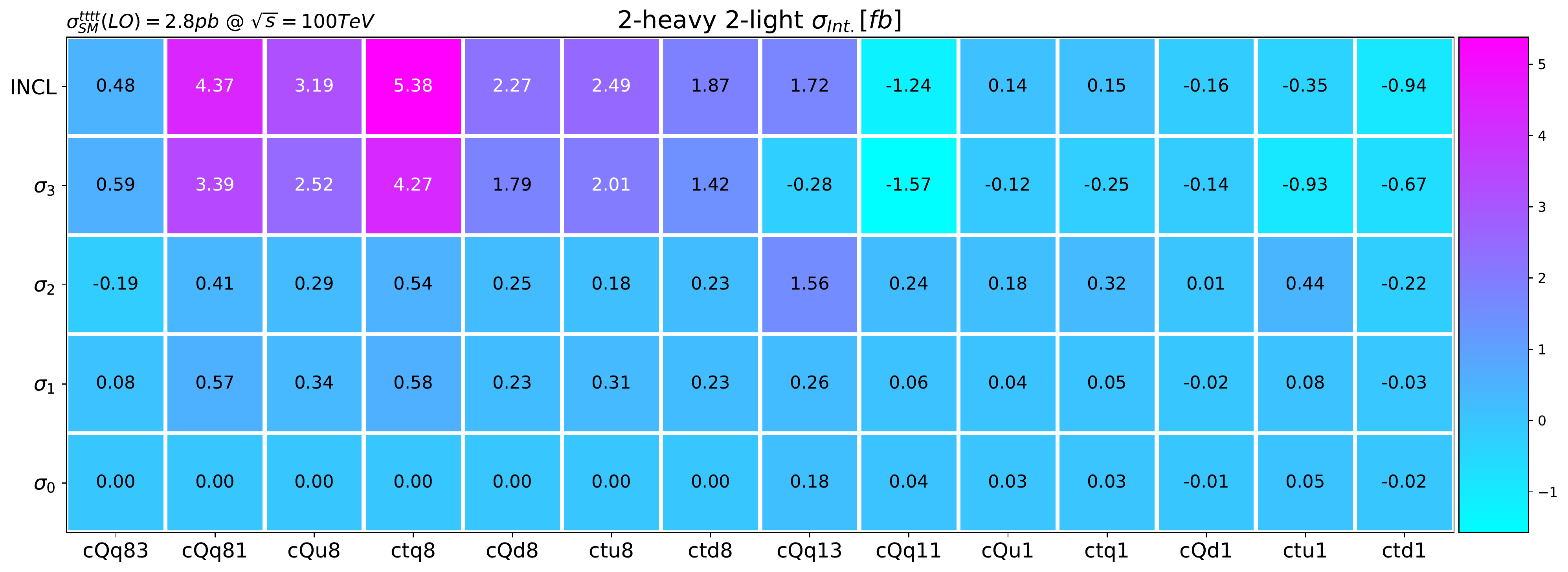}
    \caption{\label{fig:dim64f_2heavy_map_100TeV} Same as \cref{fig:dim64f_4heavy_map_100TeV} but for the 2-heavy 2-light operators.}
\end{figure}

\begin{figure}[h!]
    \centering
    \includegraphics[width=0.6\textwidth]{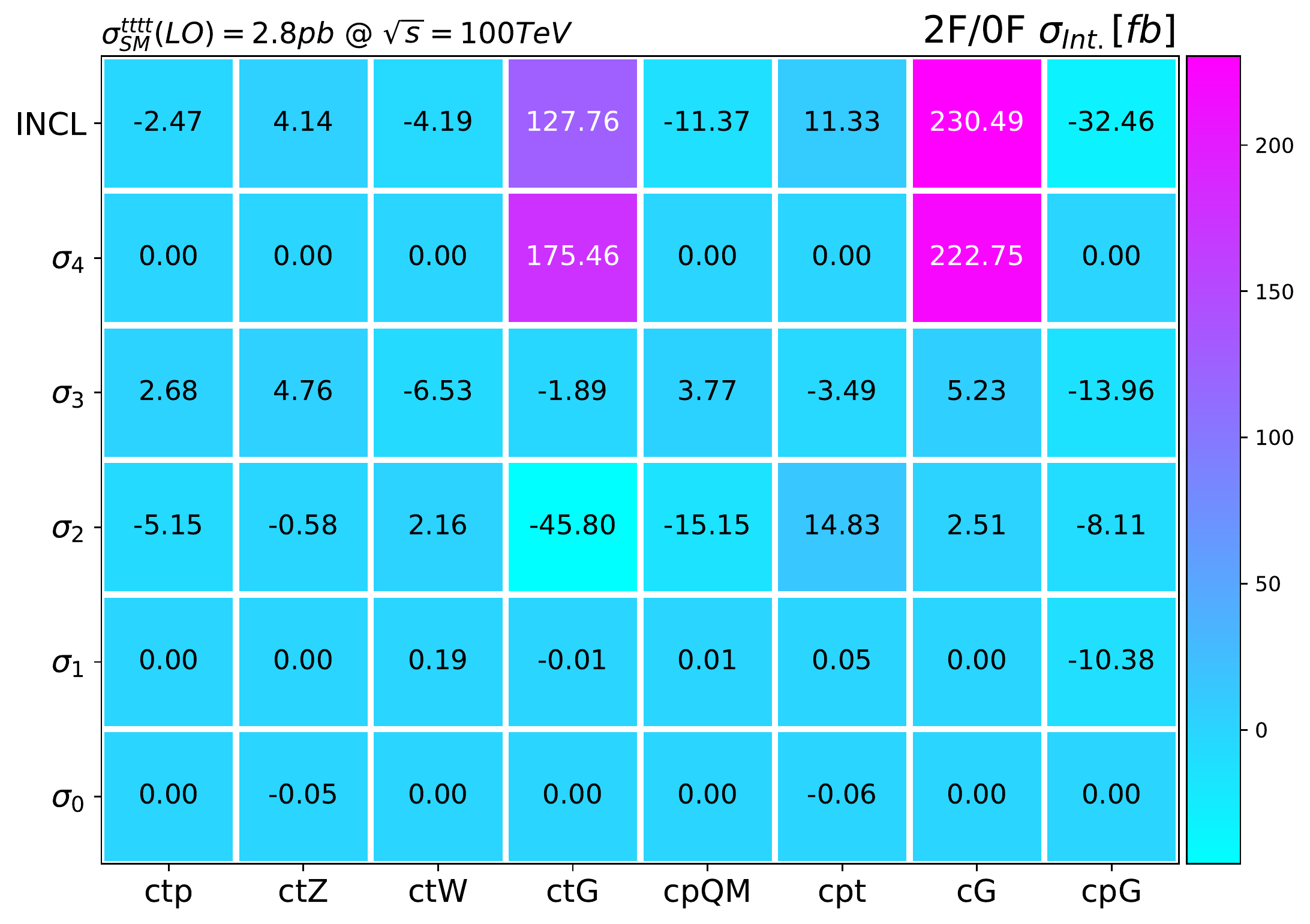}
    \caption{\label{fig:dim62f_dim60f_relevant_map_100TeV} Same as \cref{fig:dim64f_4heavy_map_100TeV} but for the contributing operators.}
\end{figure}

The total inclusive interference cross-section in the four-fermion case can be defined as follows:
\begin{equation}%
    \sigma_{\rm incl.}=\sigma_{3}+\sigma_{2}+\sigma_{1}+\sigma_{0},
    \label{eq:inclusive_cross-section}
\end{equation}%
where $\sigma_{i}$ with $i=3,2,1,0$ denotes the contributions to $\sigma_{\rm incl.}$ arising from terms with order $\alpha_{s}^{i}$ in the cross-section expansion.
For example, the $\sigma_{2}$ term denotes the interference cross-section arising \emph{only} from the formally subleading terms in $\alpha_{s}$, i.e. $\mathscr{O}(\alpha_{s}^{2}\alpha)$ and $\mathscr{O}(\alpha_{s}^{2}\alpha_{t})$ in \cref{eq:gg_xsec_4f} and \cref{eq:qq_xsec_4f}, which can be collectively written as follows:
\begin{equation}%
    \sigma_{2}\equiv\alpha_s^2\left(\,\alpha\,\sigma^{(2,1,0)}_{{\rm int}} +\alpha_t\,\sigma^{(2,0,1)}_{{\rm int}}\right).
    \label{eq:def_of_inclusive_xsec}
\end{equation}%
The interference strength is depicted in the heat maps presented in \cref{fig:dim64f_4heavy_map_13TeV}-\ref{fig:dim62f_dim60f_relevant_map_100TeV}, the columns correspond to the operators' coefficients. 
The top row shows the total inclusive interference cross-section $\sigma_{\rm incl}$ labelled \texttt{INCL}, while subsequent rows correspond to the separate contributions arranged in order of $\alpha_{s}$, in line with the example of \cref{eq:def_of_inclusive_xsec}.
\paragraph{$\mathbf{\sqrt{s}=13}$ TeV}
Starting with the 4-heavy operators in \cref{fig:dim64f_4heavy_map_13TeV}, we observe that for \emph{all} of them, the dominant interference is the one arising from formally subleading orders, $\sigma_{2}$.
This observation contrasts the ``naive'' expectation that leading (purely QCD-induced) terms would provide the highest contribution to the cross-section through $\sigma_{3}$ and consequently highlights the significance of the EW $tt \to tt$ scattering present in \tttt production at LO (see \cref{fig:4tops_LO_diags_scattering}).
The significance of such EW scattering in four-top production has been pointed out in the NLO SM computation of Ref.~\cite{Frederix:2017wme}.
It is worth noting that such naive expectation not only underestimates the interference strength of the 4-heavy operators but also generates the `wrong' sign of the interference structure. 
That is, $\sigma_{2}$ for all the 4-heavy operators has the opposite sign of $\sigma_{3}$.
The former dictates the overall sign of the inclusive predictions.
Furthermore, the lower-order-$\alpha_{s}$ cross-sections, i.e. $\sigma_{n}$ where $n<2$, are heavily suppressed, rendering the consideration of cross-section contributions only down to $\sigma_{2}$ enough to attain reliable predictions for this set of operators.
Finally, for the 4-heavy operators, the colour-singlets, i.e. \QQSi{}, \QtSi{}, and \ttSi{}, are observed to have a stronger interference compared to the colour-octets, \QQOct{} and \QtOct{}, we analyse this effect later in our discussion of differential predictions.

Moving on to the 2-heavy-2-light operators, in \cref{fig:dim64f_2heavy_map_13TeV}, we observe, except for \TriSi{}, the interference strength in this class is dominated by the formally leading $\sigma_{3}$ cross-section.
This hints at the EW scattering effects being less critical in the interference with $q\bar{q}$-initiated EFT amplitudes.
The insertions of the 2-heavy-2-light operators are only present in the $q\bar{q}$-initiated production shown in the \emph{right} diagram of \cref{fig:diags_EFT_4F}.

Finally, the interference strength of the contributing operators is presented in \cref{fig:dim62f_dim60f_relevant_map_13TeV}. 
Due to the model normalisation in \cref{eq:dim62f_op_smeftatnlo} and \cref{eq:dim60f_op_smeftatnlo}, it is immediately apparent that only \ctG{} and \cG{} have a non-vanishing $\sigma_{4}$.
This is expected because a contribution to the cross-section at this order in $\alpha_{s}$ is not available for the other contributing operators; therefore, their leading cross-sections are $\sigma_{3}$.
In contrast to the \ctW{} and \ctZ{} dipoles, the contributing two-fermion operators, \ctp{}, \cpQM{}, and \cpt{}, have formally subleading dominant cross-sections.
The \ctG{} operator even though dominating at $\sigma_{4}$, has a non-negligible $\sigma_{2}$.
Finally, and in complete contrast to \cG, the interference strength of \cpG{} tends to be inversely proportional to orders in $\alpha_{s}$, in other words, proportional to the number of EW propagators.
\paragraph{$\mathbf{\sqrt{s}=100}$ TeV}
The only difference between the LHC and the FCC-hh computations is that we fix the scale of new physics $\Lambda$ to 3 TeV for the latter.
This is intended to ensure a reliable expansion of the EFT series given the high collision energy of FCC-hh.
The interference strength at the FCC-hh scenario from the 4-heavy, 2-heavy-2-light, and the contributing operators are presented in \cref{fig:dim64f_4heavy_map_100TeV}, \cref{fig:dim64f_2heavy_map_100TeV}, and \cref{fig:dim62f_dim60f_relevant_map_100TeV}, respectively. 
Apart from the expected scaling of the cross-sections in the FCC-hh scenario, we see a similar pattern across the board when comparing to the LHC study, albeit with some slight differences:
the $\sigma_{3}$ interference of \QtSi{} has an opposite sign in the 100 TeV scenario.
The \QuSi{} and \tqSi{} operators have a slightly dominant $\sigma_{2}$ in contrast to the LHC case where the dominant cross-section is $\sigma_{3}$. 
Finally, for \cpG{}, and while $\sigma_{1}$ is still significant in parallel to the LHC scenario, the $\sigma_{3}$ interference is slightly stronger in the 100 TeV scenario. 

In summarising this section, we present \cref{tab:summary_inclusive} in which all the operators align with their most dominant cross-section contributions.
\begin{table}[h!]
    \begin{center}
\renewcommand{\arraystretch}{1.}
\begin{tabular}{lcccc}\hline
             & 4H & 2L2H & 2F & 0F \\\hline\hline
$\sigma_4$ &  $\times$ &  $\times$ & {\color{blue}$c_{tG}$}  &  $c_G$  \\\hline
\multirow{2}{*}{$\sigma_3$} & \multirow{2}{*}{-} &
$c_{Qq}^{83},c_{Qu}^{8},c_{tq}^{8},c_{Qd}^{8},c_{tu}^{8},c_{td}^{8},c_{Qq}^{81}$ & \multirow{2}{*}{$c_{t\varphi},c_{tZ},c_{tW}$}& \multirow{2}{*}{-}\\ & & \multicolumn{1}{l}{$c_{Qq}^{11},c_{Qu}^{1},c_{tq}^{1},c_{Qd}^{1},c_{tu}^{1},c_{td}^{1}$} & & \\\hline
$\sigma_2$ & $c_{QQ}^8,c_{QQ}^1,c_{Qt}^8,c_{Qt}^1,c_{tt}^1$&$c_{Qq}^{31}$
&  $c_{\varphi t},c_{\varphi Q}^{(-)}$ & - \\
$\sigma_1$ &  - & - &- &{\color{blue}$c_{\varphi G}$}\\
$\sigma_0$ & - & - & -& -\\\hline
\end{tabular}
\end{center}
    \caption{Indication of the most significant contribution to the total cross-section of each operator at $\sqrt{s}=13$ TeV. 
    Entries labelled $\times$ denote such coupling order is not allowed for the given class of operators.
    The blue colour denotes operators with contributions not only dominant at this given order in $\alpha_s$, but also other (higher or lower) orders in $\alpha_s$ are significant enough that they can alter the total rate if not considered.}
    \label{tab:summary_inclusive}
\end{table}
Furthermore, we put together all operators featuring an unexpected enhancement to their cross-sections from formally subleading terms,
\begin{equation}
    \textit{all}\,\,\text{4-heavy}\qquad \text{and}\qquad
    \{\mathcal{O}_{Qq}^{3,1},\mathcal{O}_{t\phi},\mathcal{O}_{tG},\mathcal{O}_{\phi Q}^{(-)},\mathcal{O}_{\phi t},\mathcal{O}_{\phi G}\}.
    \label{eq:good_operators}
\end{equation}
Hereafter, we refer to this group of operators as the \emph{`non-naive'} ones.
More precisely, we define non-naive operators as ones for which \emph{any} of their formally non-leading terms is significant in estimating their total interference cross-section.
\section{Differential predictions}\label{sec:differential}
\begin{figure}[h!]
    \centering
    \includegraphics[trim=1.8cm 4.2cm 0.2cm 0.0cm, clip,width=.32\textwidth]{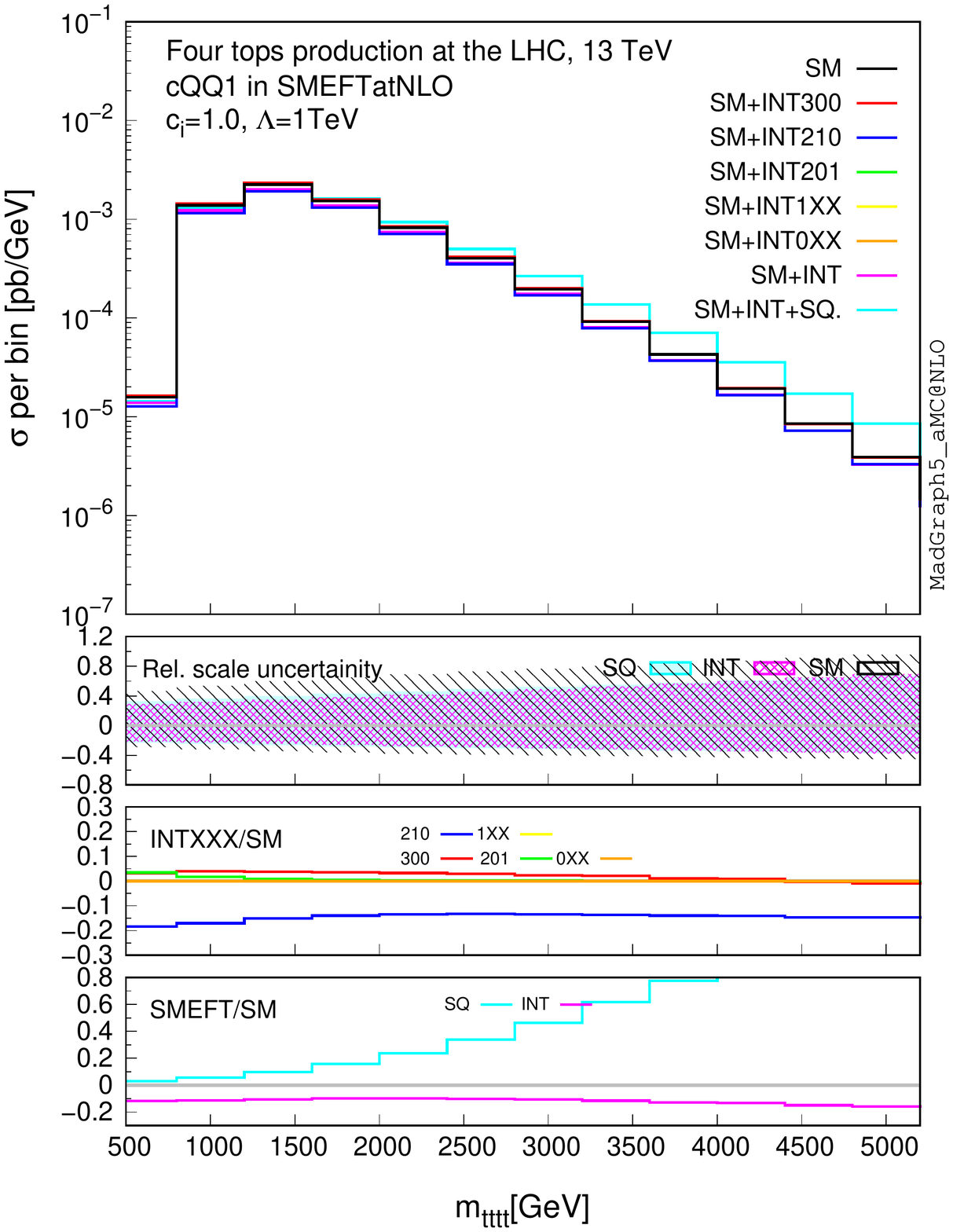}
    \includegraphics[trim=1.8cm 4.2cm 0.2cm 0.0cm, clip,width=.32\textwidth]{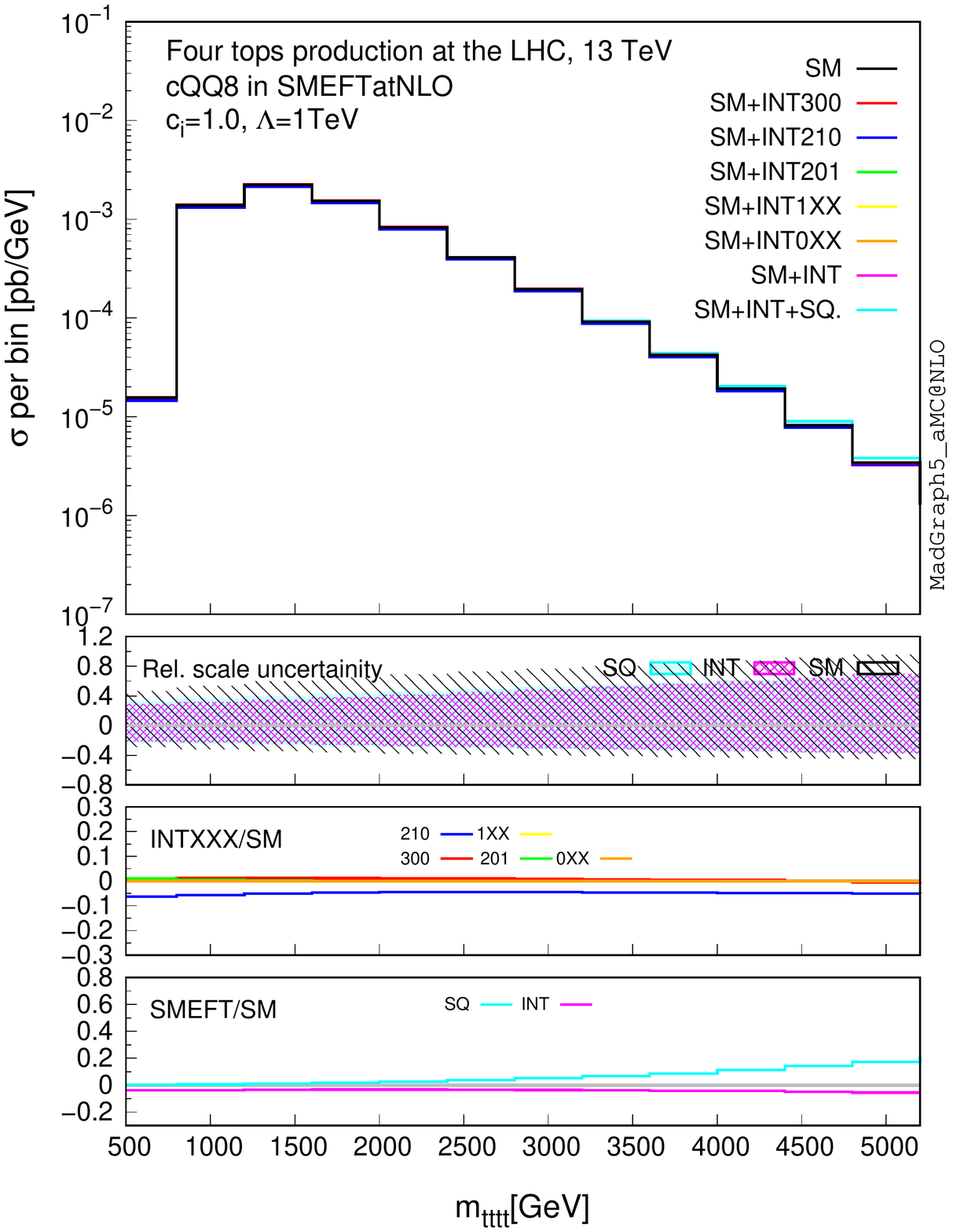}
    \includegraphics[trim=1.8cm 4.2cm 0.2cm 0.0cm, clip,width=.32\textwidth]{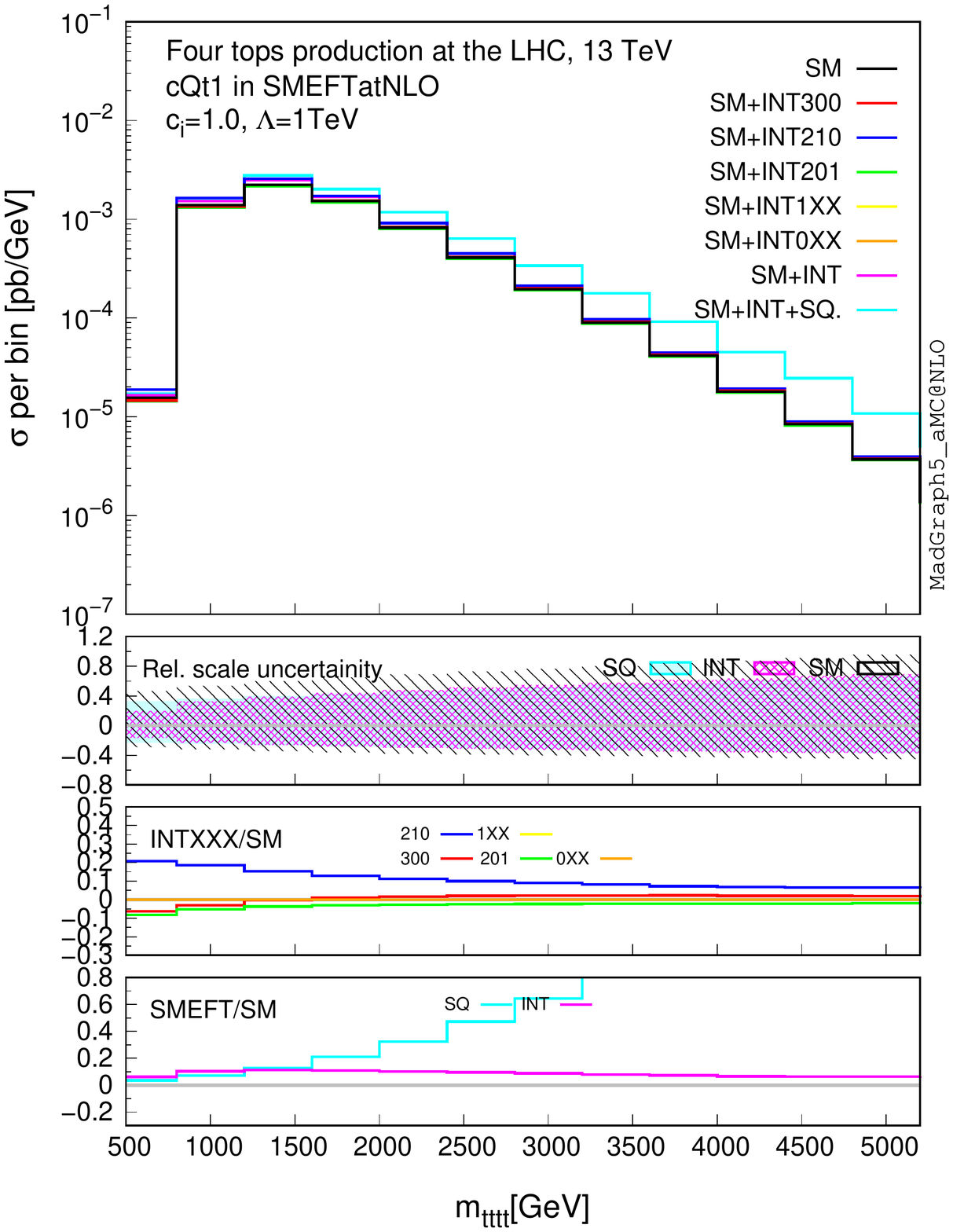}\\
    \includegraphics[trim=1.8cm 4.2cm 0.2cm 0.0cm, clip,width=.32\textwidth]{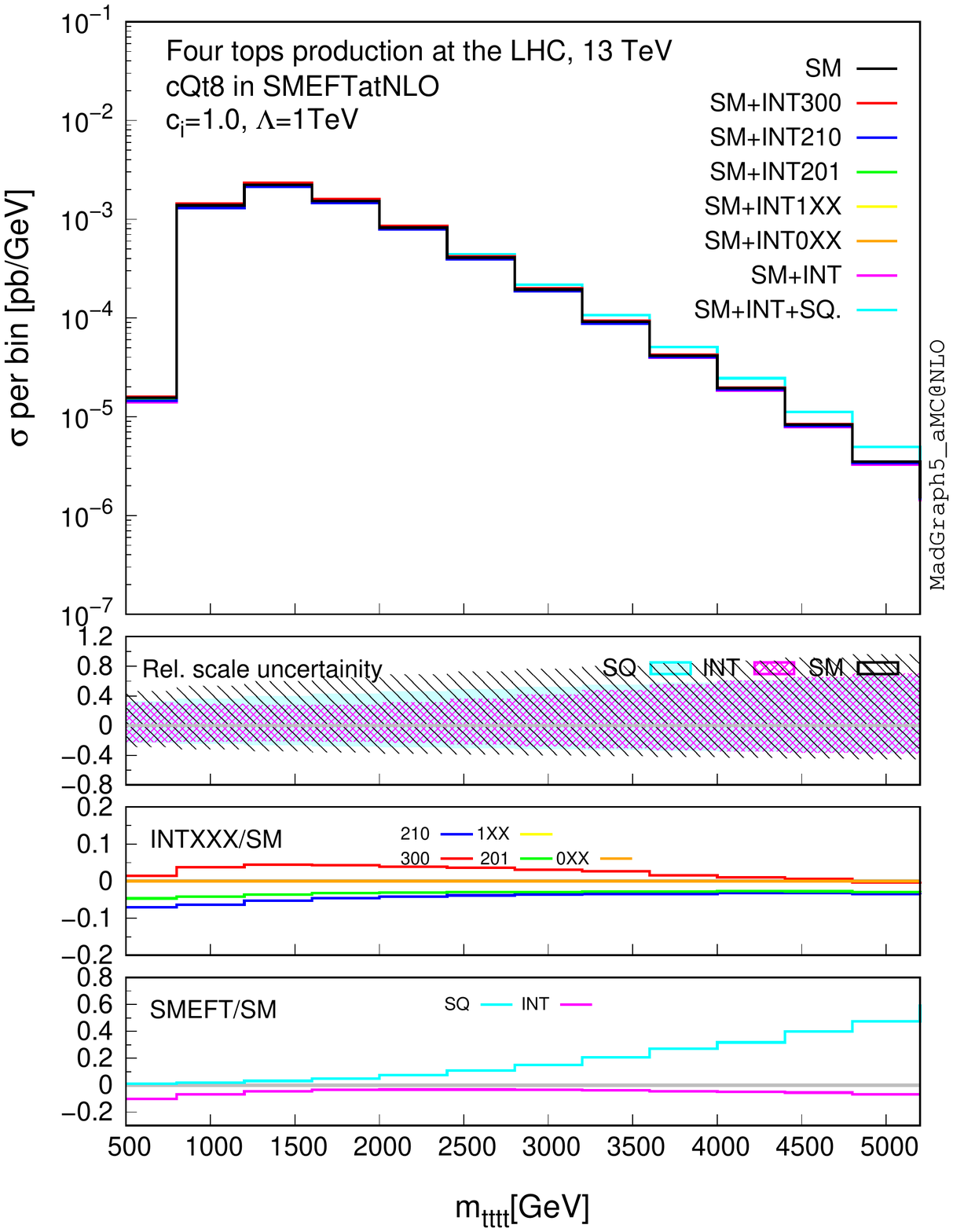}
    \includegraphics[trim=1.8cm 4.2cm 0.2cm 0.0cm, clip,width=.32\textwidth]{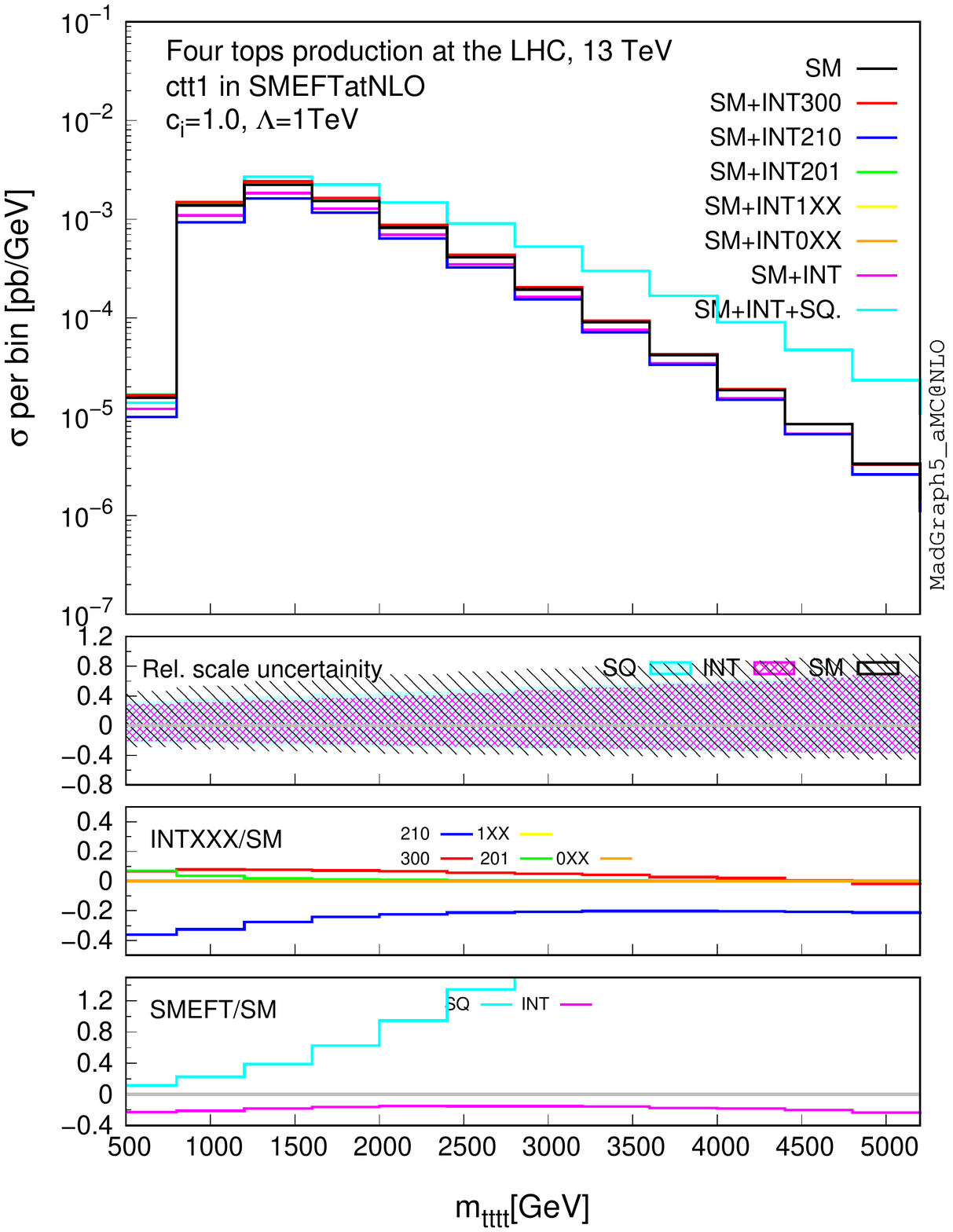}
    \includegraphics[trim=1.8cm 4.2cm 0.2cm 0.0cm, clip,width=.32\textwidth]{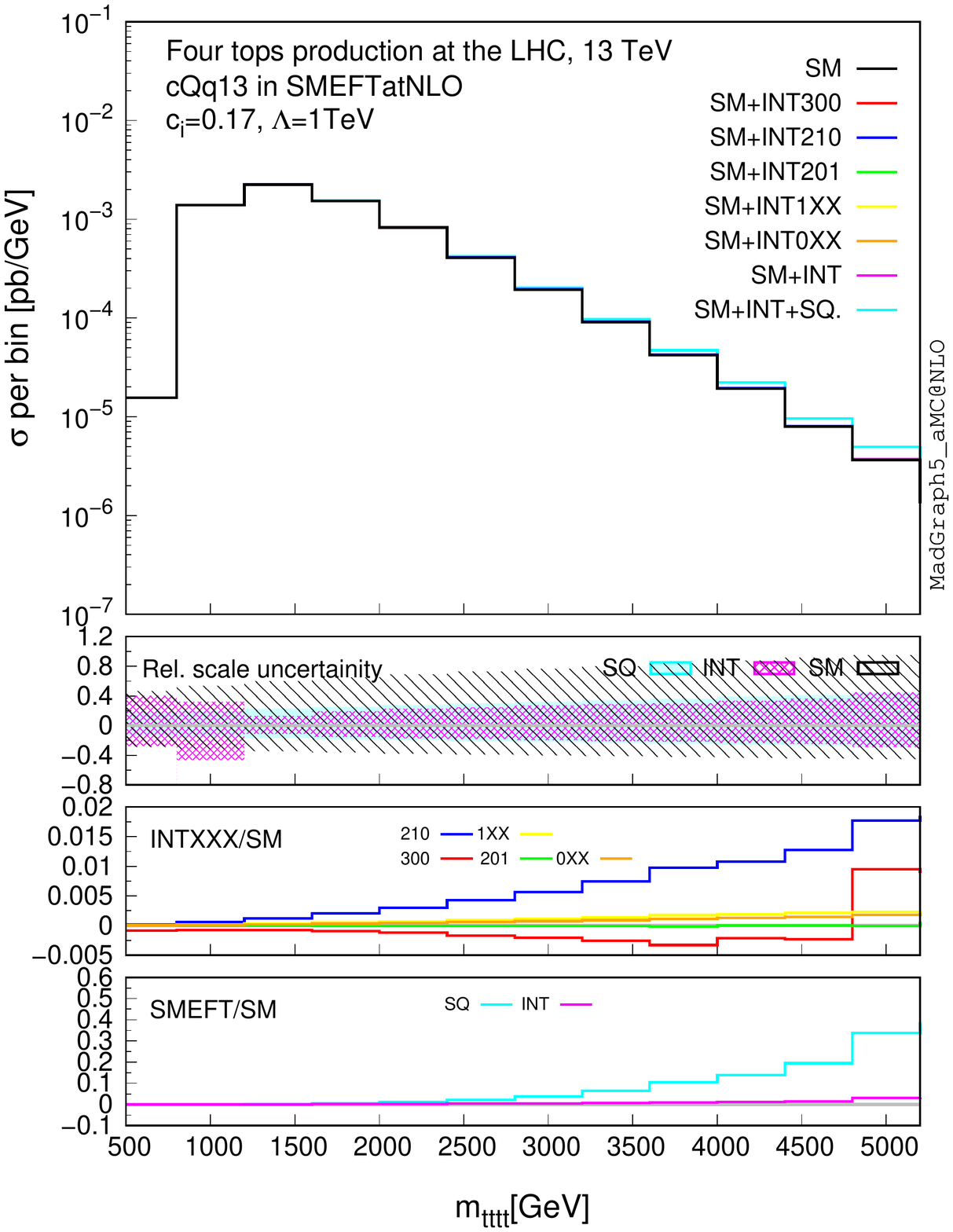}
    \caption{\label{fig:dim64f_good_diff_13tev} Differential EFT predictions at $\sqrt{s}=13$ TeV shown in the invariant mass bins of the four top quark system, $m_{tttt}$, for the set of non-naive four-fermion operators, i.e. all 4-heavy operators and the \TriSi{} operator. 
    The values of the coefficients are extracted from the global fit study of Ref.~\cite{Ethier:2021bye}, and $\Lambda$ is set to 1 TeV. 
    The first inset displays the relative scale uncertainties individually calculated for the SM, the interference, and the squared EFT contributions. 
    The second inset shows the ratio of the interference at each order in $\alpha_{s}$ to the SM prediction.
    The last inset shows the ratio of the total interference and the total squared contributions to the SM.}
\end{figure}

\begin{figure}[h!]
    \centering
    \includegraphics[trim=1.8cm 4.2cm 0.2cm 0.0cm, clip,width=.32\textwidth]{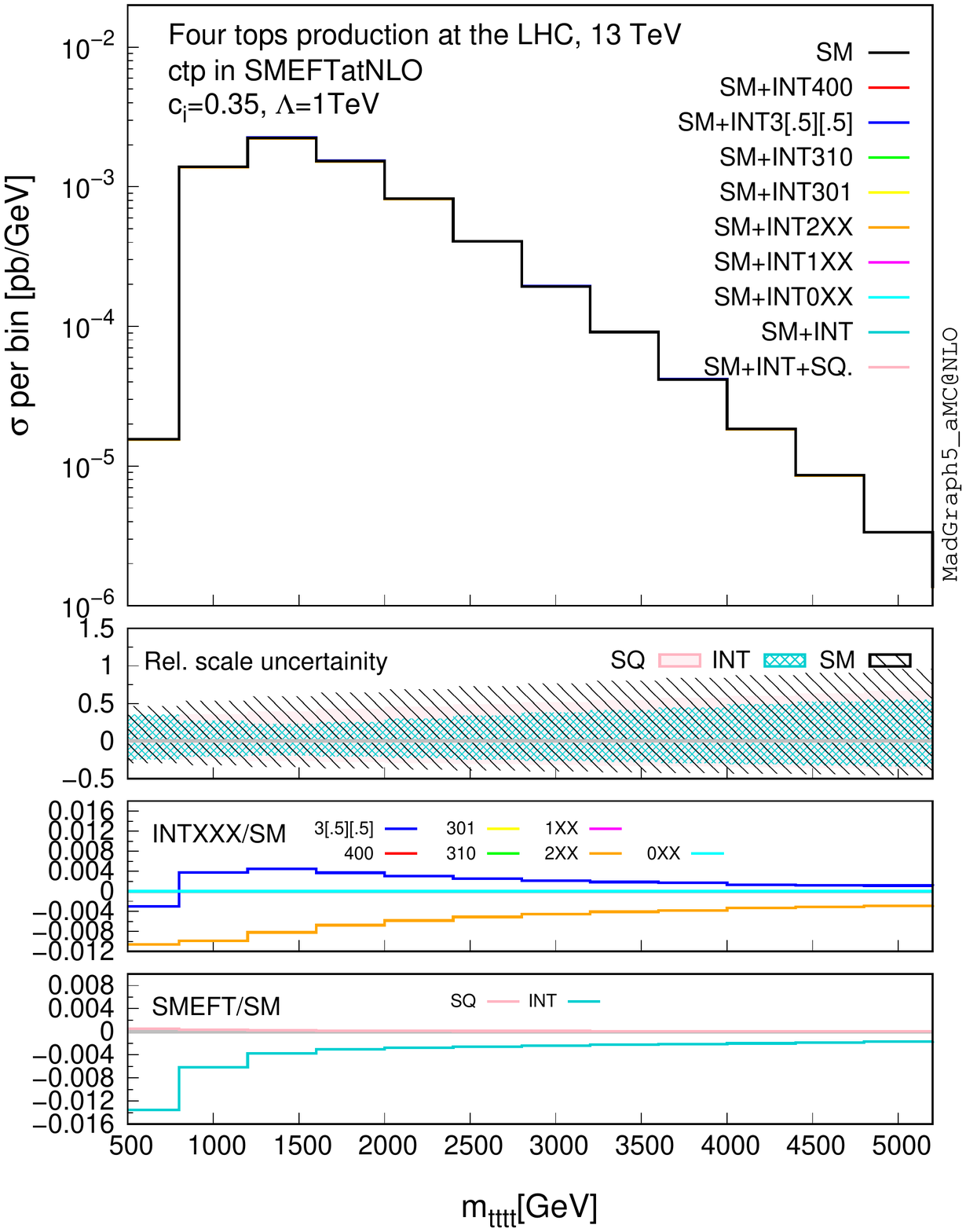}
    \includegraphics[trim=1.8cm 4.2cm 0.2cm 0.0cm, clip,width=.32\textwidth]{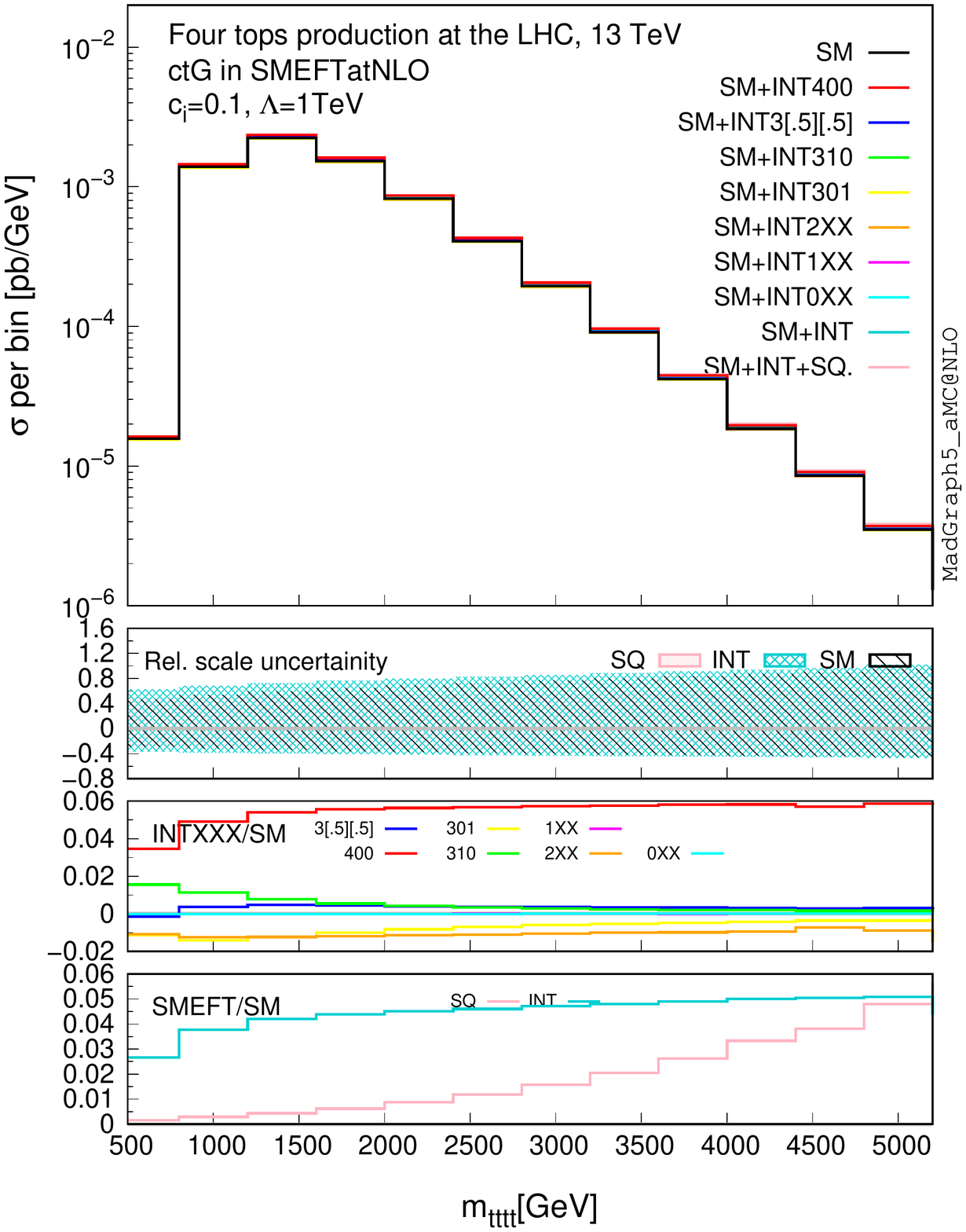}\\
    \includegraphics[trim=1.8cm 4.2cm 0.2cm 0.0cm, clip,width=.32\textwidth]{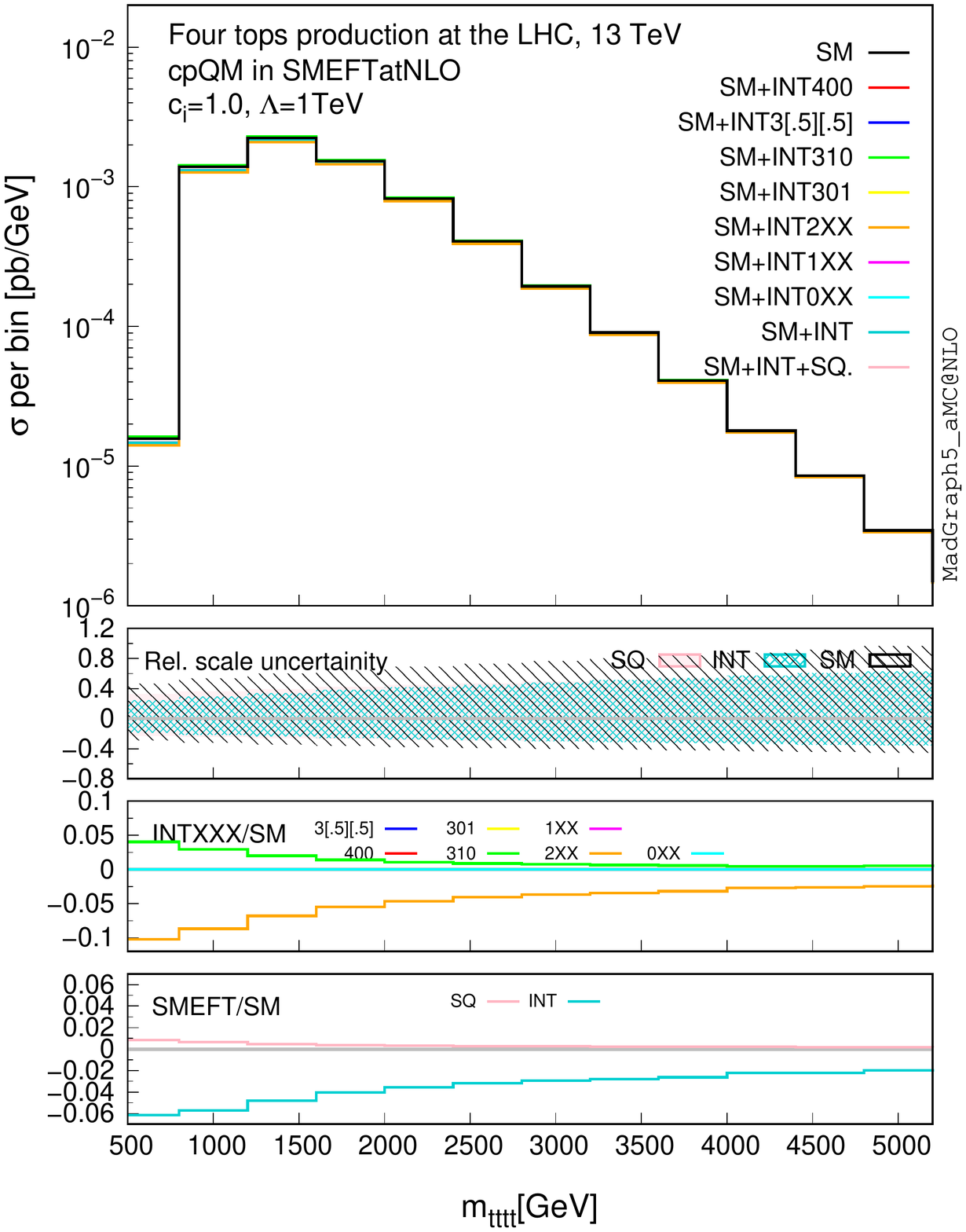}
    \includegraphics[trim=1.8cm 4.2cm 0.2cm 0.0cm, clip,width=.32\textwidth]{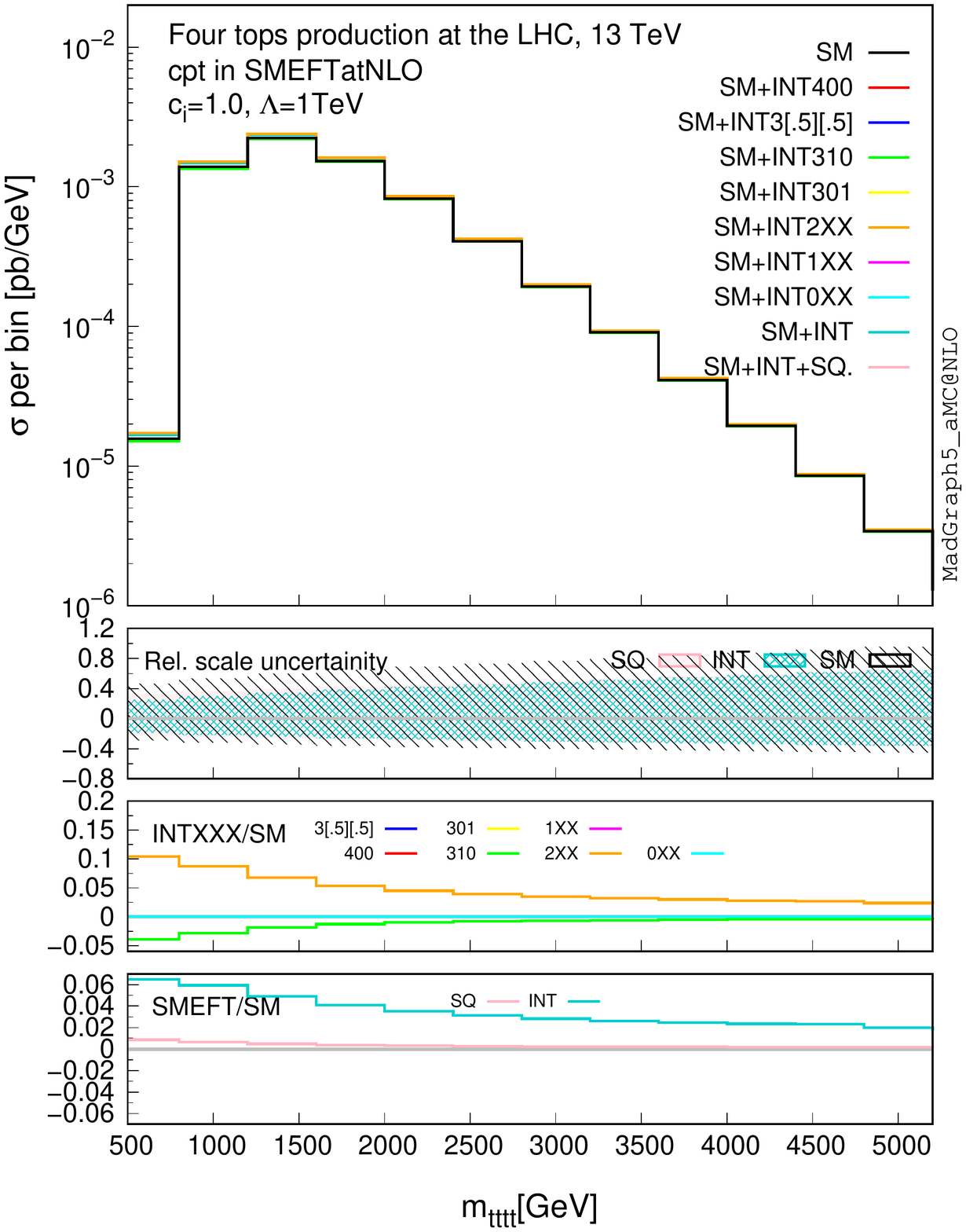}
    \includegraphics[trim=1.8cm 4.2cm 0.2cm 0.0cm, clip,width=.32\textwidth]{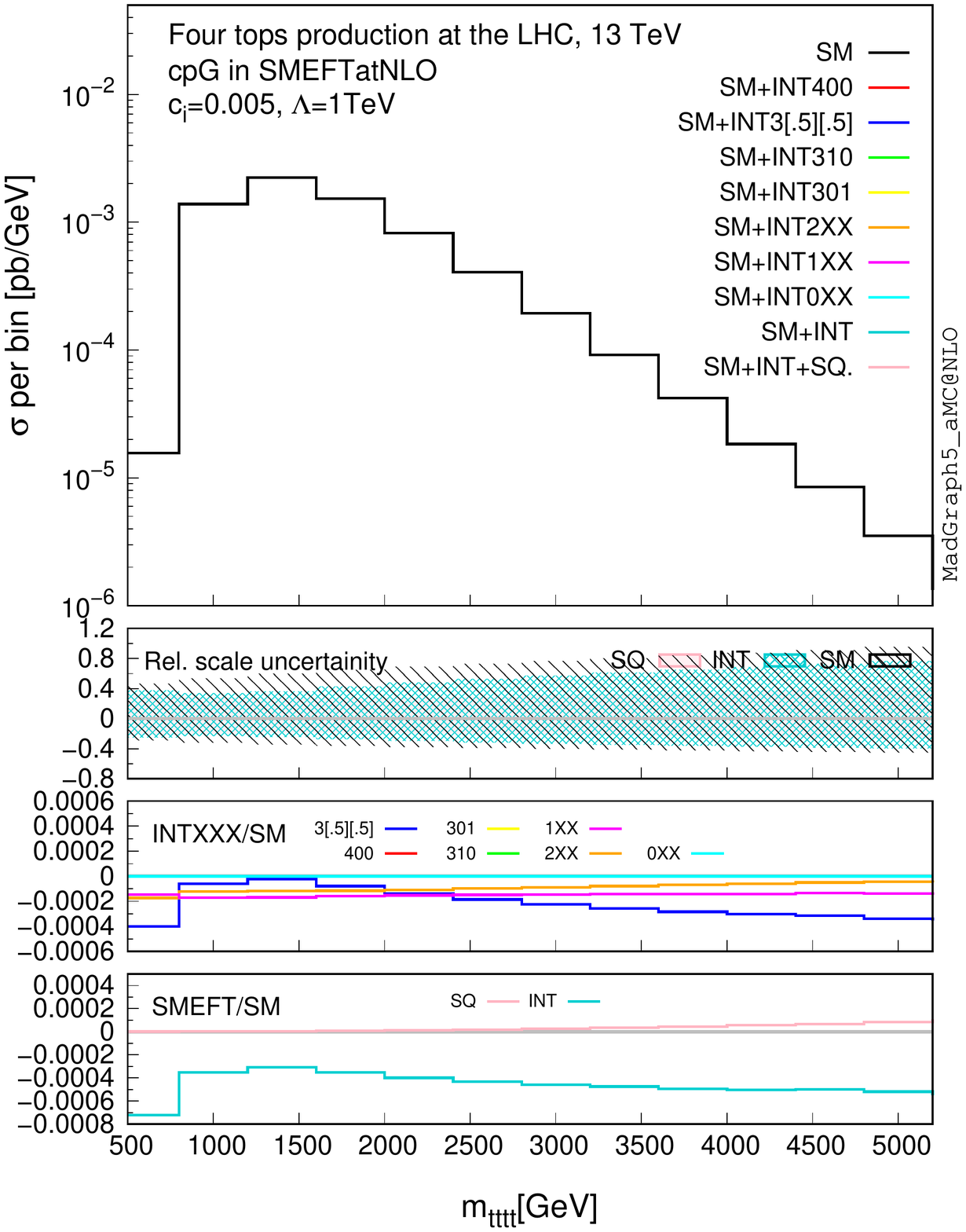}
    \caption{\label{fig:dim62f_dim60f_good_diff_13tev} Same as \cref{fig:dim64f_good_diff_13tev} but for the non-naive two-fermion and purely-bosonic operators.}
\end{figure}

\begin{figure}[h!]
    \centering
    \includegraphics[width=0.9\textwidth]{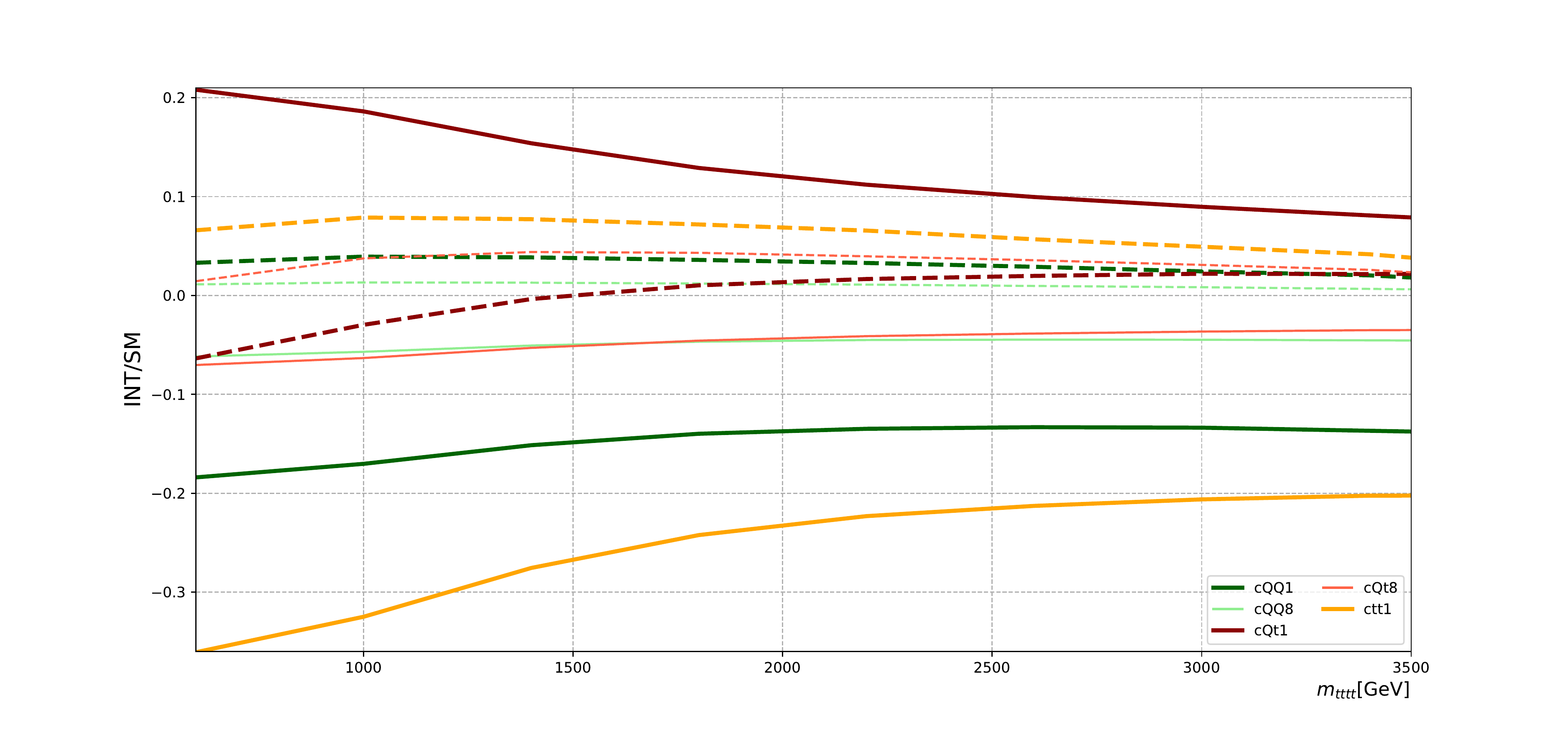}
    \caption{\label{fig:dim6heavy_int} The ratio of the linear EFT interference to the SM prediction for the non-naive 4-heavy operators. 
    Solid lines depict the $\mathscr{O}(\alpha_s^2\alpha)$ contribution, while dashed lines depict the $\mathscr{O}(\alpha_s^3)$ one.
    Thick (thin) lines represent colour-singlets (-octets).}
    \label{fig:sum_octet_singlets}
\end{figure}

\begin{figure}[h!]
    \centering
    \includegraphics[trim=1.8cm 4.2cm 0.2cm 0.0cm, clip,width=.32\textwidth]{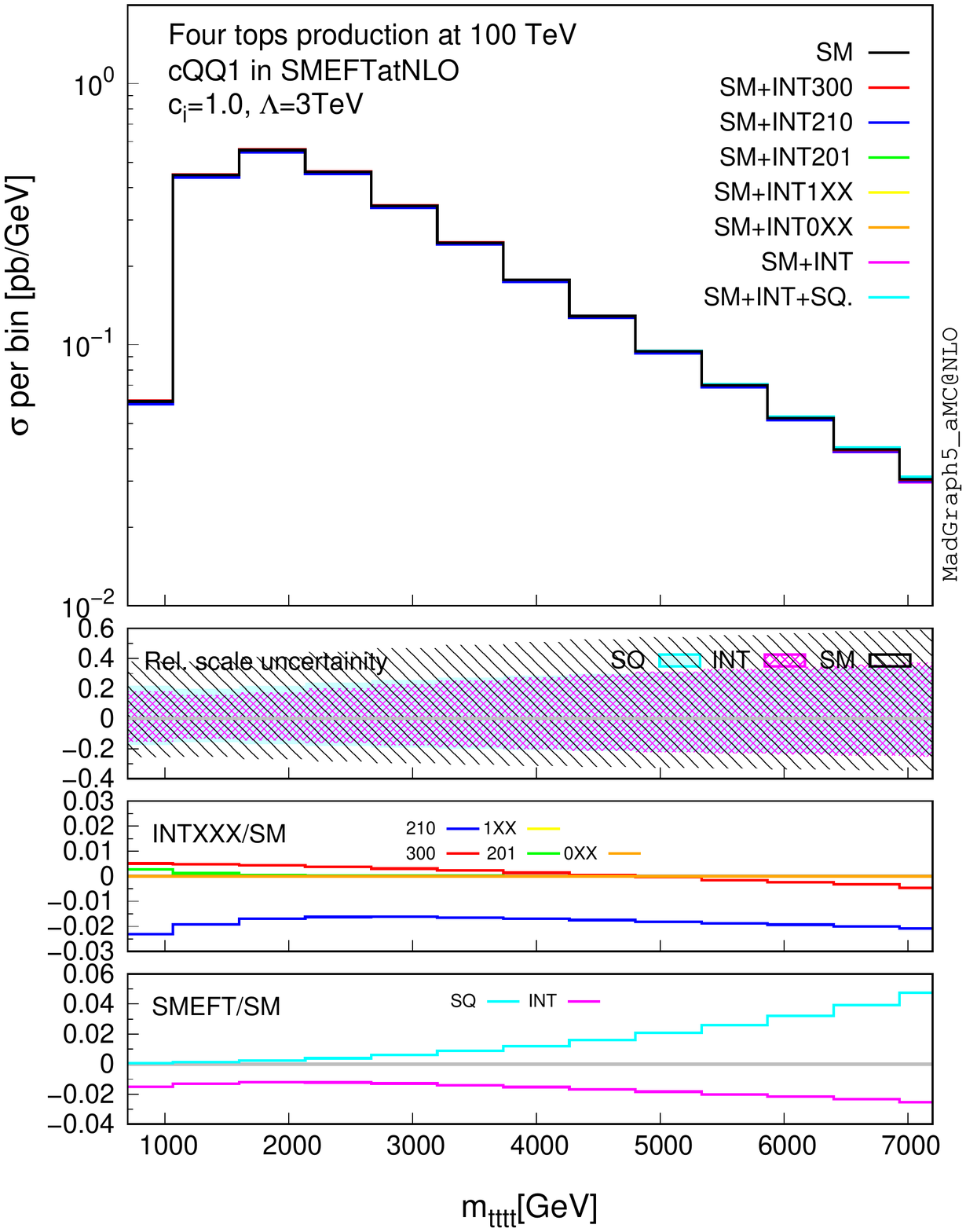}
    \includegraphics[trim=1.8cm 4.2cm 0.2cm 0.0cm, clip,width=.32\textwidth]{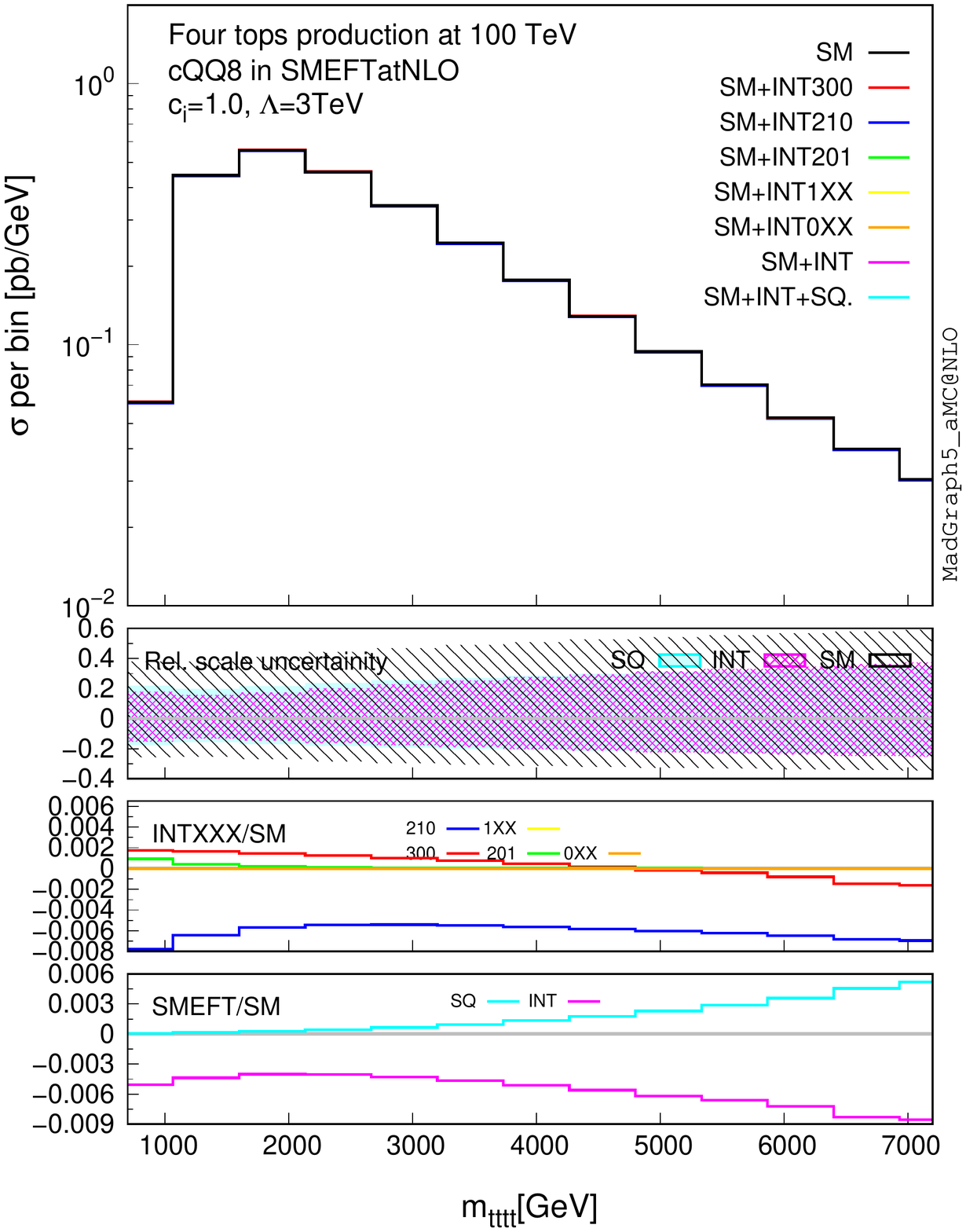}
    \includegraphics[trim=1.8cm 4.2cm 0.2cm 0.0cm, clip,width=.32\textwidth]{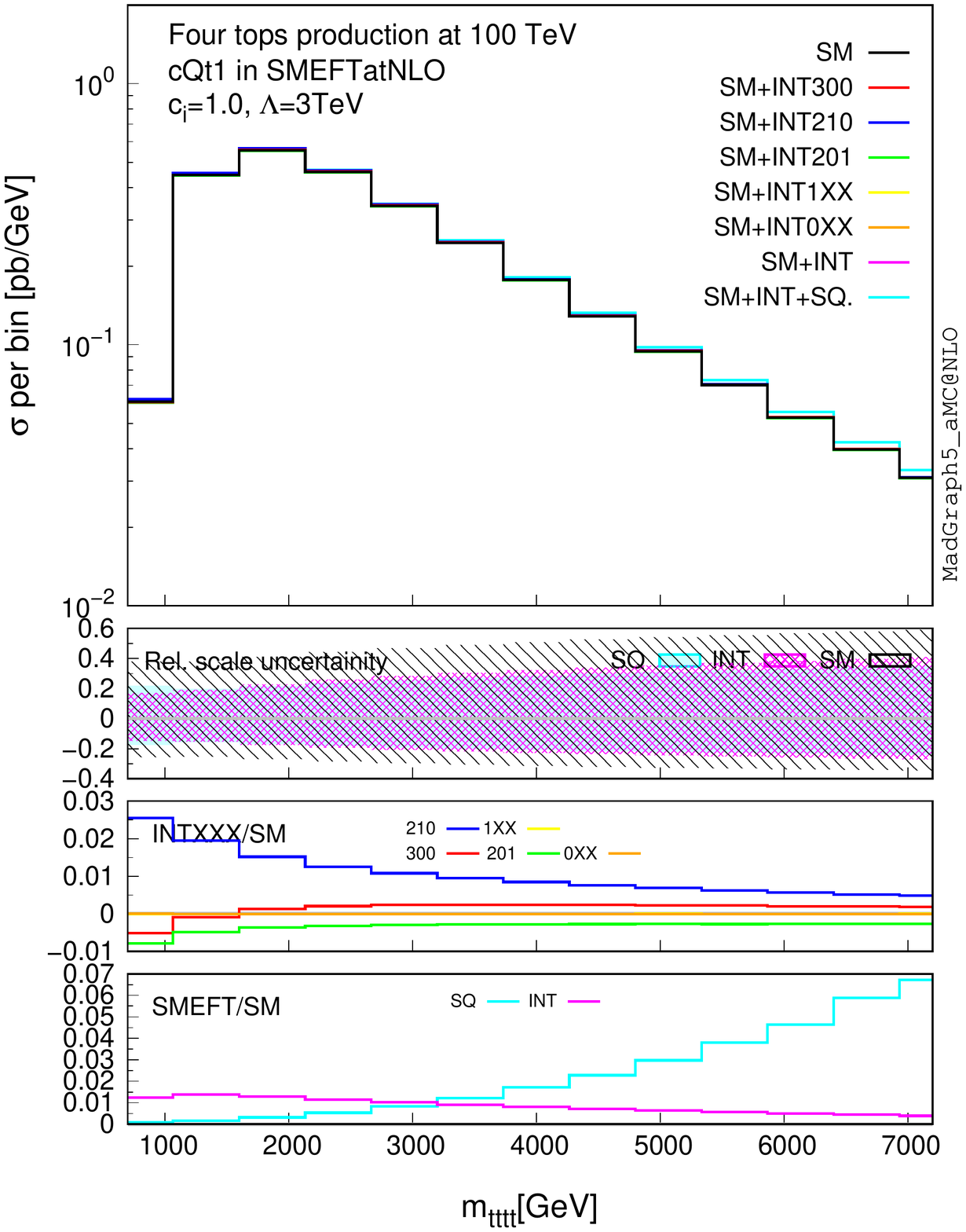}\\
    \includegraphics[trim=1.8cm 4.2cm 0.2cm 0.0cm, clip,width=.32\textwidth]{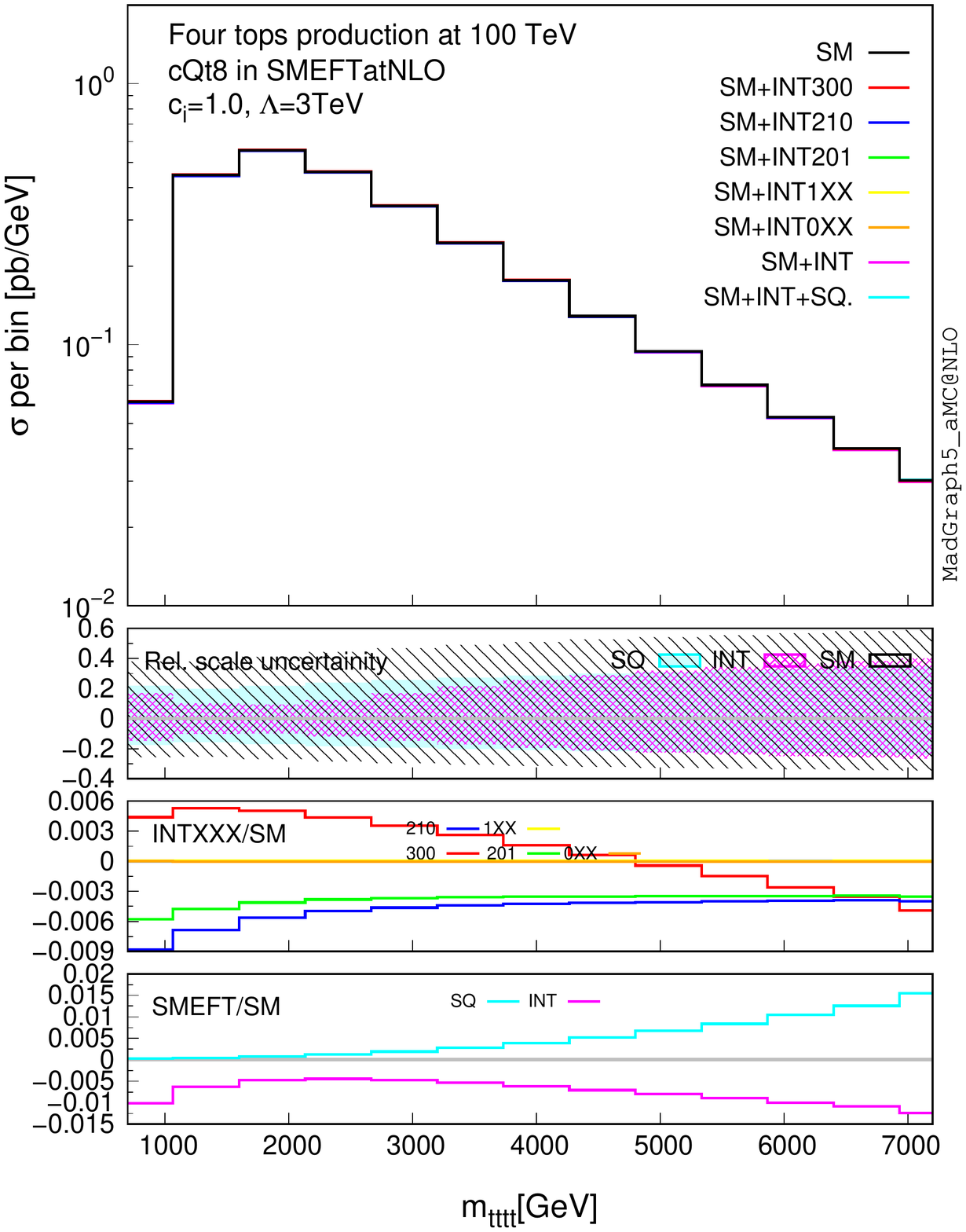}
    \includegraphics[trim=1.8cm 4.2cm 0.2cm 0.0cm, clip,width=.32\textwidth]{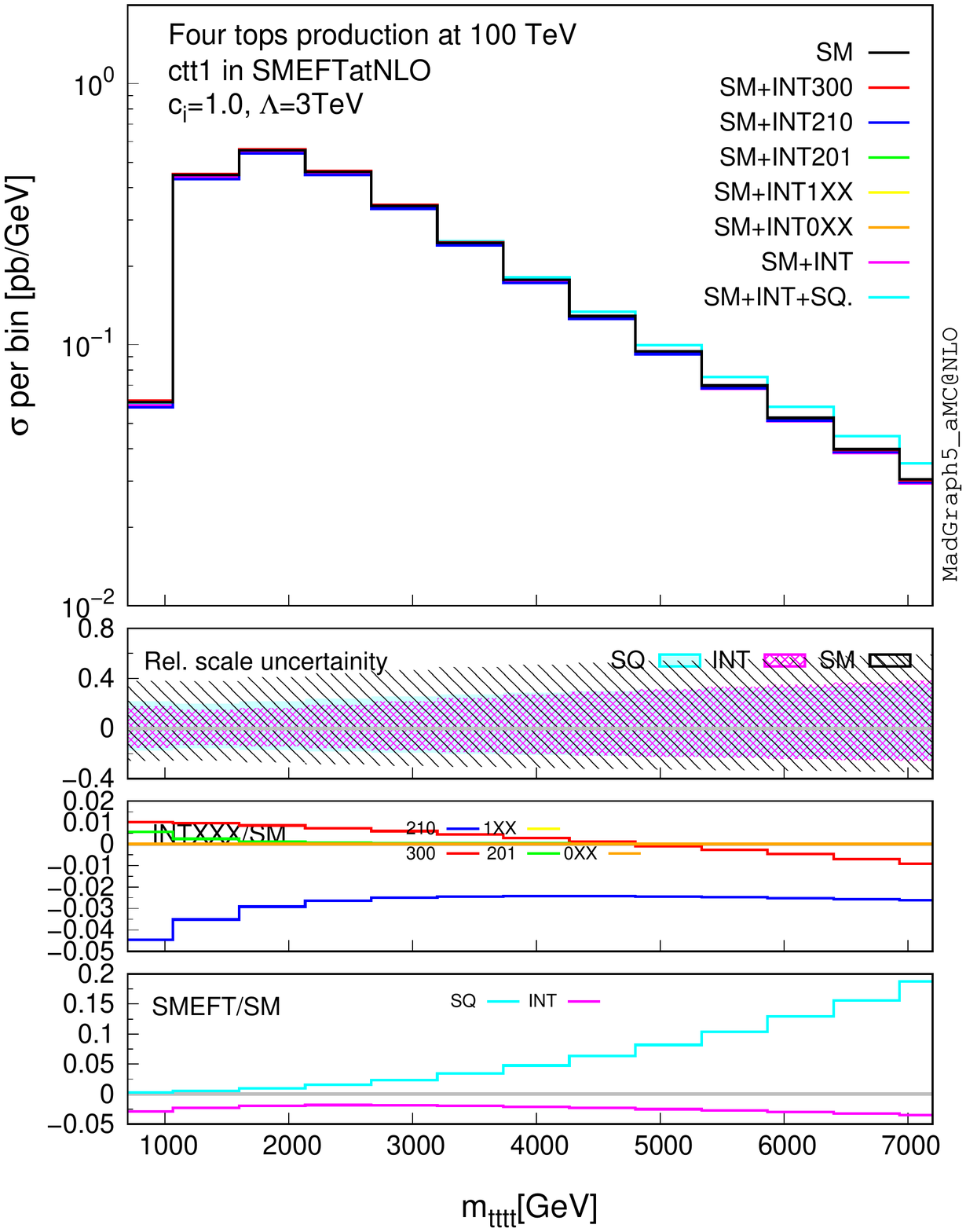}
    \includegraphics[trim=1.8cm 4.2cm 0.2cm 0.0cm, clip,width=.32\textwidth]{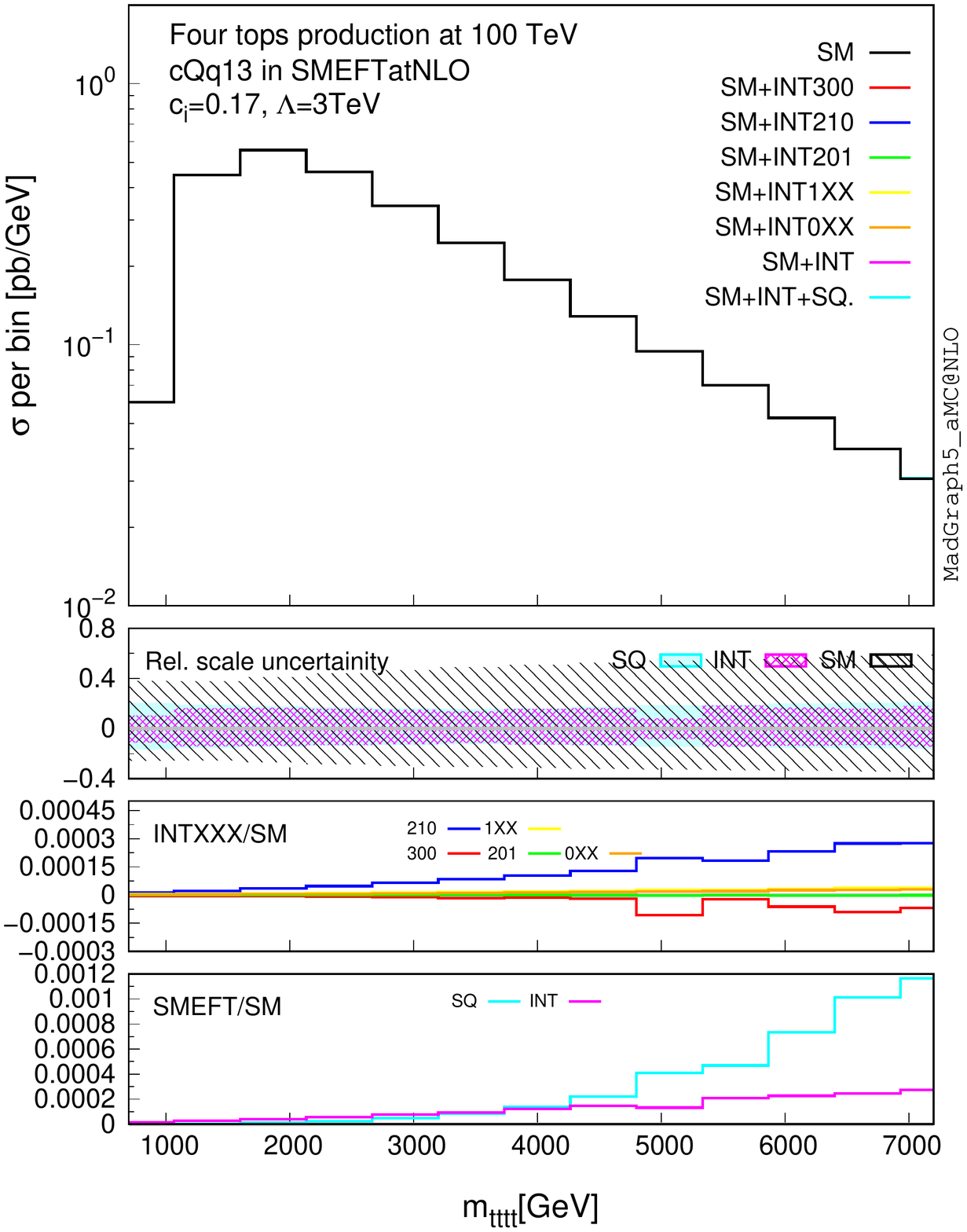}
    \caption{\label{fig:dim64f_good_diff_100tev} Same as \cref{fig:dim64f_good_diff_13tev} but for $\sqrt{s}=100$ TeV.}
\end{figure}

\begin{figure}[h!]
    \centering
    \includegraphics[trim=1.8cm 4.2cm 0.2cm 0.0cm, clip,width=.32\textwidth]{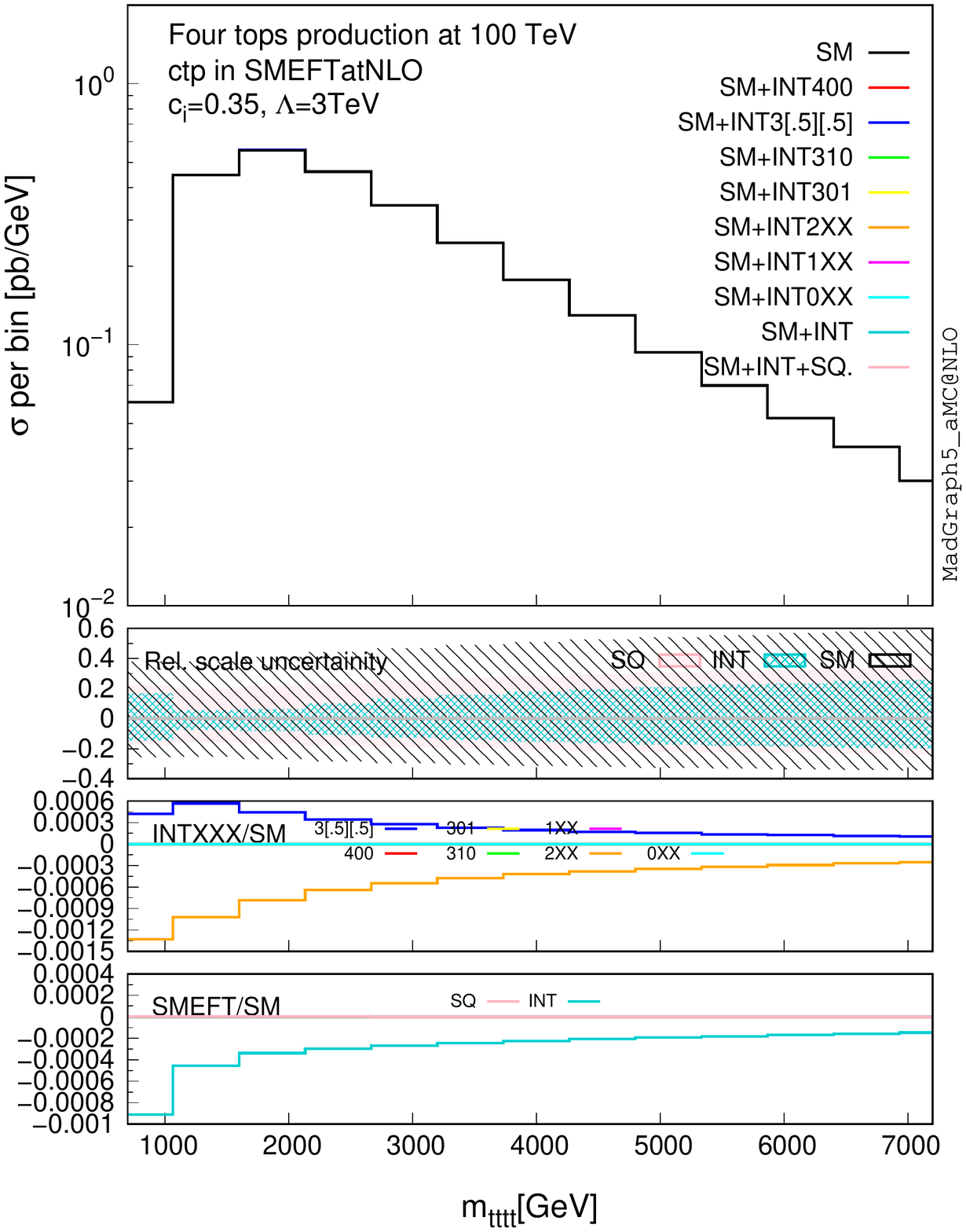}
    \includegraphics[trim=1.8cm 4.2cm 0.2cm 0.0cm, clip,width=.32\textwidth]{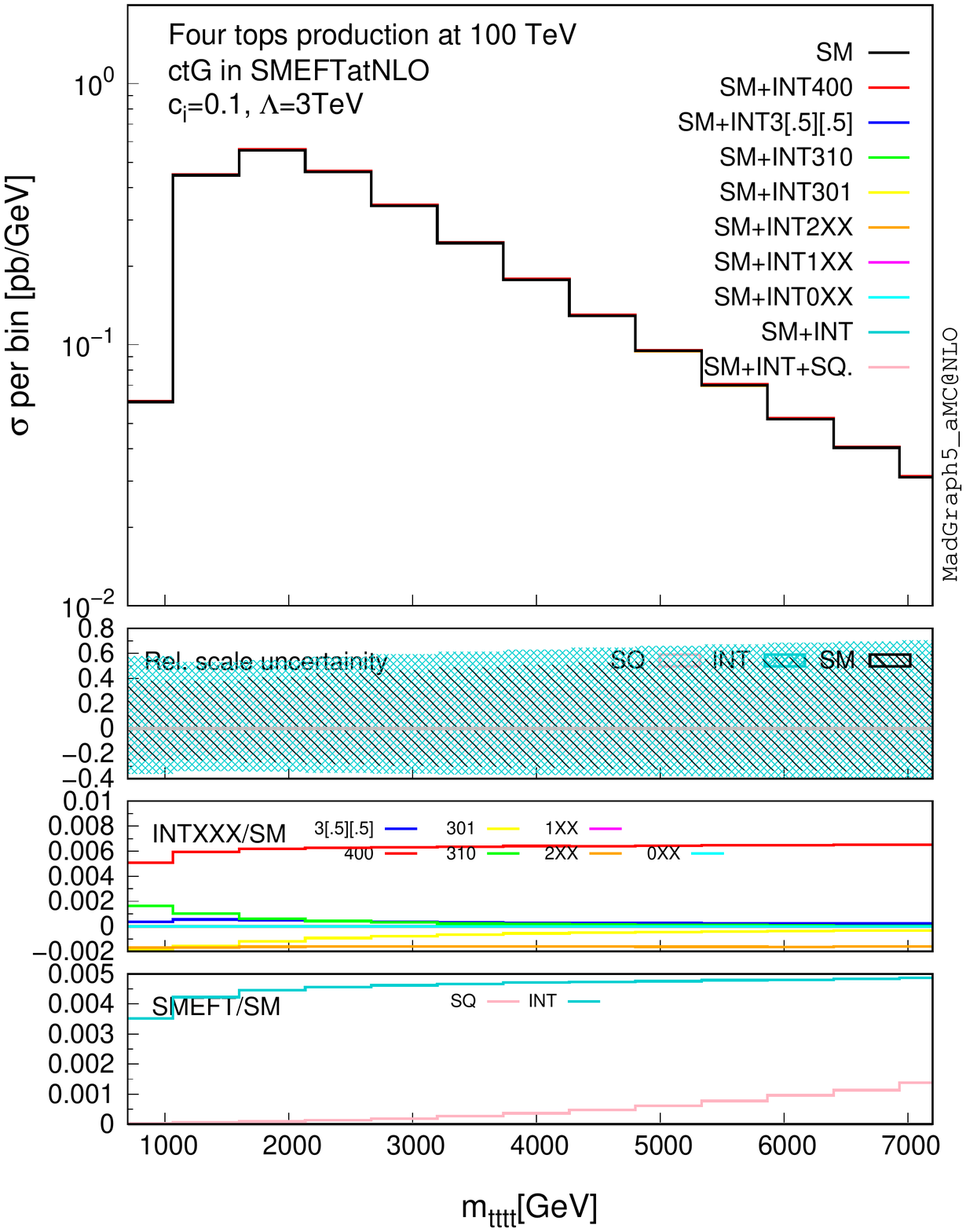}\\
    \includegraphics[trim=1.8cm 4.2cm 0.2cm 0.0cm, clip,width=.32\textwidth]{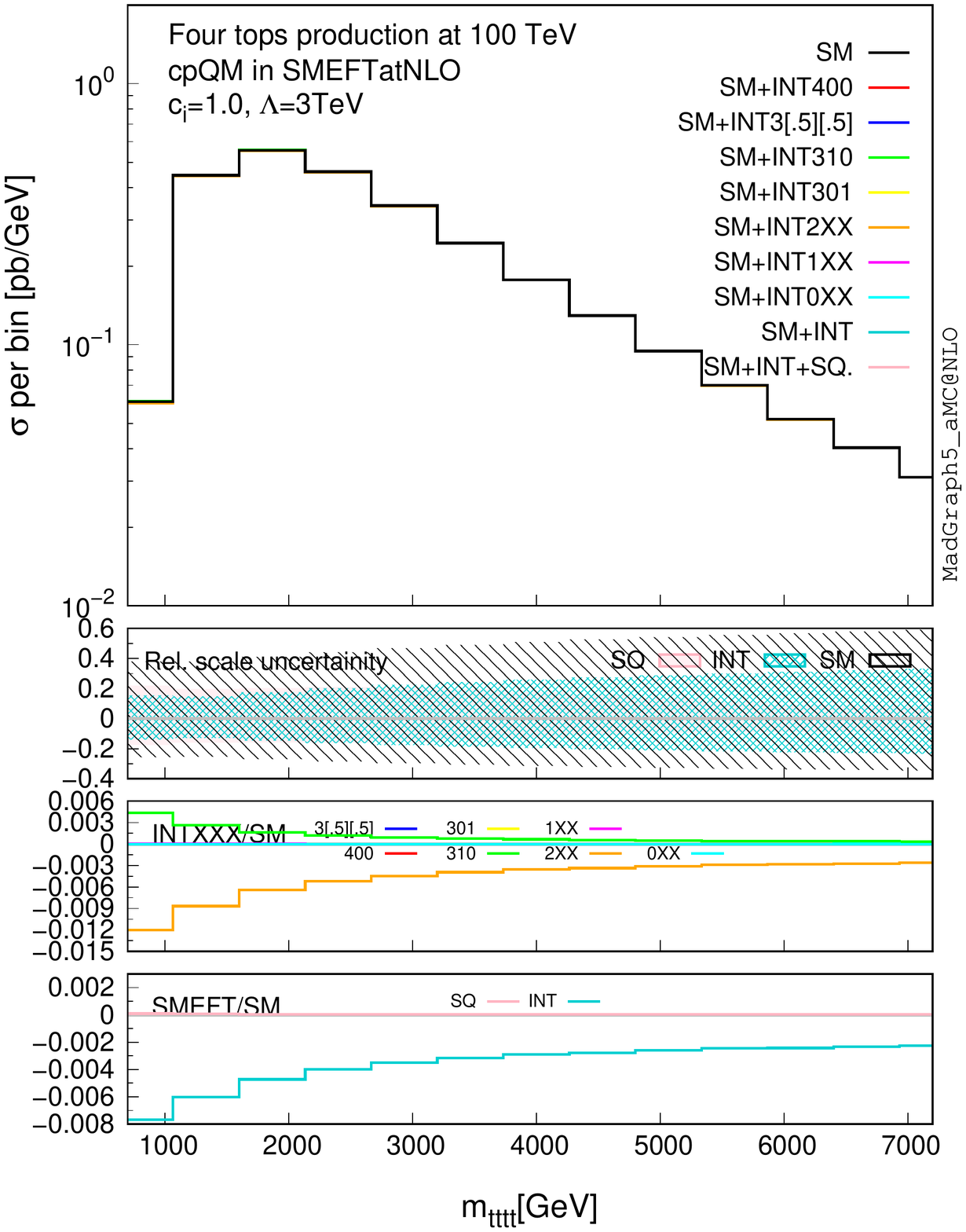}
    \includegraphics[trim=1.8cm 4.2cm 0.2cm 0.0cm, clip,width=.32\textwidth]{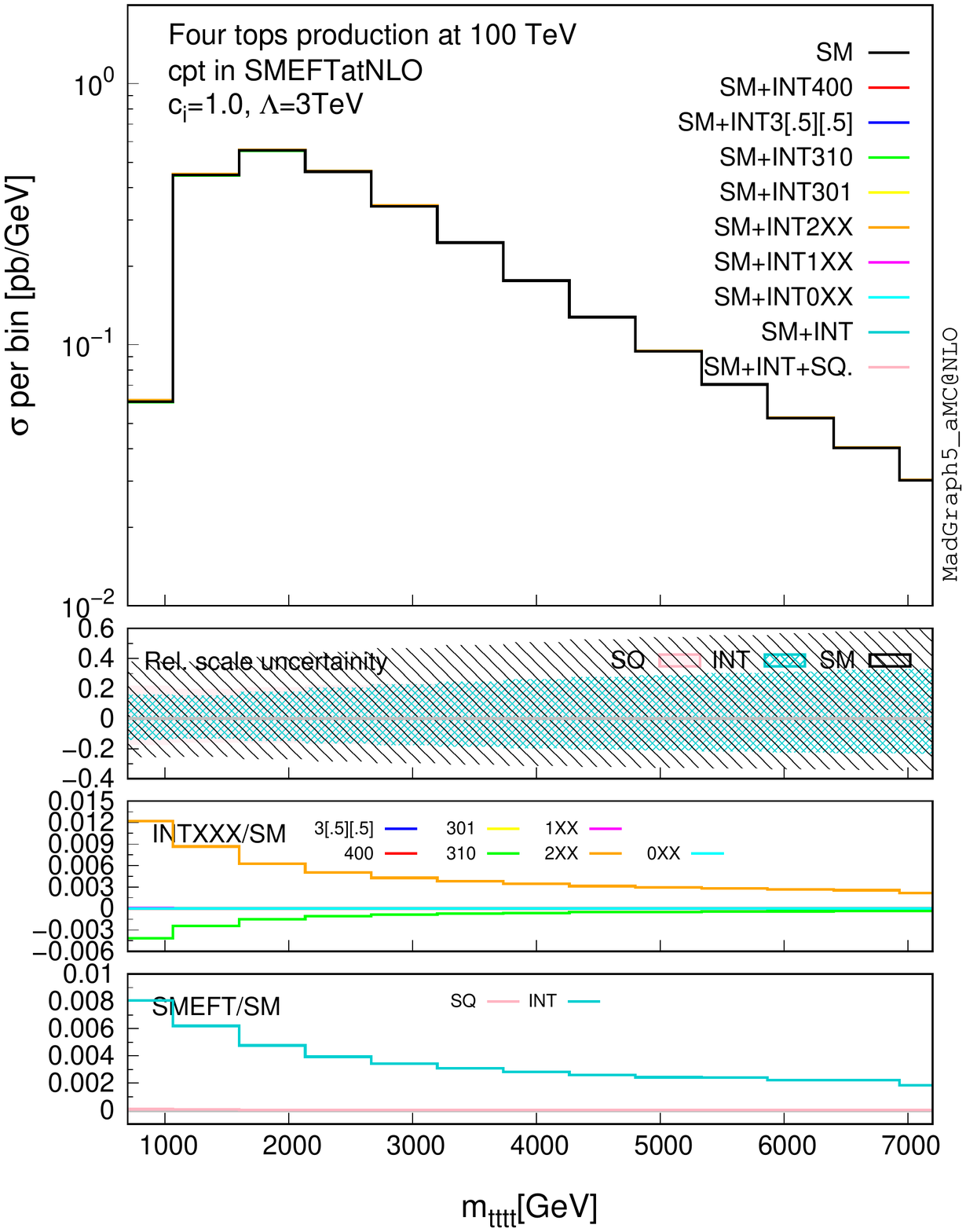}
    \includegraphics[trim=1.8cm 4.2cm 0.2cm 0.0cm, clip,width=.32\textwidth]{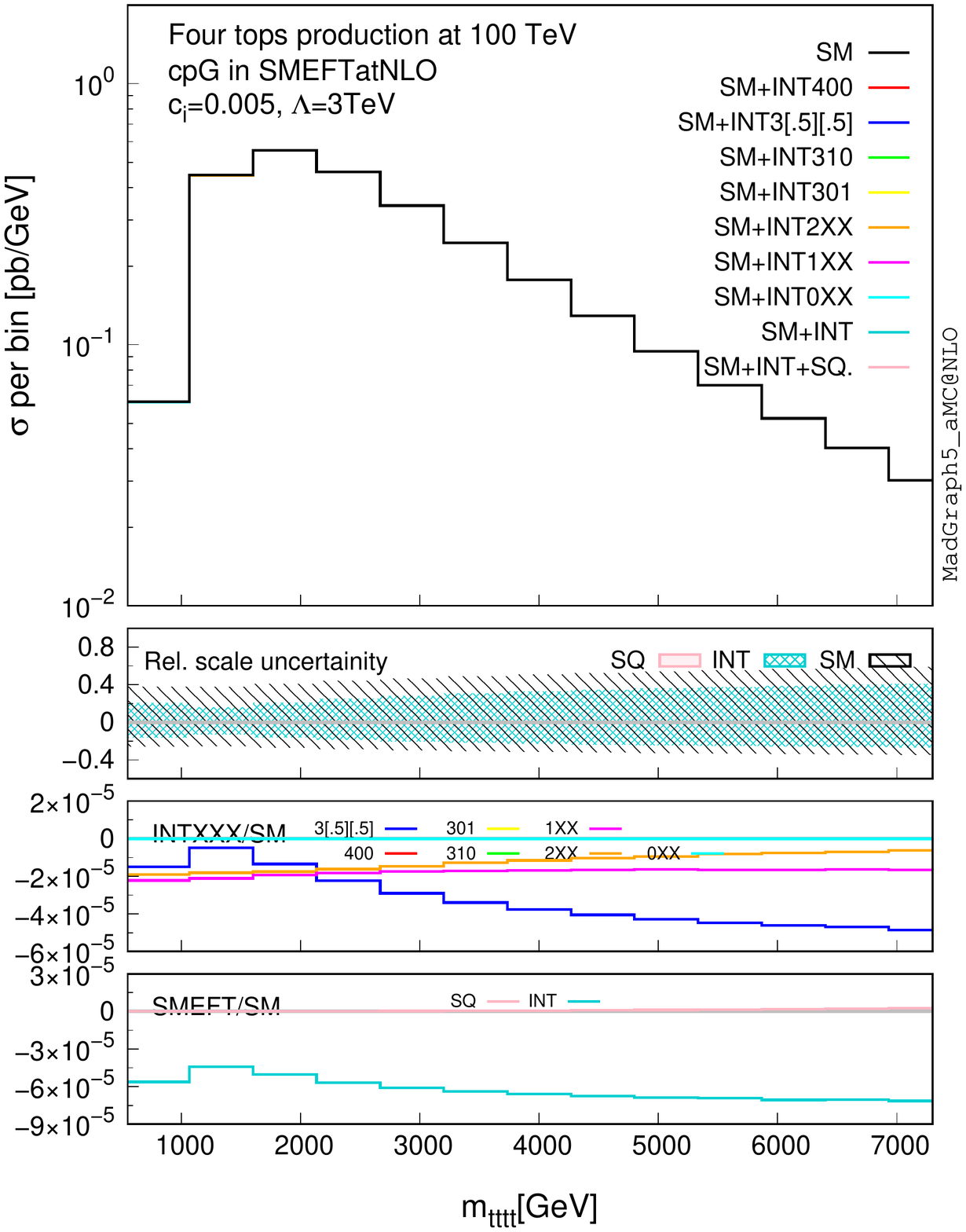}
    \caption{\label{fig:dim62f_dim60f_good_diff_100tev} Same as \cref{fig:dim62f_dim60f_good_diff_13tev} but for $\sqrt{s}=100$ TeV.}
\end{figure}

This section presents the LHC and FCC-hh differential predictions for the set of non-naive operators of \cref{eq:good_operators}.
The distributions are given in \cref{fig:dim64f_good_diff_13tev}-\ref{fig:dim62f_dim60f_good_diff_100tev}.
Differential predictions for the rest of the operators are given in \cref{sec:app_add_13_100}.
We present the distributions in the invariant mass bins of the four top quark system, $m_{tttt} \sim \sqrt{s}$, for which we also include the SMEFT diagonal\footnote{~These contributions correspond to the squared interference between the same dimension-six operators.} quadratic contributions. 
The input parameters are the same as the ones used for the inclusive results.
We show pure SM predictions, SM summed to the linear EFT interference, and SM summed to the linear EFT interference and the diagonal quadratic contributions.
Moreover, results are presented in every order of $\alpha_{s}$, e.g. \texttt{INT201} indicates the interference contribution (\texttt{INT}) induced from $\mathscr{O}(\alpha_{s}^{2}\alpha_{t})$ terms, where the first, second and third digits denote the orders of $\alpha_{s}$, $\alpha$, and $\alpha_{t}$, respectively.
For orders `below' the formal subleading one, we sum all EW-induced contributions at this given $\alpha_s$-order.
For example, \texttt{1XX} indicates summing all possible EW-induced contributions with one $\alpha_{s}$ coupling, in parallel to the notation used in \cref{sec:inclusive}; \texttt{1XX} $\equiv \sigma_{1}$.
The relative scale uncertainties are computed individually from a nine-point renormalisation and factorisation scale variation around the central scale of $340$ GeV for each EFT contribution.
Contrary to inclusive predictions, we here used WCs values extracted approximately from the global fit of Ref.~\cite{Ethier:2021bye}, except for \cG{}, for which the value of its coefficient is specified in the corresponding plot.

\paragraph{$\mathbf{\sqrt{s}=13}$ TeV}
One clear pattern in all of the 4-heavy operators' predictions is the sizable interference cross-section arising from the $\mathscr{O}(\alpha_{s}^{2}\alpha)$ term and depicted by the blue line in the second inset. 
This observation corroborates what we observed in the inclusive results section; formally subleading contributions are significant. 
For $\sigma_{2}$, we find that the dominant contribution comes from the $\mathscr{O}(\alpha_{s}^{2}\alpha)$ terms rather than those of $\mathscr{O}(\alpha_{s}^{2}\alpha_t)$.
Moreover, and similarly to the inclusive predictions, colour-singlets are observed to interfere with the SM comparatively stronger than colour-octet ones. 
We had already mentioned this pattern in \cref{sec:inclusive}; however, we relegated the detailed discussion to this section.

\paragraph{Interlude: on the interference pattern of colour-singlets and  colour-octets} 
It is expected that in formally QCD-dominated production processes, amplitudes with colour-octet insertions would exhibit a stronger interference pattern compared to colour-singlets. 
For instance, in $\bar{t}t$ production, colour-singlet contributions vanish at the tree-level when considering only purely QCD-induced SM amplitudes~\cite{Brivio:2019ius}. 
In four-top production, however, the interference between colour-singlets and the QCD SM amplitudes is non-zero due to the presence of more complicated colour structures, which allow the top-anti-top pair to be in a colour-singlet state. 
This is evident at the $\sigma_3$-level, where the colour-singlets interference can be of the same order as that of the colour-octets. 
In addition, and as discussed previously, the EW scattering effects in the $gg \to \bar{t}t\bar{t}t$ born-level amplitudes, depicted in \cref{fig:4tops_LO_diags_scattering}, provide significant contributions to the cross-section, in the SM and in the SMEFT. 
This explains the weaker interference strength of colour-octets compared to colour-singlets in the set of 4-heavy operators at the $\sigma_2$-level.
The reason comes from the different colour flow in the $tt$ $s$-channel scattering sub-amplitude;
EFT amplitudes with a colour-octet insertion would interfere with the formally leading SM amplitudes where a gluon is the mediator of the $tt \to tt$ scattering.
On the other hand, amplitudes with the insertion of a colour-singlet operator are expected to interfere with the formally subleading SM amplitudes where the $tt \to tt$ scattering happens via an EW mediator.

In contrast to the 4-heavy operators, the previously-mentioned insignificance of EW scattering effects in the $q\bar{q}$-initiated production channel explains the `typical' stronger interference strength of the 2-heavy-2-light colour-octets compared to colour-singlets. 
The only exception to this is the \TriSi{} operator.
Interestingly, the latter is also the \emph{only} 2-heavy 2-light operator in the set of non-naive operators.
As discussed above, this suggests that the enhancement from formally subleading terms is indeed intertwined with the `unusual' more potent interference from colour-singlets.
The interference pattern of colour-octets and -singlets in the set of 4-heavy operators is summarised in \cref{fig:sum_octet_singlets}.

Moving to the non-naive two-fermion and purely-bosonic operators, we observe a different EFT structure compared to 4-heavy ones. 
EFT amplitudes with insertions of four-fermion operators have effective contact terms as depicted in \cref{fig:diags_EFT_4F}.
The energy scaling of such amplitudes leads to consistent growth of the quadratic contribution as a function of $\sqrt{s}$.
This is seen in the third inset of all the distributions in \cref{fig:dim64f_good_diff_13tev}. 
On the other hand, due to their different energy scaling, the two-fermion operators exhibit suppressed squared contributions and, in most cases, a decaying interference.
Such effect is apparent in the third insets of \cref{fig:dim62f_dim60f_good_diff_13tev}, with a notable exception of the \ctG{} operator.
The latter receives enhancement in its energy scaling from gluon field strength derivatives, hence its different EFT structure: the linear contribution scales as $\sim 1/E^{3}$, while the quadratic one scales as $\sim 1/E^{2}$.
Such EFT structure is also evident for \cG{} in \cref{fig:dim62f_dim60f_rest_diff_13tev}.

In summarising the LHC results, we note that the 4-heavy operators are the most sensitive probes to four-top production.
This can be deduced from their sizable interference magnitude compared to the 2-heavy-2-light operators and the two-fermion and purely-bosonic ones.   

\paragraph{$\mathbf{\sqrt{s}=100}$ TeV}
Despite the high collision energy of the FCC-hh computation, the increased value of $\Lambda$, i.e. $\Lambda=3$ TeV, initially set to ensure a reliable EFT expansion, significantly suppresses the magnitude of the EFT contributions when compared to the ones from the LHC study. 
As a reminder, changing $\Lambda$ from 1 to 3 TeV suppresses the interference contribution by a factor of 9 and the quadratic one by a factor of $81$. 
Nevertheless, and in exploiting the predictions of $m_{tttt}$ at higher energies, we observe the expected energy growth of EFT contributions as a function of $\sqrt{s}$; inherently due to the contact term nature of those operators. 
Again, we observe colour-singlets interfering with the SM more strongly than colour-octets, drawing parallels to the LHC predictions.
However, in contrast to the LHC predictions, for almost all 2-heavy-2-light colour-octets presented in \cref{fig:dim64f_rest_diff_100tev} for the FCC-hh scenario and the choice of $\Lambda$, the linear interference contributions dominate the quadratic ones. 
Moving to two-fermion and purely-bosonic operators, and albeit a milder quadratics growth for \cG{} at FCC-hh in \cref{fig:dim62f_dim60f_rest_diff_100tev}, we see a similar behaviour between the LHC and FCC-hh predictions.
\section{Sensitivity projections at future colliders}
\label{sec:projections}
Given the current evidence of \tttt production amounting to 4.7$\sigma$ significance~\cite{ATLAS-CONF-2021-013}, \tttt is expected to be discovered at the LHC with the Run III data. 
The four-top production is induced mainly by gluons in the initial state rendering the $gg$-initiated production dominating $\sim 87\%$ of the SM \tttt cross-section at 13 TeV and $\sim 99\%$ at 100 TeV.
Since the main background contribution to the \tttt signal arises from $t\bar{t}W$ production~\cite{ATLAS:2018kxv}, which proceeds only in the quark-initiated mode, an increase in the collision energy can lead to an improvement of the signal-to-background ratio of four-top production.
Moreover, uncertainties polluting the experimental measurement can be progressively reduced as a function of an increasing integrated luminosity.
The study of Ref.~\cite{Azzi:2019yne} combined the expected \tttt experimental sensitivity at future LHC runs and the state-of-art theoretical calculations~\cite{Frederix:2017wme} to predict the total uncertainty by which the \tttt cross-section can be determined. 
Such expected uncertainties are 102\%, 58\%, and 40\%, at 95\% CL, for 13, 14, and 27 TeV runs respectively with a corresponding integrated luminosities of 300 fb$^{-1}$, 3 ab$^{-1}$, and 15 ab$^{-1}$~\cite{Azzi:2019yne}.
In this section, and using these estimated uncertainties, we reproduce the study of Ref.~\cite{Azzi:2019yne} and subsequently add predictions from FCC-hh for comparison.
It is worth noting that this study assumes the sensitivity of the effective operators to be mainly induced from inclusive measurements.
We impose an EFT validity cut on the invariant mass of the four top quarks, $m_{tttt}<3$ TeV, so the SMEFT predictions can be matched to UV models with higher energy scales.
The EFT validity cut is assumed not significantly to alter the projected sensitivity.

\begin{figure}[h!]
    \centering
    \includegraphics[width=.48\textwidth]{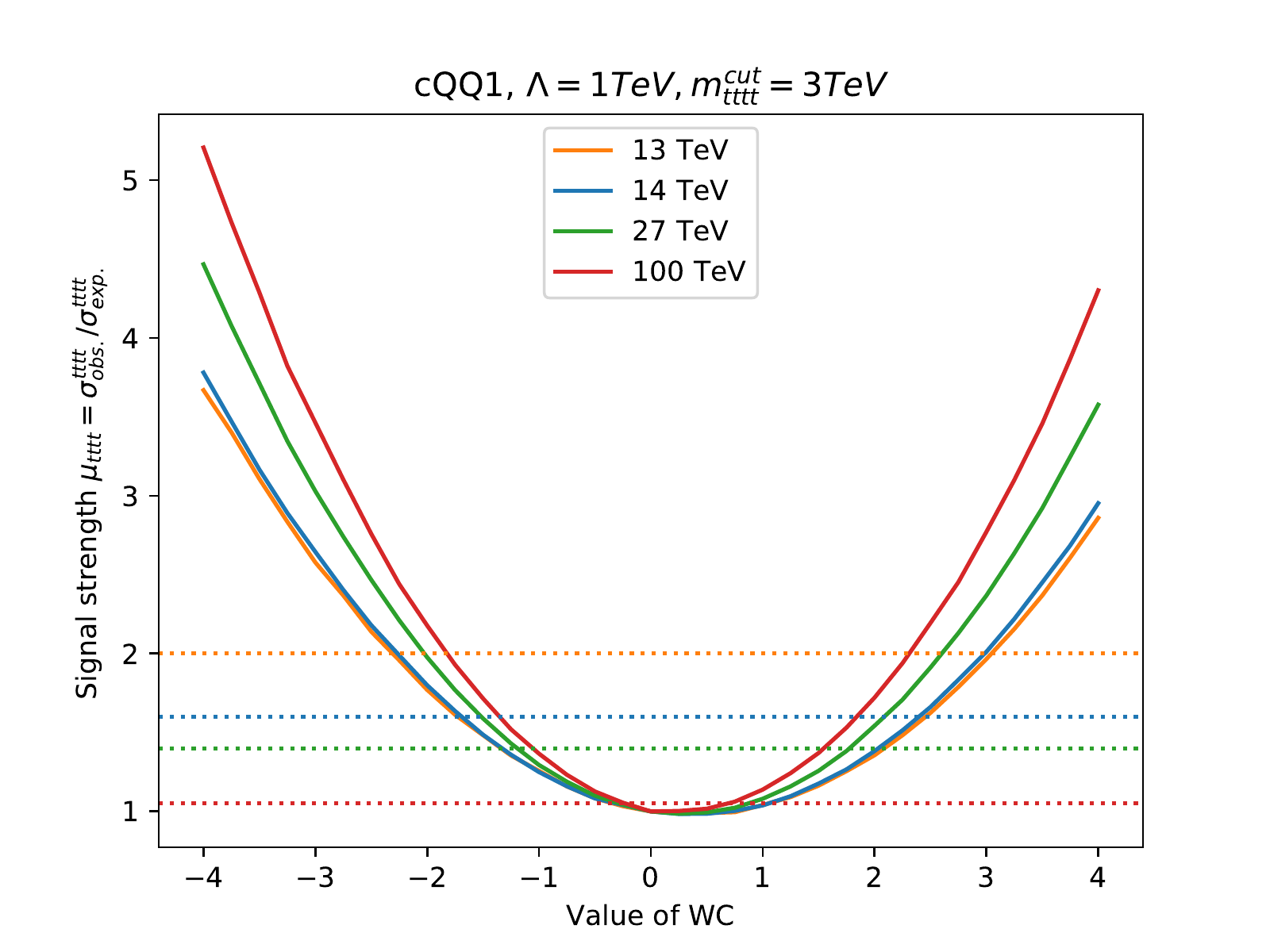}
    \includegraphics[width=.48\textwidth]{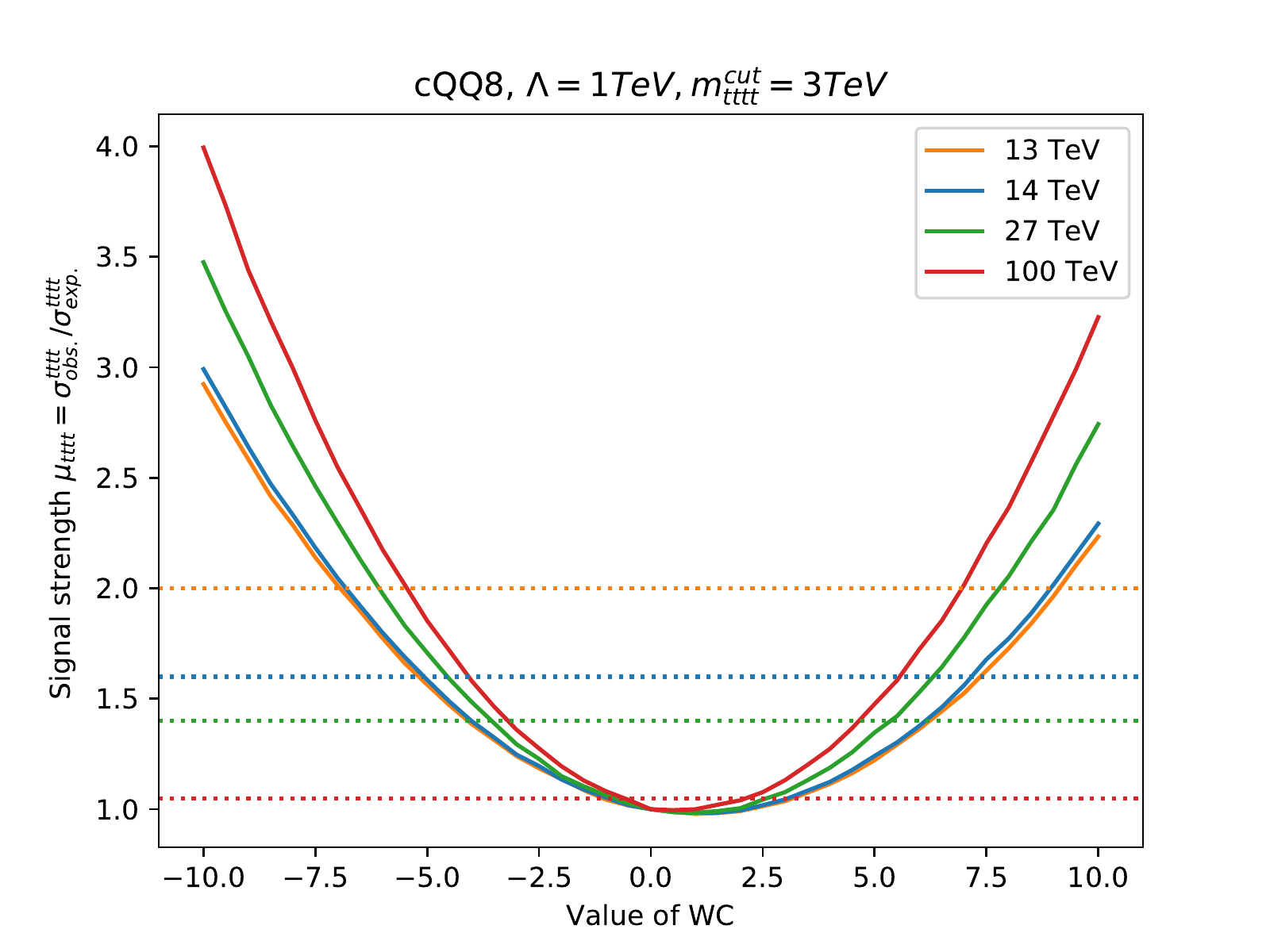}\\
    \includegraphics[width=.48\textwidth]{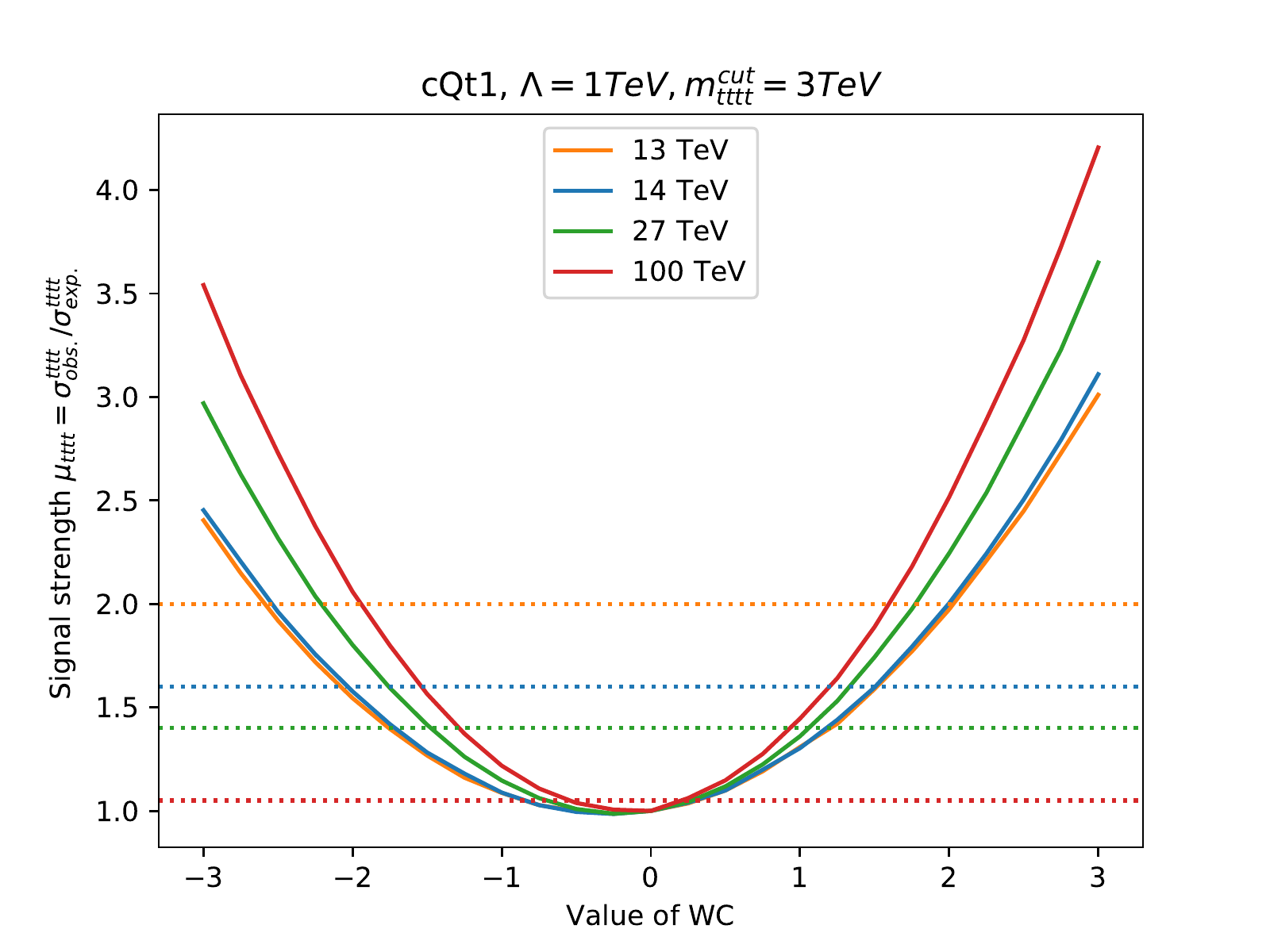}
    \includegraphics[width=.48\textwidth]{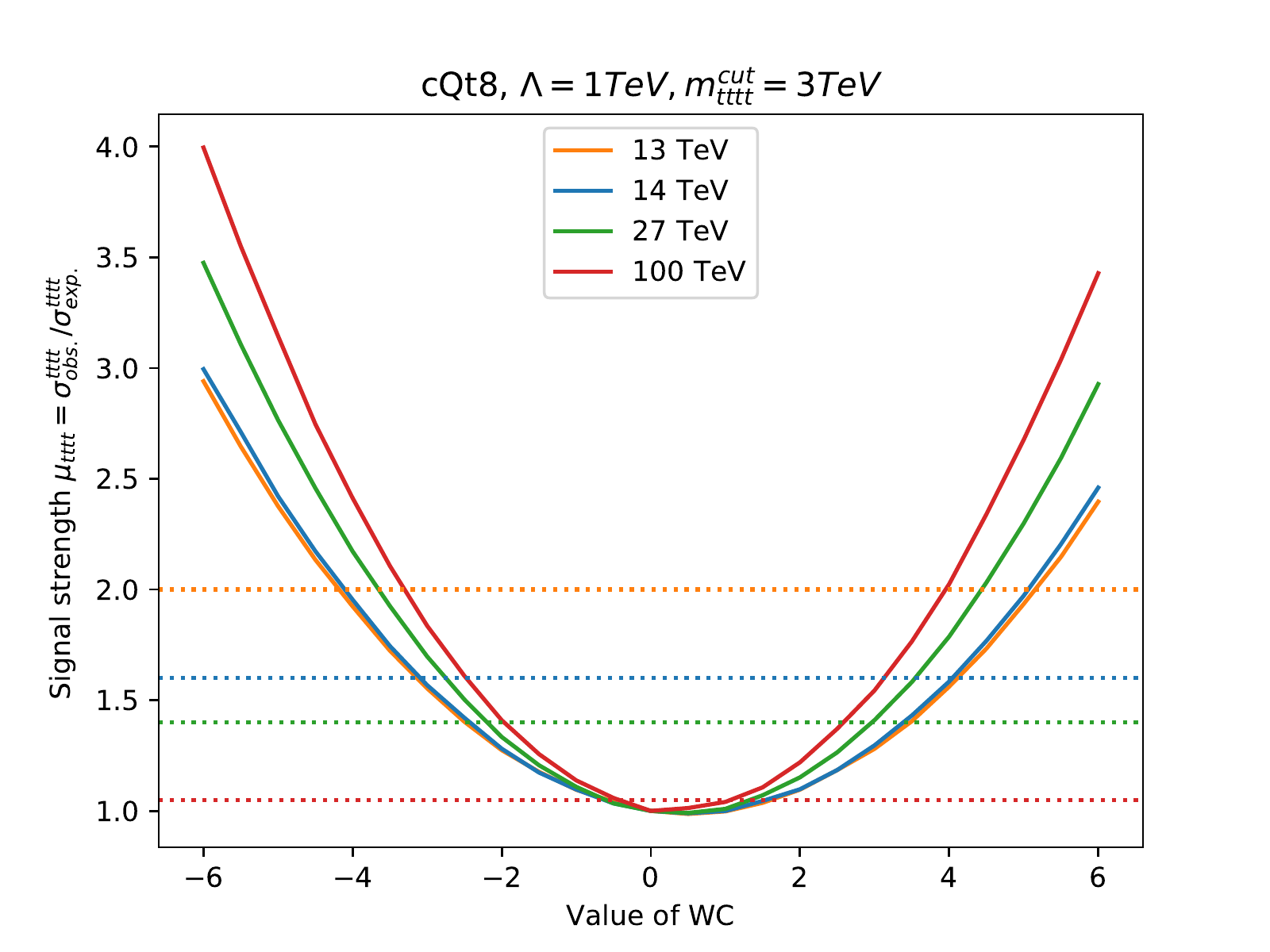}\\
    \includegraphics[width=.48\textwidth]{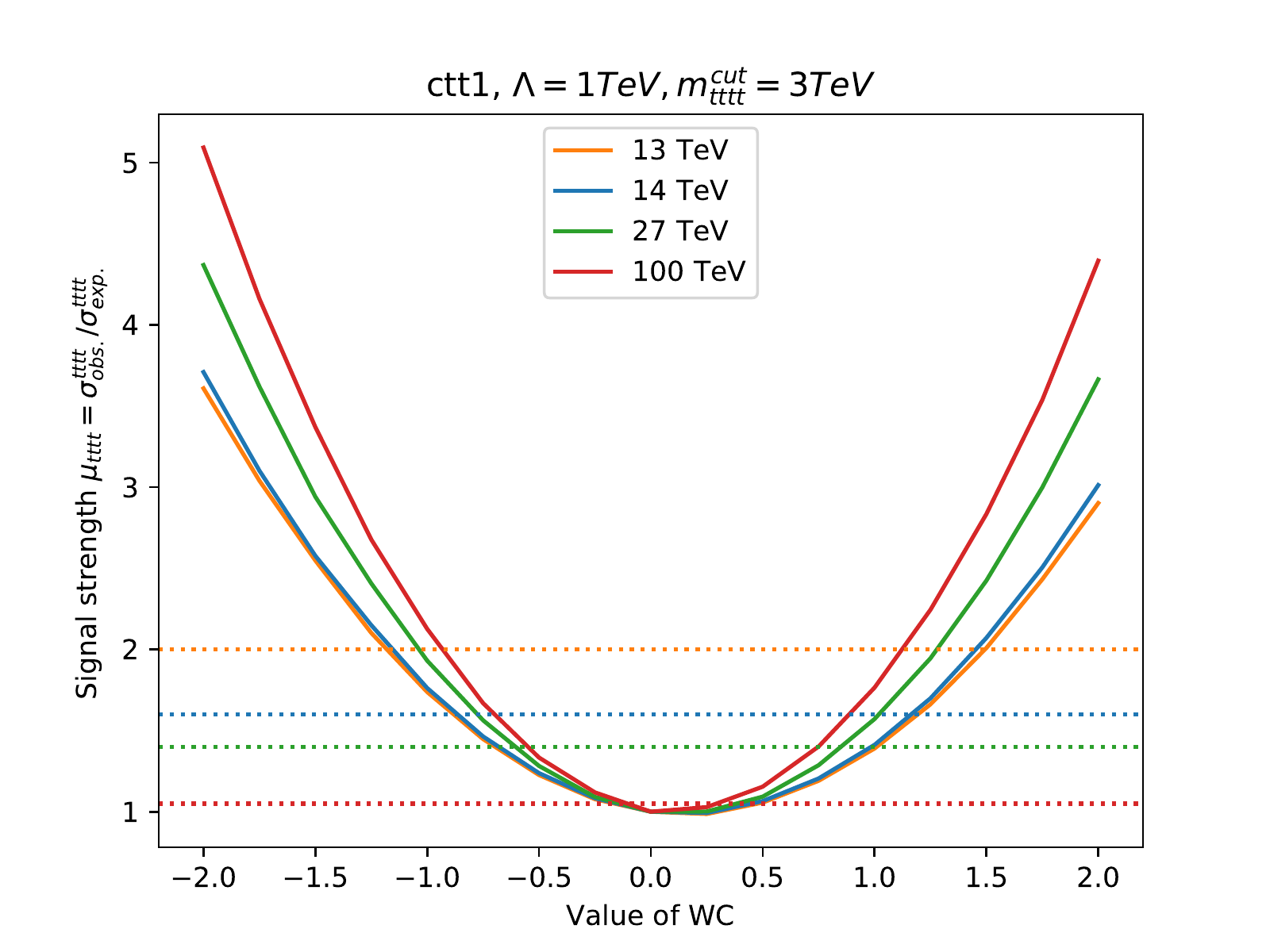}
    \caption{Four-top production signal strength as a function of the WC values for all the 4-heavy operators. 
    The EFT predictions include both linear interference and quadratic contributions.
    The horizontal lines represent the expected measurement at each collision energy derived from the expected total uncertainty.}
    \label{fig:projections_4heavy}
\end{figure}

\begin{figure}[h!]
    \centering
    \includegraphics[width=.48\textwidth]{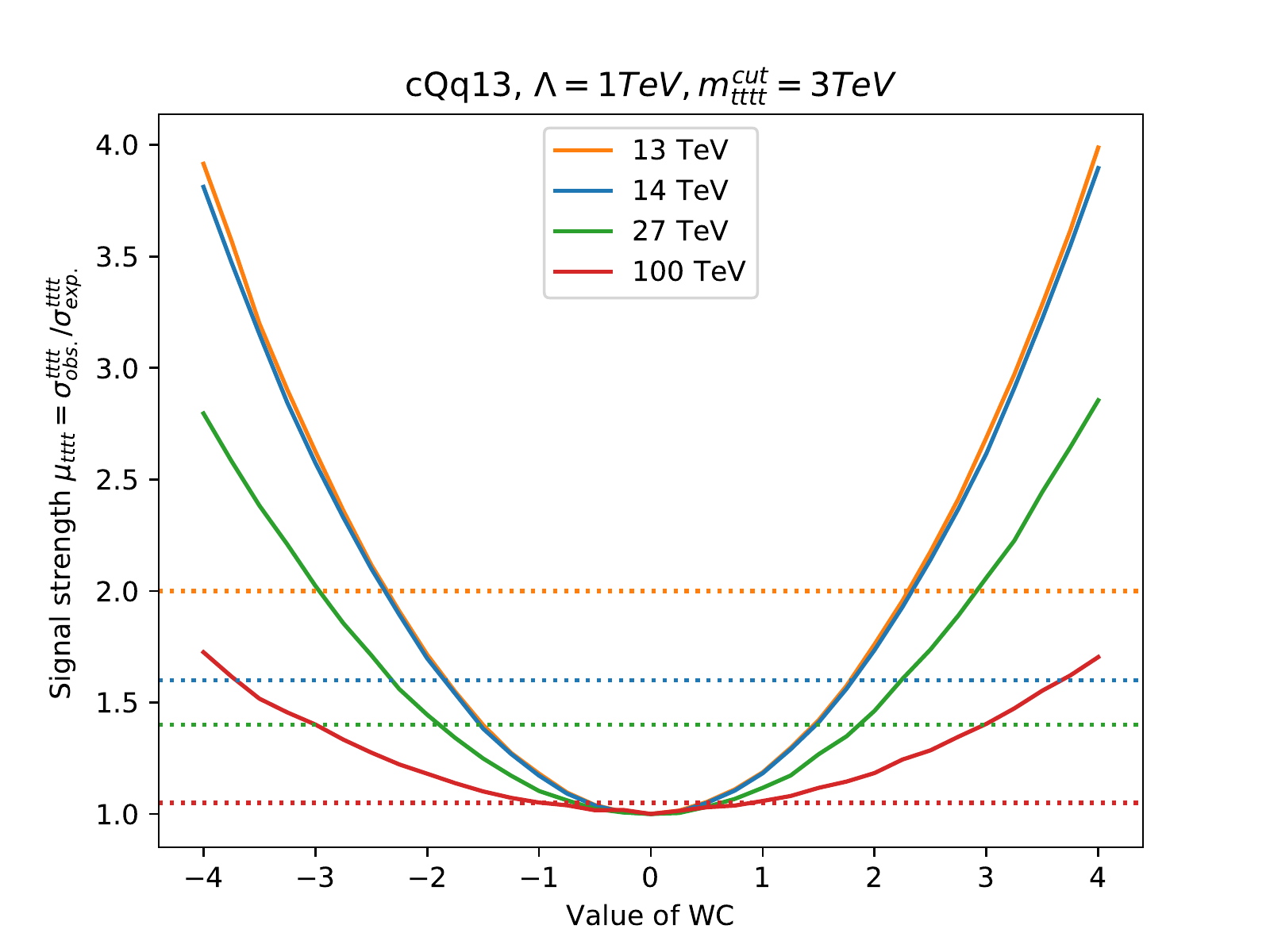}
    \includegraphics[width=.48\textwidth]{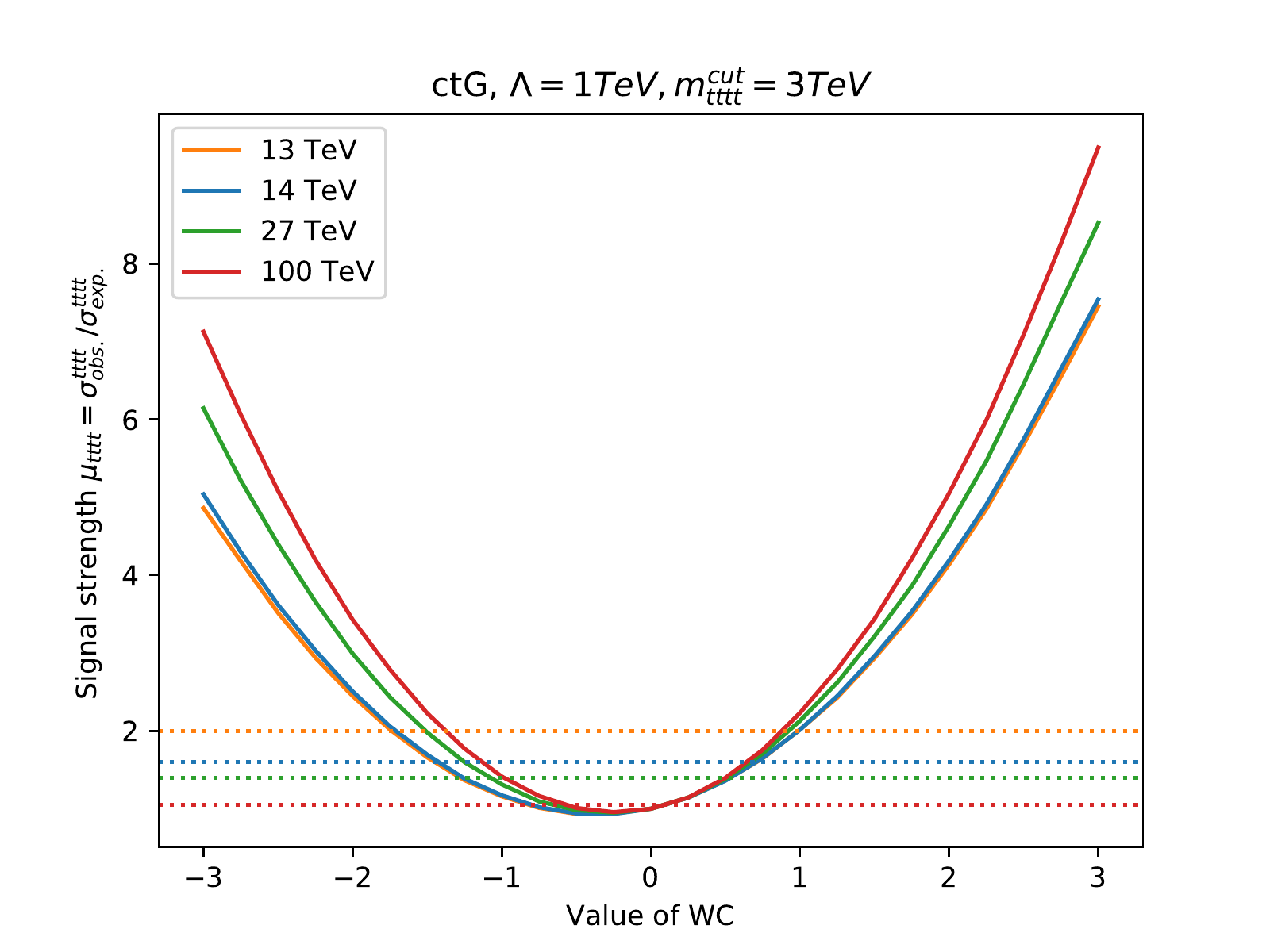}
    \caption{Same as \cref{fig:projections_4heavy} but for the \TriSi{} (\emph{left}) and \ctG{} (\emph{right}) operators.}
    \label{fig:projections_2heavy_and_ctg}
\end{figure}

\begin{figure}[h!]
    \centering
    \includegraphics[width=.48\textwidth]{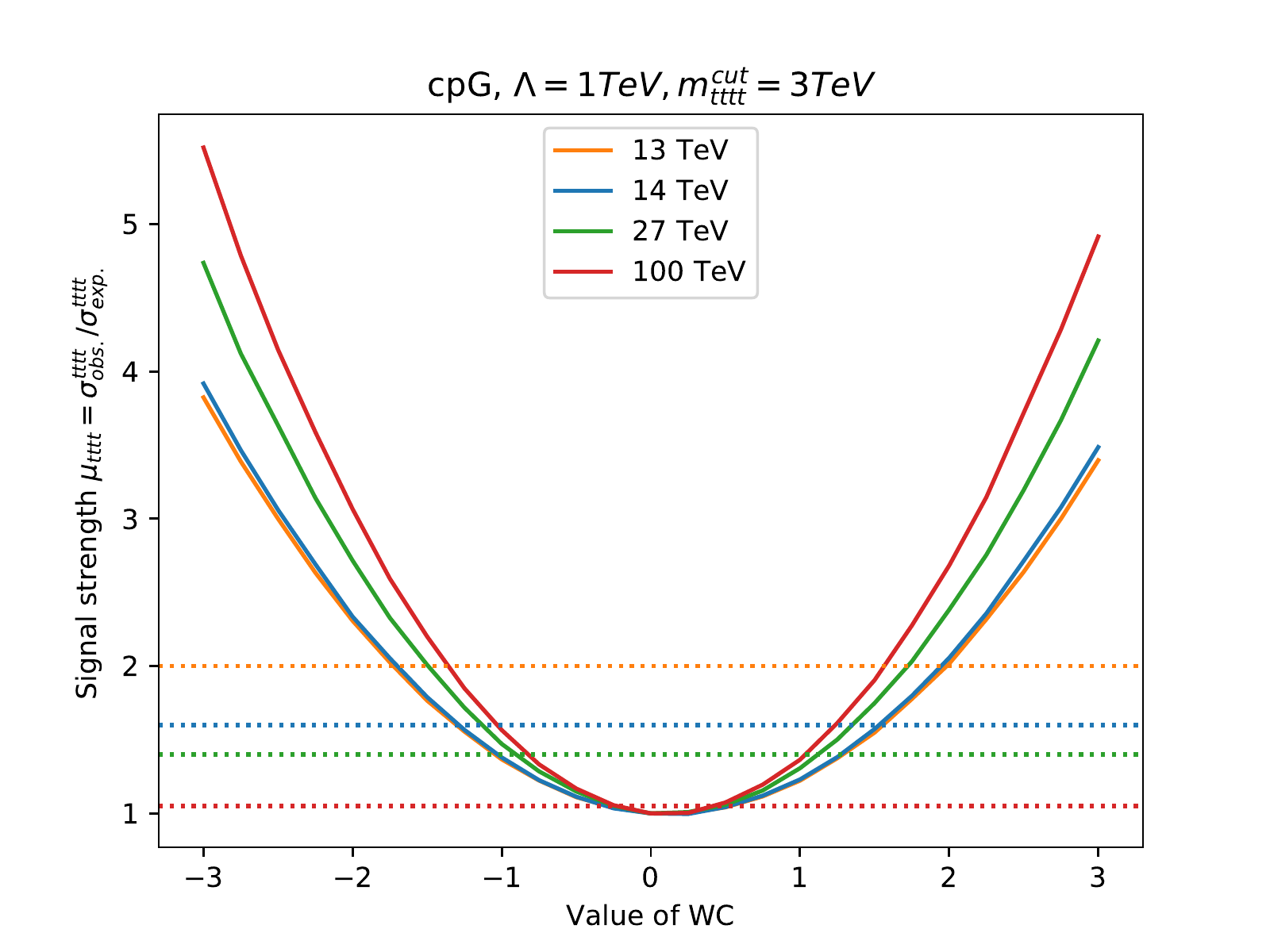}
    \includegraphics[width=.48\textwidth]{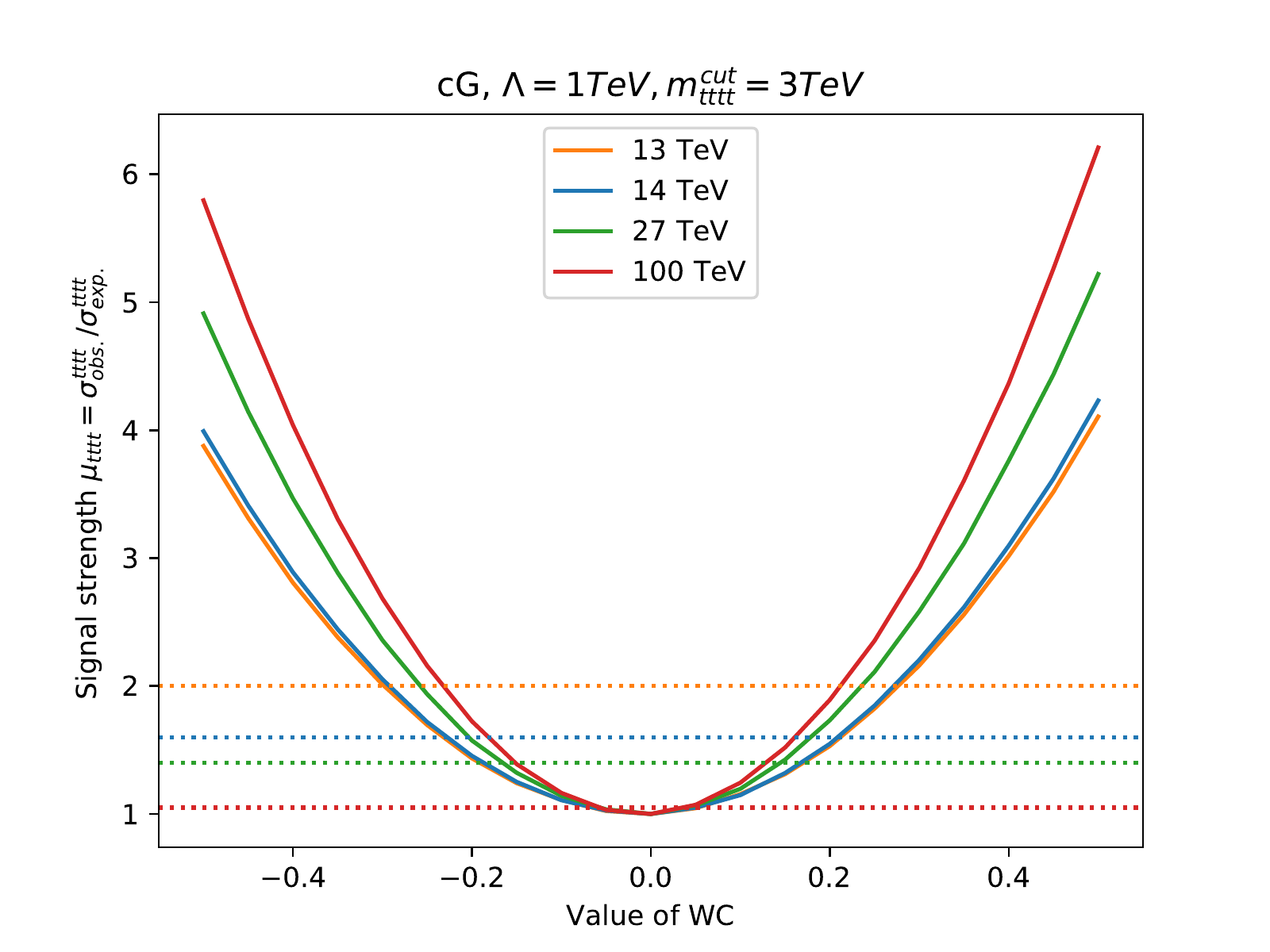}
    \caption{Same as \cref{fig:projections_4heavy} but for the \cpG{} (\emph{left}) and \cG{} (\emph{right}) operators.}
    \label{fig:projections_0f}
\end{figure}

To attain the projected sensitivity, we perform a scan over different values of the WCs, computing the \tttt signal strength, $\mu_{tttt}$, at each point. 
The signal strength is defined as $\mu_{tttt}=\sigma_{\mathrm{obs.}}^{tttt}/\sigma_{\mathrm{exp.}}^{tttt}$, where $\sigma_{\mathrm{obs.}}^{tttt}$ is the obtained SMEFT cross-section including the interference and the quadratic contributions, and $\sigma_{\mathrm{exp.}}^{tttt}$ is the \tttt cross-section assuming no EFT contributions.
The WC scans for all the 4-heavy operators presented in \cref{fig:projections_4heavy} show a significant EFT sensitivity enhancement at high collision energies, in contrast to the 2-heavy-2-light operators, of which \TriSi{} operator's projection is presented in the \emph{left} panel of \cref{fig:projections_2heavy_and_ctg}.
The reduced sensitivity of 2-heavy-2-light operators at high energies is because the $gg$-initiated production dominates over the $q\bar{q}$ one as the collision energy increases. 
Furthermore, in the \emph{right} panel of \cref{fig:projections_2heavy_and_ctg}, we show the scan of the \ctG{} operator's WC, for it being the most collision-energy-sensitive two-fermion operator in the set of contributing operators.
Finally, the projections for the purely-bosonic operators \cpG{} and \cG{} are shown in \cref{fig:projections_0f}. 

We summarise the expected individual limits on the WCs of the 4-heavy operators in \cref{tab:limits_from_projections}.
In obtaining the limits, we used the previously-mentioned expected total uncertainties at future LHC runs (represented by the horizontal dashed line in the plots), keeping an estimate of 5\% total uncertainty on the \tttt cross-section measurement at FCC-hh. 
High collision energies will certainly aid in constraining the 4-heavy effective coefficients through four-top production.
On the other hand, we expect other top quark processes to be more sensitive to the rest of the operators.
\begin{table}[h!]
\renewcommand{\arraystretch}{1.0}
\centering
\begin{tabular}{cllll}
\hline
$c_{i}$ & 13 TeV         & 14 TeV         & 27 TeV          & 100 TeV        \\
\hline\hline
$c_{QQ}^{1}$                               
& {[}-2.2,3{]}   
& {[}-1.8,2.2{]} 
& {[}-1.2,1.8{]}  
& {[}-0.25,0.7{]} \\
$c_{QQ}^{8}$                                
& {[}-6.75,9{]}  
& {[}-5,7{]}    
& {[}-3.75,5.1{]} 
& {[}-1.0,2.25{]}   \\
$c_{Qt}^{1}$                               
& {[}-2.6,2{]}   
& {[}-2,1.4{]}   
& {[}-1.4,1.1{]}  
& {[}-0.6,0.3{]}   \\
$c_{Qt}^{8}$                                
& {[}-4.2,5.3{]} 
& {[}-3.2,4{]}   
& {[}-2.1,2.7{]}  
& {[}-0.45,1.05{]} \\
$c_{tt}^{1}$                                
& {[}-1.2,1.4{]} 
& {[}-0.7,1.2{]} 
& {[}-0.6,0.8{]}  
& {[}-0.15,0.35{]}\\
\hline
\end{tabular}
    \caption{Theoretical individual limits on the 4-heavy operators' coefficients for 13, 14, 27, and 100 TeV hadron colliders at the 95\%CL level.
    Predictions are obtained including both linear and quadratic contributions in SMEFT with $\Lambda=1$ TeV.}
    \label{tab:limits_from_projections}
\end{table}
\section{Toy fits} 
\label{sec:toy}
\begin{figure}[h!]
    \centering
    \includegraphics[width=0.6\textwidth]{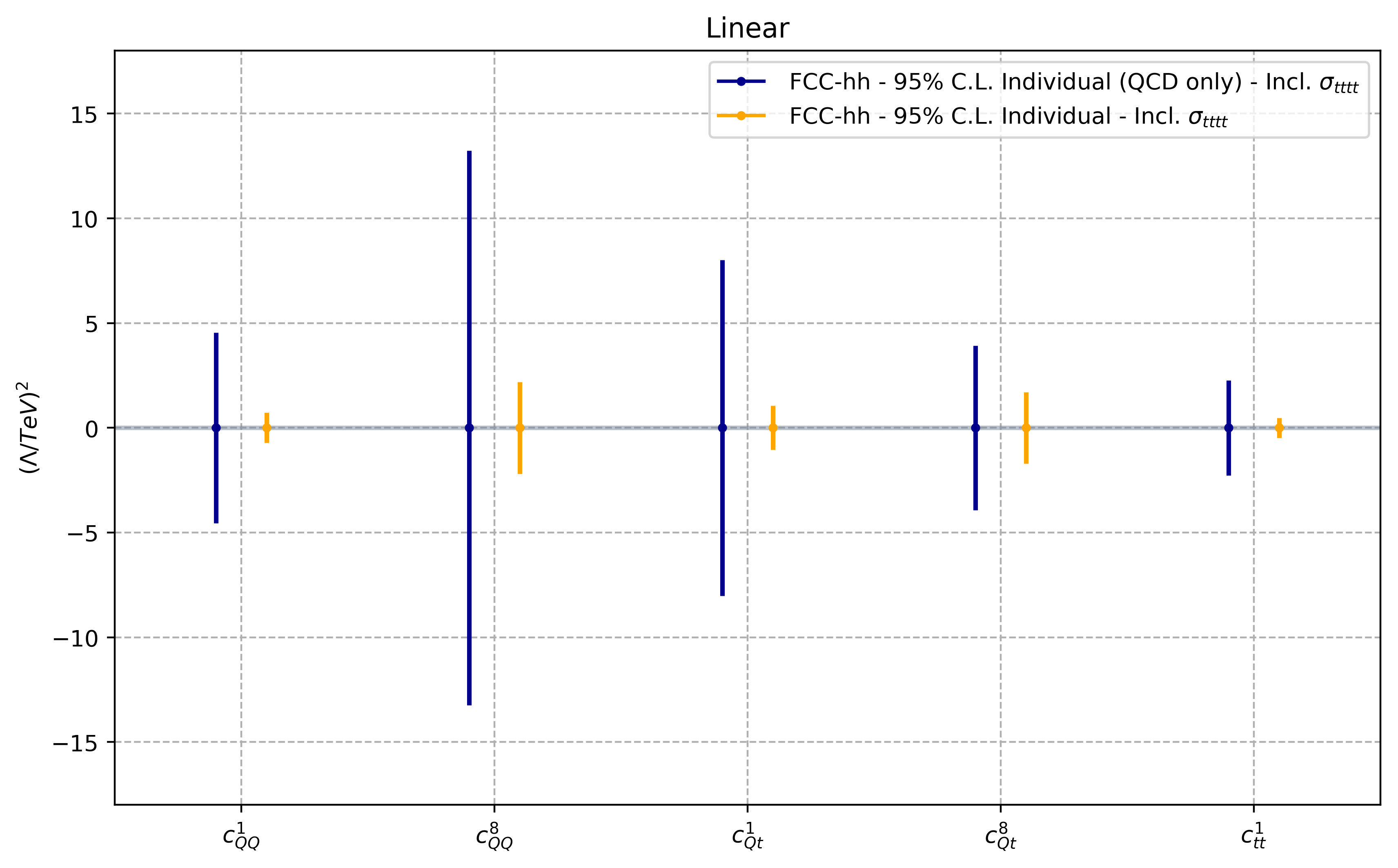}
    \caption{95\%CL limits on the 4-heavy operators' coefficients at the FCC-hh scenario from a $\rchi^{2}$ fit.
    The limits are shown when only considering leading QCD terms and when considering all the terms in the mixed QCD-EW cross-section expansion.
    The fit uses the inclusive \tttt cross-section, $\sigma_{tttt}$.
    EFT predictions were obtained at the interference level.}
    \label{fig:fit_results_bar_plot_4H_INT}
\end{figure}

In this section, we present limits on effective operators' coefficients from simplified $\rchi^{2}$ individual fits in various collider scenarios: the LHC, FCC-hh and the HL-LHC. 
We explore the impact of (\emph{i}) subleading EW terms, (\emph{ii}) differential information and (\emph{iii}) the collider energy on the WCs bounds. 
\paragraph{Impact of subleading EW terms}
We start by considering the relevance of the subleading terms in the interference cross-section expansion of the 4-heavy operators.
In \cref{fig:fit_results_bar_plot_4H_INT}, the individual limits on the 4-heavy coefficients at the FCC-hh are presented in two cases: (I) with only QCD-induced (leading) terms taken into account, and (II) when contributions to the cross-section from all tree-level terms in the mixed QCD-EW expansion are included.
For the SM prediction at the FCC-hh, we use the results of \cref{tab:sm_decomposition} with a 20\% theoretical (systematic) uncertainty.
EFT predictions include only the linear interference contributions.
For simplicity, we assume the experimental measurement to be that of the SM cross-section reported in \cref{tab:sm_decomposition} with a 5\% total (statistical and systematic) uncertainty.
The importance of the subleading terms is evident when considering \emph{only} the contributions from linear interference.
However, and since quadratic contributions of four-heavy operators are \emph{only} QCD-induced, including them in the fit would reduce the sensitivity to the subleading terms.

\begin{figure}[h!]
    \centering
    \includegraphics[width=0.6\textwidth]{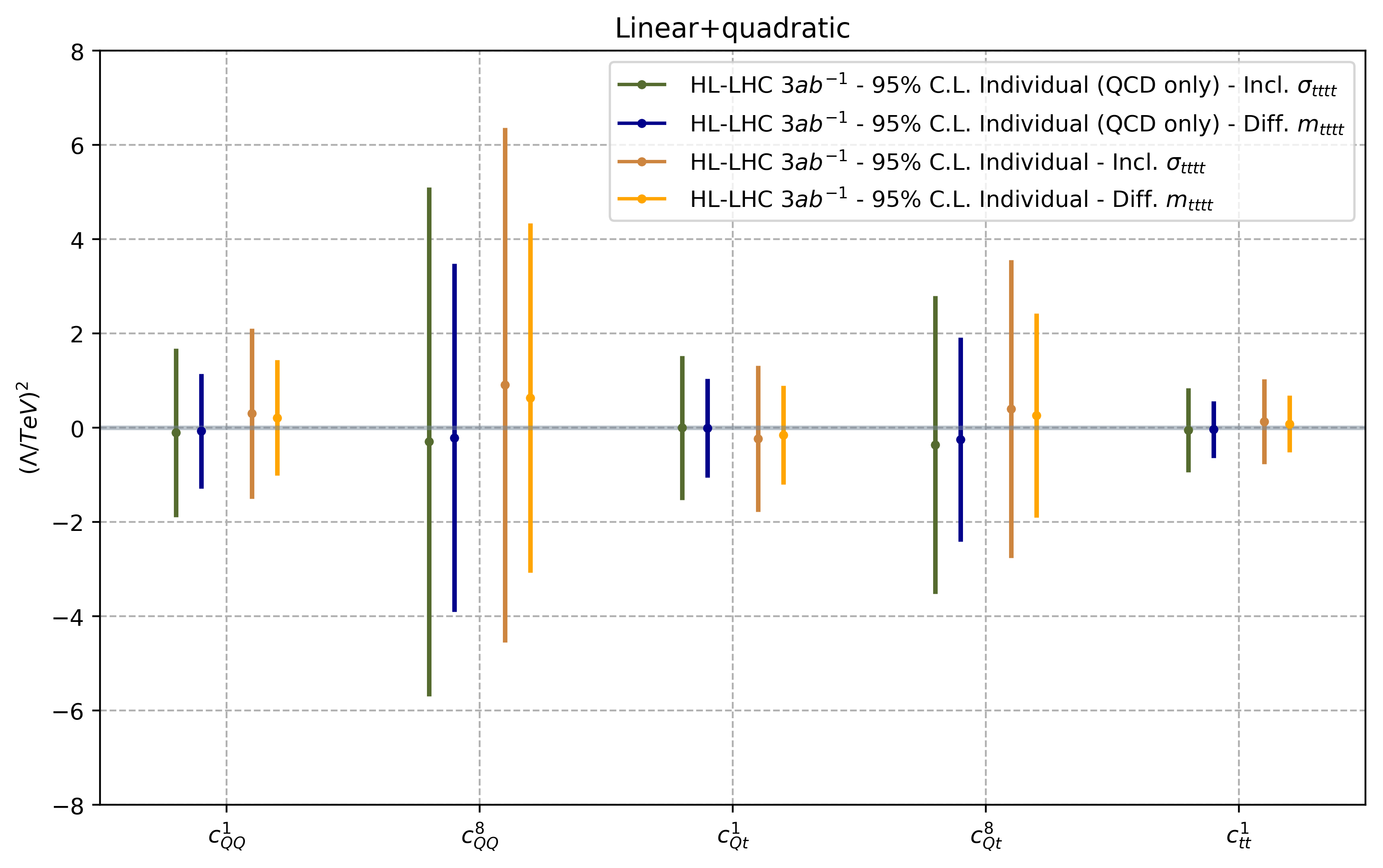}
    \caption{95\%CL limits on the 4-heavy operators' coefficients at the HL-LHC scenario from a $\rchi^{2}$ fit.
    The limits are shown for when only considering leading QCD terms and when considering all the terms, in using only inclusive information from $\sigma_{tttt}$ and when adding differential information from $m_{tttt}$.
    EFT predictions were obtained for the linear and quadratic contributions.}
    \label{fig:fit_results_bar_plot_HLHC}
\end{figure}

\paragraph{Impact of differential information} 
The HL-LHC will run at $\sqrt{s}=14$ TeV with 3 $\mathrm{ab}^{-1}$ of integrated luminosity; therefore, it is expected to obtain differential information for the four-top process experimentally.
Motivated by the larger impact of the EFT operators in the tails of distributions, as illustrated in \cref{fig:dim64f_good_diff_13tev}, we examine the impact of adding the invariant mass distribution of the four-top in our toy fit for the HL-LHC. 
\Cref{fig:fit_results_bar_plot_HLHC} displays the individual limits for the same two cases used previously (QCD-only and mixed QCD-EW) and compares the use of only inclusive information from $\sigma_{tttt}$ to when also adding differential information in the fit from $m_{tttt}$.
We use the HL-LHC SM prediction calculated at LO, $\sigma_{tttt}^{\rm HL}=9.0$ fb, with a 20\% theoretical uncertainty.
The EFT predictions include the linear and quadratic contributions.
We assume the experimental measurement to be that of the SM within the expected 28\% experimental total uncertainty~\cite{Azzi:2019yne}; $\sigma_{tttt}^{\rm HL}=9.0\pm2.52$ fb.
The $m_{tttt}$ distribution is organised in three bins: [600-1500], [1500-2500], [2500-6000] GeV, with total experimental uncertainties amounting to 28\% for each of the first two bins, and 60\% for the latter to account for the degradation of the statistical uncertainty based on the number of events expected in each bin. 
Even though very much simplified and not based on a detailed analysis of how observables could provide most of the sensitivity, our results indicate that differential information improves the sensitivity and should be used whenever possible. 

\begin{figure}[h!]
    \centering
    \includegraphics[width=1.0\textwidth]{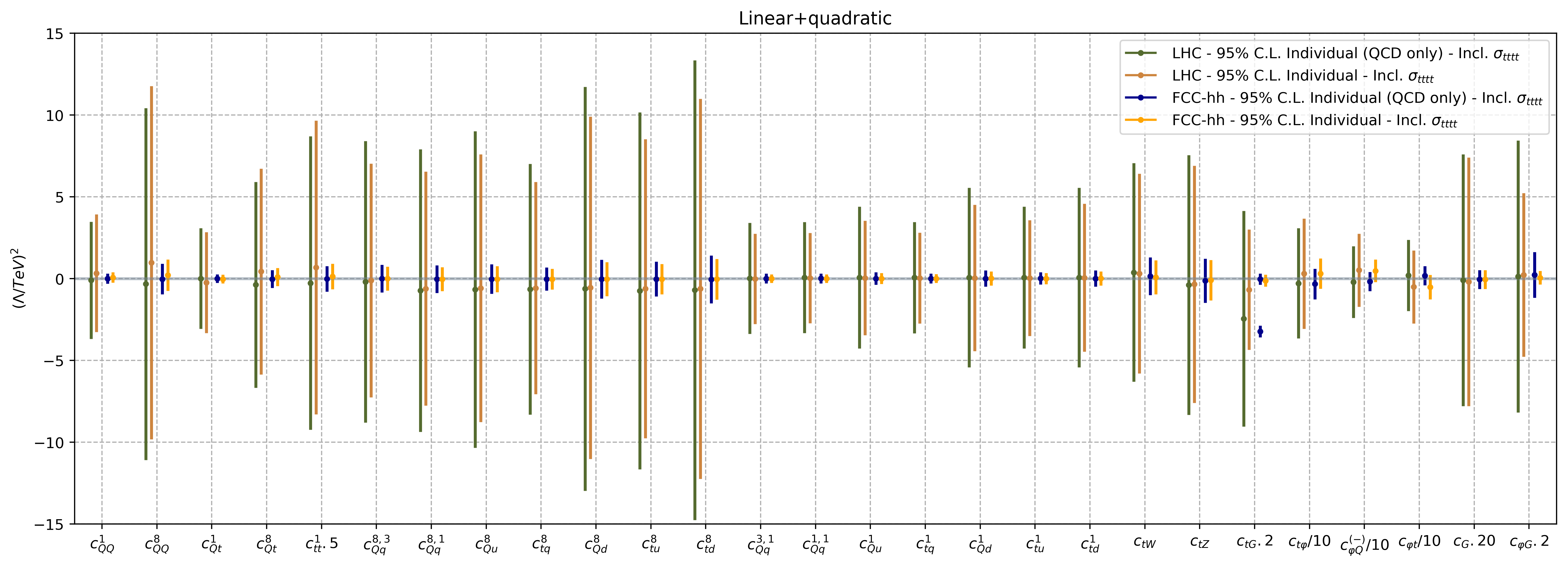}
    \caption{Limits on all four-fermion and relevant operators used in this study obtained from the $\rchi^{2}$ fit to the ATLAS~\cite{ATLAS:2020hpj} and CMS~\cite{CMS:2019rvj} inclusive measurements and using the SM prediction of Ref.~\cite{Frederix:2017wme} as well as FCC-hh projections.}
    \label{fig:fit_results_bar_plot}
\end{figure}

\paragraph{Comparison of different collider setups} 
To fully appreciate the impact of collider energy in constraining the relevant coefficients, we compare the results from current LHC measurements with the FCC-hh bounds.
For simplicity, we only use the inclusive cross-section.
The limits obtained from the fit are presented in \cref{fig:fit_results_bar_plot}. 
For both scenarios, EFT predictions include the linear and quadratic contributions.
For the LHC, we use the SM prediction at NLO in QCD of Ref.~\cite{Frederix:2017wme}, and we fit the theoretical predictions to the inclusive ATLAS~\cite{ATLAS:2020hpj} and CMS~\cite{CMS:2019rvj} measurements.
For the FCC-hh, we use the same theoretical and experimental inputs used for the previous case of \cref{fig:fit_results_bar_plot_4H_INT}.
The results from this fit show the significant constraining power that the FCC-hh will be able to provide for the SMEFT coefficients. 
Again, the effects from the subleading terms are diluted by including the quadratic contributions in the predictions. 
Finally, we note that it is expected that with the high-energy reach of the FCC-hh, differential distributions extending well into the multi-TeV range will become available and further improve the bounds beyond those expected from the inclusive cross-section.   
\section{Double insertion}\label{sec:double_insertions}
In this section, we critically assess Ref.~\cite{Zhang:2017mls}, where it was suggested that 2-heavy-2-light operators could be better constrained in \tttt than in $t\bar t$ production.
This suggestion was spurred by the results of Ref.~\cite{CMS:2017tec} reporting an upper limit on the \tttt cross-section to be 4.6 times that of the SM. 
Due to the high-energy scale related to the \tttt process, its cross-section depending on the fourth power of the operators' coefficients scales as $\sim (cE^{2}/\Lambda^{2})^{4}$, an order that double insertion of dimension-six operators can probe. 
Ref.~\cite{Zhang:2017mls} argued these terms enhance the EFT sensitivity of the 2-heavy-2-light operators to a level at which four-top can compete with top pair production in constraining said operators.

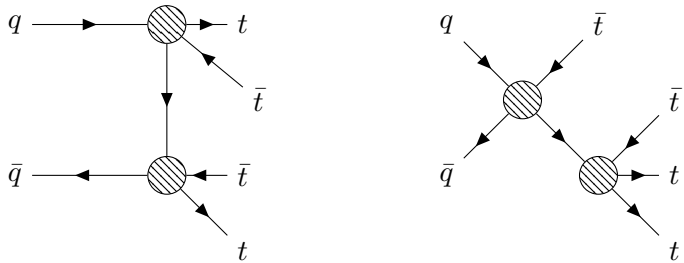
\begin{figure}[h!]%
    \centering
    \begin{tikzpicture}
\begin{feynman}[small]
\vertex (a0) {};
\vertex[right = of a0] (a1) {};
\vertex[right = of a1] (a2) {};
\vertex[right = of a2] (a3) {};
\vertex[right = of a3] (a4) {};
\vertex[below = of a0] (c0) {$q$};
\vertex[right = of c0] (c1) {};
\vertex [blob] (c2) at (2,-1){};
\vertex[right = of c2] (c3) {$t$};
\vertex[right = of c3] (c4) {};
\vertex[right = of c4] (c5) {};
\vertex[below = of c0] (b0) {};
\vertex[right = of b0] (b1) {};
\vertex[right = of b1] (b2);
\vertex[right = of b2] (b3) {$\bar{t}$};
\vertex[right = of b3] (b4) {};
\vertex[right = of b4] (b5) {};
\vertex[below = of b0] (d0) {$\bar{q}$};
\vertex[right = of d0] (d1){};
\vertex [blob] (d2) at (2,-3){};
\vertex[right = of d2] (d3) {$\bar{t}$};
\vertex[right = of d3] (d4) {};
\vertex[right = of d4] (d5) {};
\vertex[below = of d0] (e0) {};
\vertex[right = of e0] (e1) {};
\vertex[right = of e1] (e2) {};
\vertex[right = of e2] (e3) {$t$};
\vertex[right = of e3] (e4) {};
\diagram*{
(c0) -- [fermion] (c2),
(d0) -- [anti fermion] (d2),
(c2) -- [fermion] (d2),
(c2) -- [anti fermion] (b3),
(c2) -- [fermion] (c3),
(d2) -- [anti fermion] (d3),
(d2) -- [fermion] (e3),
};
\end{feynman}
\end{tikzpicture}
\begin{tikzpicture}
\begin{feynman}[small]
\vertex (a0) {};
\vertex[right = of a0] (a1) {};
\vertex[right = of a1] (a2) {};
\vertex[right = of a2] (a3) {};
\vertex[right = of a3] (a4) {};
\vertex[below = of a0] (c0) {$q$};
\vertex[right = of c0] (c1) {};
\vertex[right = of c1] (c2) {$\bar{t}$};
\vertex[right = of c2] (c3) {};
\vertex[right = of c3] (c4) {};
\vertex[right = of c4] (c5) {};
\vertex[below = of c0] (b0) {};
\vertex[right = of b0] (b1) {};
\vertex [blob] (b1) at (1,-2){};
\vertex[right = of b1] (b2) {};
\vertex[right = of b2] (b3) {$\bar{t}$};
\vertex[right = of b3] (b4) {};
\vertex[right = of b4] (b5) {};
\vertex[below = of b0] (d0) {$\bar{q}$};
\vertex[right = of d0] (d1){};
\vertex [blob] (d2) at (2,-3){};
\vertex[right = of d2] (d3) {$t$};
\vertex[right = of d3] (d4) {};
\vertex[right = of d4] (d5) {};
\vertex[below = of d0] (e0) {};
\vertex[right = of e0] (e1) {};
\vertex[right = of e1] (e2) {};
\vertex[right = of e2] (e3) {$t$};
\vertex[right = of e3] (e4) {};
\diagram*{
(c0) -- [fermion] (b1),
(d0) -- [anti fermion] (b1),
(b1) -- [anti fermion] (c2),
(b1) -- [fermion] (d2),
(d2) -- [fermion] (e3),
(d2) -- [fermion] (d3),
(d2) -- [anti fermion] (b3),
};
\end{feynman}
\end{tikzpicture}
    \caption{Representative diagrams of four-top production with two EFT insertions represented by the shaded blobs.}
    \label{fig:double_insertions_diags}
\end{figure}

Our study investigates the strength of the double-insertion contributions in four-top production. 
In particular, we compare the EFT sensitivity from double-insertion to that from the squared single-insertion of the same 2-heavy-2-light operator.
As previously discussed, schematically, these contributions at $\mathscr{O}(\Lambda^{-4})$ can be respectively written as follows:
\begin{align}
    d\sigma_{\rm dbl} \sim |\mathcal{A}_{\rm SM}\,\mathcal{A}_{\rm (d6)^2}|
    \qquad
    \text{vs.}
    \qquad
    d\sigma_{\rm quad} \sim |\mathcal{A}_{\rm (d6)}|^2.
\end{align}
The Feynman diagrams depicting the amplitudes with two dimension-six EFT insertions are shown in \cref{fig:double_insertions_diags}.
 \vspace{-0.5cm}

As a first step, we proceed by reproducing the predictions of Ref.~\cite{Zhang:2017mls}.
We indeed observed that the double-insertion contributions enhance the cross-section compared to the squared ones.
However, this is only true given the loose constraints on WCs, of order $\mathscr{O}(5-10)$, that Ref.~\cite{Zhang:2017mls} considered.
Given the current comparatively stringent bounds from the global study of Ref.~\cite{Ethier:2021bye}, we do not find an enhanced EFT sensitivity due to double insertion. 
Our results are presented in \cref{tab:double_insertions_final_results} where we fixed the value of $c$ to unity and denoted amplitudes with one EFT insertion with $\mathcal{A}_1$ and those with two insertions with $\mathcal{A}_2$.
\begin{table}[h!]
    \centering
\renewcommand{\arraystretch}{1}
\centering
\begin{tabular}{cccc|ccc}
\hline
\multirow{2}{*}{}     & \multicolumn{5}{c}{2-heavy 2-light at $c_{i}$=1}    
&  
\\ 
& \multicolumn{3}{c}{$\sqrt{s}=13$ TeV}                                                        
& \multicolumn{3}{c}{$\sqrt{s}=100$ TeV}
\\ \hline
$\mathcal{O}_{i}$     & $|\mathcal{A}_{1}|^2$ {[}fb{]} & $\sum_{k}\mathscr{O}(\mathcal{A}_{2})_{k}$ {[}fb{]} & ratio & $|\mathcal{A}_{1}|^2$ {[}fb{]} & $\sum_{k}\mathscr{O}(\mathcal{A}_{2})_{k}$ {[}fb{]} & ratio 
\\ \hline\hline
\OctTr 
& 0.27                                    & 0.01                                                 & 0.04
& 6.40                                    & 0.40                                                 & 0.06     
\\
\OctSi 
& 0.28                                    & 0.05                                                 & 0.18            
& 6.36                                    & 0.63                                                 & 0.10      
\\
\QuOct 
& 0.21                                    & 0.03                                                 & 0.14    
& 5.34                                    & 0.50                                                 & 0.09      
\\
\tqOct 
& 0.34                                    & 0.06                                                 & 0.18    
& 8.44                                    & 0.76                                                 & 0.09       
\\
\QdOct 
& 0.13                                    & 0.03                                                 & 0.23      
& 3.13                                    & 0.35                                                 & 0.11       
\\
\tuOct 
& 0.17                                    & 0.03                                                 & 0.18      
& 3.97                                    & 0.41                                                 & 0.10       
\\
\tdOct 
& 0.10                                    & 0.02                                                 & 0.20     
& 2.18                                    & 0.27                                                 & 0.12       
\\
\hline
\TriSi 
& 1.84                                    & 0.15                                                 & 0.08      
& 46.98                                    & 5.49                                                 & 0.12       
\\
\SiSi  
& 1.84                                    & 0.08                                                 & 0.04      
& 47.35                                    & 0.81                                                 & 0.02      
\\
\QuSi  
& 1.14                                    & 0.06                                                 & 0.05      
& 29.94                                    & 2.83                                                 & 0.09       
\\
\tqSi  
& 1.80                                    & 0.14                                                 & 0.08      
& 46.54                                    & 6.33                                                 & 0.14       
\\
\QdSi  
& 0.70                                    & 0.08                                                 & 0.11      
& 17.55                                    & 2.15                                                 & 0.12       
\\
\tuSi  
& 1.11                                    & 0.04                                                 & 0.04       
& 29.10                                    & 2.48                                                 & 0.09       
\\
\tdSi  
& 0.68                                    & 0.05                                                 & 0.07      
& 17.44                                    & 1.79                                                 & 0.10      
\\
\hline
\end{tabular}

    \caption{Cross-section predictions from the diagonal squared single-insertion contributions at $\mathscr{O}(\Lambda^{-4})$ are denoted by $|\mathcal{A}_{1}|^{2}$ and compared to the sum of all double-insertion contributions, $\mathscr{O}(\mathcal{A}_{2})$, up to $\mathscr{O}(\Lambda^{-8})$. 
    The ratio column is that of double-insertion contributions to squared single-insertion ones.
    The results are presented for the LHC and FCC-hh collider setups.}
    \label{tab:double_insertions_final_results}
\end{table}
We fixed $\Lambda$ to 1(3) TeV for the 13(100) TeV predictions, and summed all contributions arising from double insertion up to $\mathscr{O}(\Lambda^{-8})$.
We see that for all the 2-heavy-2-light operators, the contributions from the double insertion are negligible compared to ones from squared single-insertion, namely $\mathscr{O}(10)$ smaller. 
A similar pattern is observed for the FCC-hh case.

It is worth mentioning that while the squared dimension-six single-insertion in the EFT expansion is invariant under field transformation, the term corresponding to a double-insertion of dimension-six is not invariant unless dimension-eight operators are taken into account.
\section{Summary and conclusions}
\label{sec:conclusions}
This work presented a complete analysis of four top quark production at hadron colliders within the SMEFT framework. 
We have based our studies on predictions at the tree-level, yet including all the possible QCD- and EW-coupling order contributions, keeping gauge and top-Yukawa couplings separate. 
Observables were computed in the SMEFT by considering linear, quadratic, and in specific instances, double-insertion contributions of dimension-six operators. 
Within the large set of SMEFT operators possibly contributing to four-top production, we have identified a subclass, named  \emph{non-naive}, consisting of all the four heavy operators and the following subset of two-fermion and bosonic operators $\{\mathcal{O}_{Qq}^{3,1},\mathcal{O}_{t\phi},\mathcal{O}_{tG},\mathcal{O}_{\phi Q}^{(-)},\mathcal{O}_{\phi t},\mathcal{O}_{\phi G}\}$, i.e. the operators whose leading contributions at the linear level to four-top production cross-section arise from formally subleading terms in the QCD-EW expansion.

We have then analysed the operators' contributions to four-top production at the LHC and the FCC-hh colliders. 
Three main conclusions can be drawn. 
First, the 4-heavy operators provide the most significant contribution through the $\mathscr{O}(\alpha_s^2\alpha_{\mathrm{w}})$ terms. 
The same happens for the remaining six operators in the non-naive set.
Second, $\mathscr{O}(\alpha_s^2\alpha)$ is dominant compared to $\mathscr{O}(\alpha_s^2\alpha_t)$ coupling orders for the four-heavy operators in the non-naive set. 
Third, from $t\bar{t}$ production, one would naively expect that colour-octets would provide the dominant contributions. 
However, we observe the opposite in \tttt production; within the non-naive set, the colour-singlets $c_{QQ}^1,c_{Qt}^1, c_{tt}^1$ have larger linear interference cross-sections compared to their colour-octet counterparts. 
The summary of these results can be found in \cref{tab:summary_inclusive}. 
Apart from some slight differences associated with $\OO_{Qt}^1, \OO_{Qu}^1, \OO^1_{tq}$ and $\OO_{\varphi G}$, the pattern for all the other 19 operators remains unchanged when the energy is increased to 100 TeV in the FCC-hh scenario.
In \cref{sec:differential}, we have considered the four-top invariant mass differential distributions, $m_{tttt}$, where the energy growth of each operator can be studied. 
The results on the differential level corroborate the conclusions drawn at the inclusive one.
Ideally, a complete NLO calculation should be performed as in Ref.~\cite{Frederix:2018nkq}, where significant (and exact) accidental cancellations between NLO EW contributions were observed in the SM. 
While the size of these cancellations or the lack thereof could change the expectations of the size of the corrections based on simple coupling scaling, we do not expect they would significantly alter the conclusions drawn here.

Looking ahead, we have considered in \cref{sec:projections} and \cref{sec:toy} the sensitivity attainable at future colliders, i.e.,  at 13, 14 and 27 TeV for the LHC and the FCC-hh at 100 TeV. 
As our previous LHC studies hinted, all the 4-heavy operators have a sensitivity enhancement at high collision energies, in contrast to the 2-heavy-2-light operators. 
Consequently, the expected limits on their Wilson coefficients are the most stringent, especially for colour-singlets.

We presented a study of the double insertion of dimension-six operators in four-top production.
The aim was to scrutinise the claim of enhanced sensitivity in four-top production from multiple insertions of 2-heavy-2-light operators. 
Given the current bounds on this set of operators, we find that the sensitivity is not enhanced. 
This finding supports the conclusion that 2-heavy-2-light operators are better constrained elsewhere than in four-top production. 

We stress the importance of our results in summarising the four-top production SMEFT predictions and analysing them in each order of the QCD-EW expansion, underlining the significance of subleading terms for the four-heavy quark operators.
Moreover, within all the dimension-six SMEFT operators, we have shown that four top quark production provides robust constraints on the four-heavy coefficients.

Finally, our analysis motivates an effort towards a systematic study of subleading effects in other processes, such as e.g. $t\bar{t}Z$ and, in particular, $t\bar{t}W(+$jets). 
Once technically possible, such studies should be upgraded to the NLO accuracy, including QCD and EW corrections, to control the uncertainties fully. 

\section*{Acknowledgments}
We thank Olivier Mattelaer, Ken Mimasu, Alexander Ochirov, Davide Pagani, and Marco Zaro for valuable discussions. 
This work has received funding from the European Union's Horizon 2020 research and innovation program as part of the Marie Skłodowska-Curie Innovative Training Network MCnetITN3 (grant agreement no. 722104) and by the F.R.S.-FNRS under the ``Excellence of Science'' EOS be.h project no. 30820817. 
R.A.’s research is funded by the F.R.S-FNRS project no. 40005600.
E.V. has received funding from the European Research Council (ERC) under the European Union's Horizon 2020 research and innovation programme (grant agreement No. 949451) and from a Royal Society University Research Fellowship through grant URF/R1/201553.
Computational resources have been provided by the supercomputing facilities of the Université catholique de Louvain (CISM/UCL) and the Consortium des Équipements de Calcul Intensif en Fédération Wallonie Bruxelles (CÉCI) funded by the Fond de la Recherche Scientifique de Belgique (F.R.S.-FNRS) under convention 2.5020.11 and by the Walloon Region.

\appendix
\section{Translations and constraints}\label{sec:app_4f_def_constraints}
\Cref{tab:dim64f_smeftatnlo_basis} presents the definitions of the \texttt{SMEFTatNLO} 4F operators, $\mathcal{O}_{i}$, in terms of the Warsaw basis coefficients.  
Respectively, \cref{tab:dim64f_wc_bounds} and \cref{tab:contributing_operators_wc_bounds} present the bounds on the 4F and contributing operators (except for \cG{}) obtained from the global fit of Ref.~\cite{Ethier:2021bye}.
\begin{table}[h!]
\centering
\resizebox{\textwidth}{!}{
\renewcommand{\arraystretch}{1.0}
\centering
\begin{tabular}{cccccc}
\hline
\multicolumn{6}{c}{\textbf{2-heavy 2-light}}                                 
\\ \hline\hline
    \multicolumn{1}{c}{\textbf{$\Op{i}$}} 
&   \multicolumn{1}{c}{\textbf{UFO}}     
&   \multicolumn{1}{c}{\textbf{Translation}}                                                                                   
&   \multicolumn{1}{c}{\textbf{$\Op{i}$}} 
&   \multicolumn{1}{c}{\textbf{UFO}}     
&   \multicolumn{1}{c}{\textbf{Translation}}                          
\\ \hline
    \multicolumn{1}{l}{\SiSi}   
&   \multicolumn{1}{l}{$\texttt{cQq11}$} 
&   \multicolumn{1}{l}{$\sum\limits_{i=1,2}[C_{qq}^{(1)}]^{ii33}+\frac{1}{6}[C_{qq}^{(1)}]^{i33i}+\frac{1}{2}[C_{qq}^{(3)}]^{i33i}$} 
&   \multicolumn{1}{l}{\OctSi}   
&   \multicolumn{1}{l}{$\texttt{cQq18}$} 
&   $\sum\limits_{i=1,2}[C_{qq}^{(1)}]^{i33i}+3[C_{qq}^{(3)}]^{i33i}$ 
\\ \hline
    \multicolumn{1}{l}{\TriSi}   
&   \multicolumn{1}{l}{$\texttt{cQq31}$} 
&   \multicolumn{1}{l}{$\sum\limits_{i=1,2}[C_{qq}^{(3)}]^{ii33}+\frac{1}{6}[C_{qq}^{(1)}]^{i33i}-\frac{1}{6}[C_{qq}^{(3)}]^{i33i}$} 
&   \multicolumn{1}{l}{$\Opp{Qq}{3,8}$}   
&   \multicolumn{1}{l}{$\texttt{cQq38}$} 
&   $\sum\limits_{i=1,2}[C_{qq}^{(1)}]^{i33i}-[C_{qq}^{(3)}]^{i33i}$  
\\ \hline
    \multicolumn{1}{l}{\tuSi}     
&   \multicolumn{1}{l}{$\texttt{ctu1}$}  
&   \multicolumn{1}{l}{$\sum\limits_{i=1,2}[C_{uu}]^{ii33}+\frac{1}{3}[C_{uu}]^{i33i}$}         
&   \multicolumn{1}{l}{\tuOct}     
&   \multicolumn{1}{l}{$\texttt{ctu8}$}  
&   $\sum\limits_{i=1,2}2[C_{uu}]^{i33i}$                             
\\ \hline
    \multicolumn{1}{l}{\tdSi}     
&   \multicolumn{1}{l}{$\texttt{ctd1}$}  
&   \multicolumn{1}{l}{$\sum\limits_{i=1,2(,3)}[C_{ud}^{(1)}]^{33ii}$}     
&   \multicolumn{1}{l}{\tdOct}     
&   \multicolumn{1}{l}{$\texttt{ctd8}$}  
&   $\sum\limits_{i=1,2(,3)}[C_{ud}^{(8)}]^{33ii}$                    
\\ \hline
    \multicolumn{1}{l}{\tqSi}     
&   \multicolumn{1}{l}{$\texttt{ctq1}$}  
&   \multicolumn{1}{l}{$\sum\limits_{i=1,2}[C_{qu}^{(1)}]^{ii33}$}                              
&   \multicolumn{1}{l}{\tqOct}     
&   \multicolumn{1}{l}{$\texttt{ctq8}$}  
&   $\sum\limits_{i=1,2}[C_{qu}^{(8)}]^{ii33}$                        
\\ \hline
    \multicolumn{1}{l}{\QuSi}     
&   \multicolumn{1}{l}{$\texttt{cQu1}$}  
&   \multicolumn{1}{l}{$\sum\limits_{i=1,2}[C_{qu}^{(1)}]^{33ii}$}                                               
&   \multicolumn{1}{l}{\QuOct}     
&   \multicolumn{1}{l}{$\texttt{cQu8}$}  
&   $\sum\limits_{i=1,2}[C_{qu}^{(8)}]^{33ii}$                        
\\ \hline
    \multicolumn{1}{l}{\QdSi}     
&   \multicolumn{1}{l}{$\texttt{cQd1}$}  
&   \multicolumn{1}{l}{$\sum\limits_{i=1,2,(3)}[C_{qd}^{(1)}]^{33ii}$}                                           
&   \multicolumn{1}{l}{\QdOct}     
&   \multicolumn{1}{l}{$\texttt{cQd8}$}  
&   $\sum\limits_{i=1,2,(3)}[C_{qd}^{(8)}]^{33ii}$                        
\\ 
\multicolumn{6}{c}{\textbf{4-heavy}}                                             
\\ \hline\hline
    \multicolumn{1}{l}{\QQSi}     
&   \multicolumn{1}{l}{$\texttt{cQQ1}$}  
&   \multicolumn{1}{l}{$2[C_{qq}^{(1)}]^{3333}-\frac{2}{3}[C_{qq}^{(3)}]^{3333}$}              
&   \multicolumn{1}{l}{\QQOct}     
&   \multicolumn{1}{l}{$\texttt{cQQ8}$}  
&   $8[C_{qq}^{(3)}]^{3333}$                                          
\\ \hline
    \multicolumn{1}{l}{\QtSi}    
&   \multicolumn{1}{l}{$\texttt{cQt1}$} 
&   \multicolumn{1}{l}{$[C_{qu}^{(1)}]^{3333}$}                                                
&   \multicolumn{1}{l}{\QtOct}     
&   \multicolumn{1}{l}{$\texttt{cQt8}$}  
&   $[C_{qu}^{(8)}]^{3333}$                                           
\\ \hline
    \multicolumn{1}{l}{\ttSi}         
&   \multicolumn{1}{l}{$\texttt{ctt1}$}   
&   \multicolumn{1}{l}{$[C_{uu}^{(1)}]^{3333}$}                                                
&   \multicolumn{3}{l}{}                                                                        
\\\hline
\end{tabular}
}
\caption{\label{tab:dim64f_smeftatnlo_basis}
The translation of four-fermion operators from the Warsaw basis to the top-basis. 
The \texttt{UFO} column shows the notation of the corresponding WCs in the \texttt{SMEFTatNLO} model.}
\end{table}
\begin{table}[h!]
\centering
\resizebox{\textwidth}{!}{
\renewcommand{\arraystretch}{1.0}
\centering
\begin{tabular}{llllllllll}
\hline
\multicolumn{10}{c}{\textbf{2-heavy 2-light}}                                                       
\\ \hline\hline
\multicolumn{1}{c}{\multirow{2}{*}{\textbf{UFO}}} 
& \multicolumn{2}{c}{\textbf{$\mathscr{O}(\Lambda^{-2})$}}                                   
& \multicolumn{2}{c}{\textbf{$\mathscr{O}(\Lambda^{-4})$}}                                   
& \multicolumn{1}{c}{\multirow{2}{*}{\textbf{UFO}}} 
& \multicolumn{2}{c}{\textbf{$\mathscr{O}(\Lambda^{-2})$}}                                    
& \multicolumn{2}{c}{\textbf{$\mathscr{O}(\Lambda^{-4})$}}                                   
\\ \cline{2-5} \cline{7-10} 
\multicolumn{1}{c}{}                                   
& \multicolumn{1}{c}{\textbf{Individual}}
& \multicolumn{1}{c}{\textbf{Marginalised}} 
& \multicolumn{1}{c}{\textbf{Individual}} 
& \multicolumn{1}{c}{\textbf{Marginalised}} 
& \multicolumn{1}{c}{}                                   
& \multicolumn{1}{c}{\textbf{Individual}}  
& \multicolumn{1}{c}{\textbf{Marginalised}} 
& \multicolumn{1}{c}{\textbf{Individual}} 
& \multicolumn{1}{c}{\textbf{Marginalised}} 
\\ \hline
\multicolumn{1}{l}{\textbf{$\texttt{cQq11}$}}           
& \multicolumn{1}{l}{{[}-3.603,0.307{]}}  
& \multicolumn{1}{l}{{[}-8.047,9.400{]}}    
& \multicolumn{1}{l}{{[}-0.303,0.225{]}}  
& \multicolumn{1}{l}{{[}-0.354,0.249{]}}    
& \multicolumn{1}{l}{\textbf{$\texttt{cQq18}$}}           
& \multicolumn{1}{l}{{[}-0.273,0.509{]}}   
& \multicolumn{1}{l}{{[}-2.258,4.822{]}}    
& \multicolumn{1}{l}{{[}-0.373,0.309{]}}  
& {[}-0.555,0.236{]}                         
\\ \hline
\multicolumn{1}{l}{\textbf{$\texttt{cQq31}$}}           
& \multicolumn{1}{l}{{[}-0.099,0.155{]}}  
& \multicolumn{1}{l}{{[}-0.163,0.296{]}}    
& \multicolumn{1}{l}{{[}-0.088,0.166{]}}  
& \multicolumn{1}{l}{{[}-0.167,0.197{]}}    
& \multicolumn{1}{l}{\textbf{$\texttt{cQq38}$}}           
& \multicolumn{1}{l}{{[}-1.813,0.625{]}}   
& \multicolumn{1}{l}{{[}-3.014,7.365{]}}    
& \multicolumn{1}{l}{{[}-0.470,0.439{]}}  
& {[}-0.462,0.497{]}                         
\\ \hline
\multicolumn{1}{l}{\textbf{$\texttt{ctu1}$}}             
& \multicolumn{1}{l}{{[}-6.046,0.424{]}}  
& \multicolumn{1}{l}{{[}-15.565,15.379{]}}  
& \multicolumn{1}{l}{{[}-0.380,0.293{]}}  
& \multicolumn{1}{l}{{[}-0.383,0.331{]}}    
& \multicolumn{1}{l}{\textbf{$\texttt{ctu8}$}}             
& \multicolumn{1}{l}{{[}-0.774,0.607{]}}   
& \multicolumn{1}{l}{{[}-16.952,0.368{]}}   
& \multicolumn{1}{l}{{[}-0.911,0.347{]}}  
& {[}-1.118,0.260{]}                         
\\ \hline
\multicolumn{1}{l}{\textbf{$\texttt{ctd1}$}}             
& \multicolumn{1}{l}{{[}-9.504,-0.086{]}} 
& \multicolumn{1}{l}{{[}-27.673,11.356{]}}  
& \multicolumn{1}{l}{{[}-0.449,0.371{]}}  
& \multicolumn{1}{l}{{[}-0.474,0.347{]}}    
& \multicolumn{1}{l}{\textbf{$\texttt{ctd8}$}}             
& \multicolumn{1}{l}{{[}-1.458,1.365{]}}   
& \multicolumn{1}{l}{{[}-5.494,25.358{]}}   
& \multicolumn{1}{l}{{[}-1.308,0.638{]}}  
& {[}-1.329,0.643{]}                         
\\ \hline
\multicolumn{1}{l}{\textbf{$\texttt{ctq1}$}}             
& \multicolumn{1}{l}{{[}-0.784,2.771{]}}  
& \multicolumn{1}{l}{{[}-12.382,6.626{]}}   
& \multicolumn{1}{l}{{[}-0.205,0.271{]}}  
& \multicolumn{1}{l}{{[}-0.222,0.226{]}}    
& \multicolumn{1}{l}{\textbf{$\texttt{ctq8}$}}             
& \multicolumn{1}{l}{{[}-0.396,0.612{]}}   
& \multicolumn{1}{l}{{[}-4.035,4.394{]}}    
& \multicolumn{1}{l}{{[}-0.483,0.393{]}}  
& {[}-0.687,0.186{]}                         
\\ \hline
\multicolumn{1}{l}{\textbf{$\texttt{cQu1}$}}             
& \multicolumn{1}{l}{{[}-0.938,2.462{]}}  
& \multicolumn{1}{l}{{[}-16.996,1.072{]}}   
& \multicolumn{1}{l}{{[}-0.281,0.371{]}}  
& \multicolumn{1}{l}{{[}-0.207,0.339{]}}    
& \multicolumn{1}{l}{\textbf{$\texttt{cQu8}$}}             
& \multicolumn{1}{l}{{[}-1.508,1.022{]}}   
& \multicolumn{1}{l}{{[}-12.745,13.758{]}}  
& \multicolumn{1}{l}{{[}-1.007,0.521{]}}  
& {[}-1.002,0.312{]}                         
\\ \hline
\multicolumn{1}{l}{\textbf{$\texttt{cQd1}$}}             
& \multicolumn{1}{l}{{[}-0.889,6.459{]}}  
& \multicolumn{1}{l}{{[}-3.239,34.632{]}}   
& \multicolumn{1}{l}{{[}-0.332,0.436{]}}  
& \multicolumn{1}{l}{{[}-0.370,0.384{]}}    
& \multicolumn{1}{l}{\textbf{$\texttt{cQd8}$}}             
& \multicolumn{1}{l}{{[}-2.393,2.042{]}}   
& \multicolumn{1}{l}{{[}-24.479,11.233{]}}  
& \multicolumn{1}{l}{{[}-1.615,0.888{]}}  
& {[}-1.256,0.715{]}                         
\\
\multicolumn{10}{c}{\textbf{4-heavy}}                                            \\ \hline\hline
\multicolumn{1}{l}{\textbf{$\texttt{cQQ1}$}}            
& \multicolumn{1}{l}{{[}-6.132,23.281{]}} 
& \multicolumn{1}{l}{{[}-190,189{]}}        
& \multicolumn{1}{l}{{[}-2.229,2.019{]}}  
& \multicolumn{1}{l}{{[}-2.995,3.706{]}}    
& \multicolumn{1}{l}{\textbf{$\texttt{cQQ8}$}}             
& \multicolumn{1}{l}{{[}-26.471,57.778{]}} 
& \multicolumn{1}{l}{{[}-190,170{]}}        
& \multicolumn{1}{l}{{[}-6.812,5.834{]}}  
& {[}-11.177,8.170{]}                        
\\ \hline
\multicolumn{1}{l}{\textbf{$\texttt{cQt1}$}}             
& \multicolumn{1}{l}{{[}-195,159{]}}      
& \multicolumn{1}{l}{{[}-190,189{]}}        
& \multicolumn{1}{l}{{[}-1.830,1.862{]}}  
& \multicolumn{1}{l}{{[}-1.391,1.251{]}}    
& \multicolumn{1}{l}{\textbf{$\texttt{cQt8}$}}             
& \multicolumn{1}{l}{{[}-5.722,20.105{]}}  
& \multicolumn{1}{l}{{[}-190,162{]}}        
& \multicolumn{1}{l}{{[}-4.213,3.346{]}}  
& {[}-3.040,2.202{]}                         
\\ \hline
\multicolumn{1}{l}{\textbf{$\texttt{ctt1}$}}                 
& \multicolumn{1}{l}{{[}-2.782,12.114{]}} 
& \multicolumn{1}{l}{{[}-115,153{]}}        
& \multicolumn{1}{l}{{[}-1.151,1.025{]}}  
& \multicolumn{1}{l}{{[}-0.791,0.714{]}}    
& \multicolumn{5}{l}{}   
\\ \hline
\end{tabular}
}

\caption{\label{tab:dim64f_wc_bounds} 
Bounds on four-fermion WCs from the global analysis of Ref.~\cite{Ethier:2021bye}.}
\end{table}
\begin{table}[h!]
\centering
\resizebox{\textwidth}{!}{
\renewcommand{\arraystretch}{1.0}
\centering
\begin{tabular}{llllllllll}
\hline
\multicolumn{10}{c}{\textbf{Contributing operators}}                                                
\\ \hline\hline
\multicolumn{1}{c}{\multirow{2}{*}{\textbf{UFO}}} 
& \multicolumn{2}{c}{\textbf{$\mathscr{O}(\Lambda^{-2})$}}                                   
& \multicolumn{2}{c}{\textbf{$\mathscr{O}(\Lambda^{-4})$}}                                   
& \multicolumn{1}{c}{\multirow{2}{*}{\textbf{UFO}}} 
& \multicolumn{2}{c}{\textbf{$\mathscr{O}(\Lambda^{-2})$}}                                    
& \multicolumn{2}{c}{\textbf{$\mathscr{O}(\Lambda^{-4})$}}                                   
\\ \cline{2-5} \cline{7-10} 
\multicolumn{1}{c}{}                                   
& \multicolumn{1}{c}{\textbf{Individual}}
& \multicolumn{1}{c}{\textbf{Marginalised}} 
& \multicolumn{1}{c}{\textbf{Individual}} 
& \multicolumn{1}{c}{\textbf{Marginalised}} 
& \multicolumn{1}{c}{}                                   
& \multicolumn{1}{c}{\textbf{Individual}}  
& \multicolumn{1}{c}{\textbf{Marginalised}} 
& \multicolumn{1}{c}{\textbf{Individual}} 
& \multicolumn{1}{c}{\textbf{Marginalised}} 
\\ \hline
\multicolumn{1}{l}{\textbf{$\texttt{ctp}$}}           
& \multicolumn{1}{l}{{[}-1.331,0.355{]}}  
& \multicolumn{1}{l}{{[}-5.739,3.435{]}}    
& \multicolumn{1}{l}{{[}-1.286,0.348{]}}  
& \multicolumn{1}{l}{{[}-2.319,2.797{]}}    
& \multicolumn{1}{l}{\textbf{$\texttt{ctZ}$}}           
& \multicolumn{1}{l}{{[}-0.039,0.099{]}}   
& \multicolumn{1}{l}{{[}-15.869,5.636{]}}    
& \multicolumn{1}{l}{{[}-0.044,0.094{]}}  
& {[}-1.129,0.856{]}                         
\\ \hline
\multicolumn{1}{l}{\textbf{$\texttt{ctW}$}}           
& \multicolumn{1}{l}{{[}-0.093,0.026{]}}  
& \multicolumn{1}{l}{{[}-0.313,0.123{]}}    
& \multicolumn{1}{l}{{[}-0.084,0.029{]}}  
& \multicolumn{1}{l}{{[}-0.241,0.086{]}}    
& \multicolumn{1}{l}{\textbf{$\texttt{ctG}$}}           
& \multicolumn{1}{l}{{[}0.007,0.111{]}}   
& \multicolumn{1}{l}{{[}-0.127,0.403{]}}    
& \multicolumn{1}{l}{{[}0.006,0.107{]}}  
& {[}0.062,0.243{]}                         
\\ \hline
\multicolumn{1}{l}{\textbf{$\texttt{cpQM}$}}             
& \multicolumn{1}{l}{{[}-0.998,1.441{]}}  
& \multicolumn{1}{l}{{[}-1.690,11.569{]}}  
& \multicolumn{1}{l}{{[}-1.147,1.585{]}}  
& \multicolumn{1}{l}{{[}-2.250,2.855{]}}    
& \multicolumn{1}{l}{\textbf{$\texttt{cpt}$}}             
& \multicolumn{1}{l}{{[}-2.087,2.463{]}}   
& \multicolumn{1}{l}{{[}-3.270,18.267{]}}   
& \multicolumn{1}{l}{{[}-3.028,2.195{]}}  
& {[}-13.260,3.955{]}                         
\\ \hline
\multicolumn{1}{l}{\textbf{$\texttt{cpG}$}}             
& \multicolumn{1}{l}{{[}-0.002,0.005{]}} 
& \multicolumn{1}{l}{{[}-0.043,0.012{]}}  
& \multicolumn{1}{l}{{[}-0.002,0.005{]}}  
& \multicolumn{1}{l}{{[}-0.019,0.003{]}}    
\\ \hline
\end{tabular}
}
\caption{\label{tab:contributing_operators_wc_bounds} 
Bounds on the contributing two-fermion and purely-bosonic WCs from the global analysis of Ref.~\cite{Ethier:2021bye}, except for the \cG{} operator.}
\end{table}

\section{Additional results for the LHC and FCC-hh}\label{sec:app_add_13_100}
\begin{figure}[h!]
    \centering
    \includegraphics[trim=1.8cm 4.2cm 0.2cm 0.0cm, clip,width=.24\textwidth]{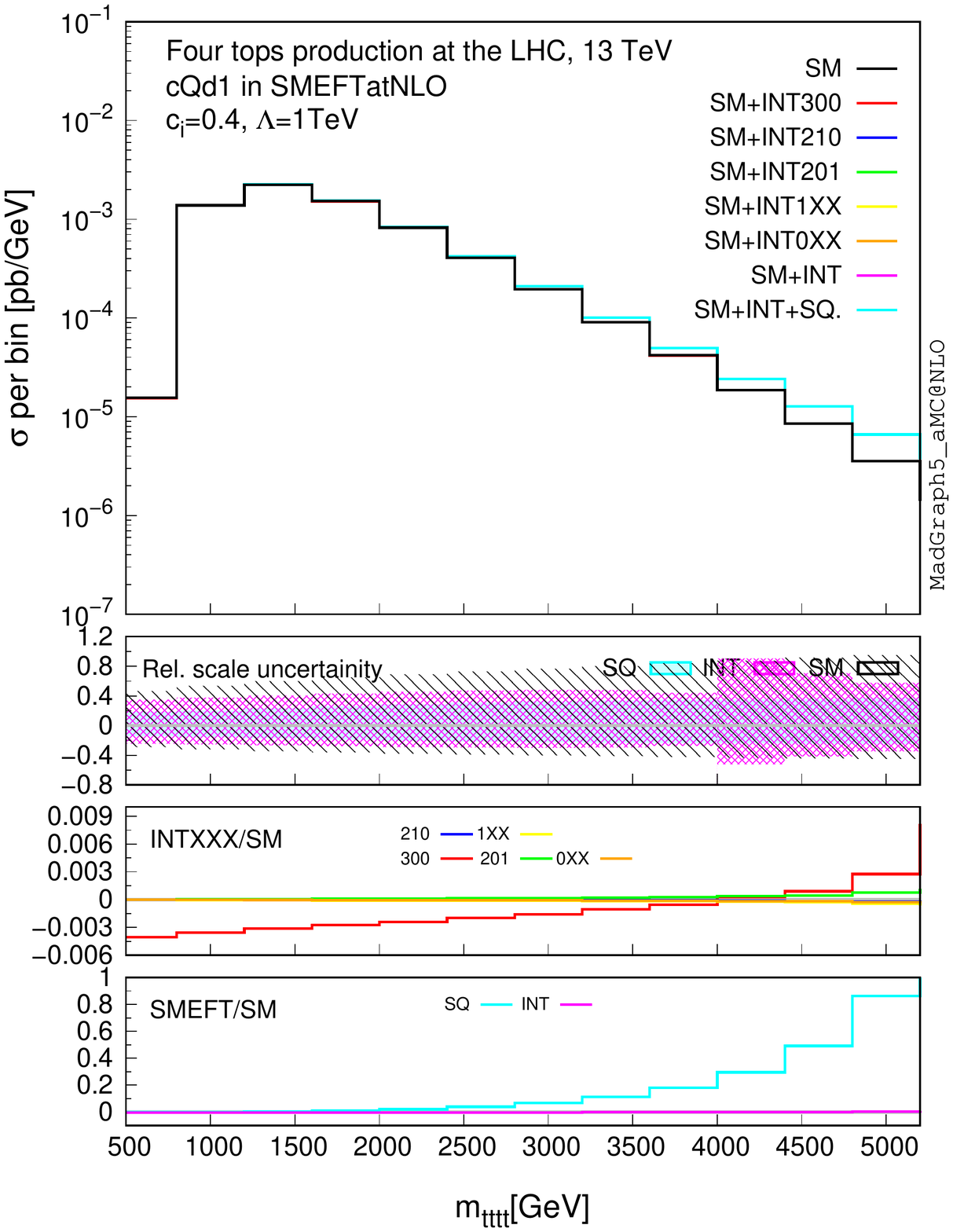}
    \includegraphics[trim=1.8cm 4.2cm 0.2cm 0.0cm, clip,width=.24\textwidth]{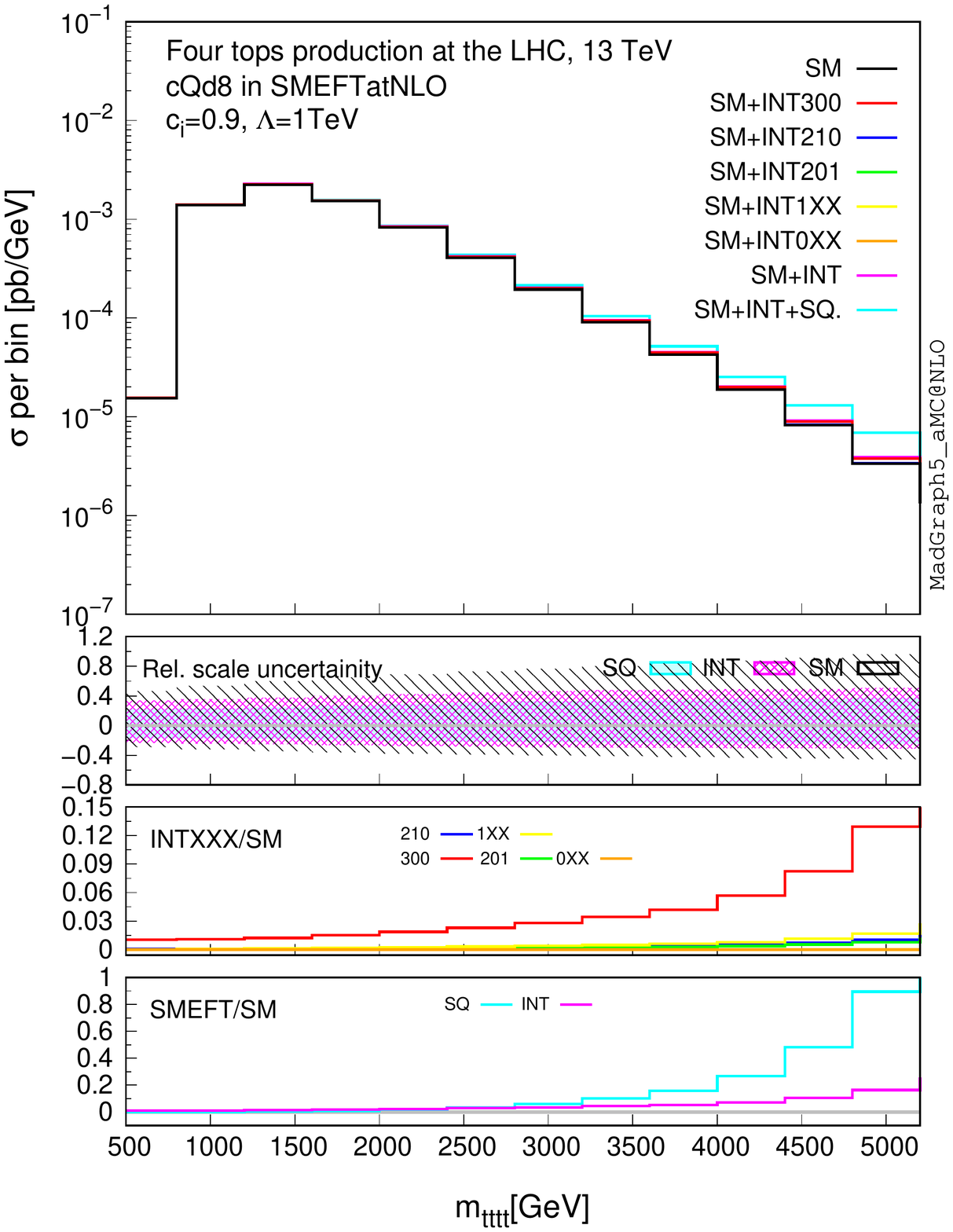}
    \includegraphics[trim=1.8cm 4.2cm 0.2cm 0.0cm, clip,width=.24\textwidth]{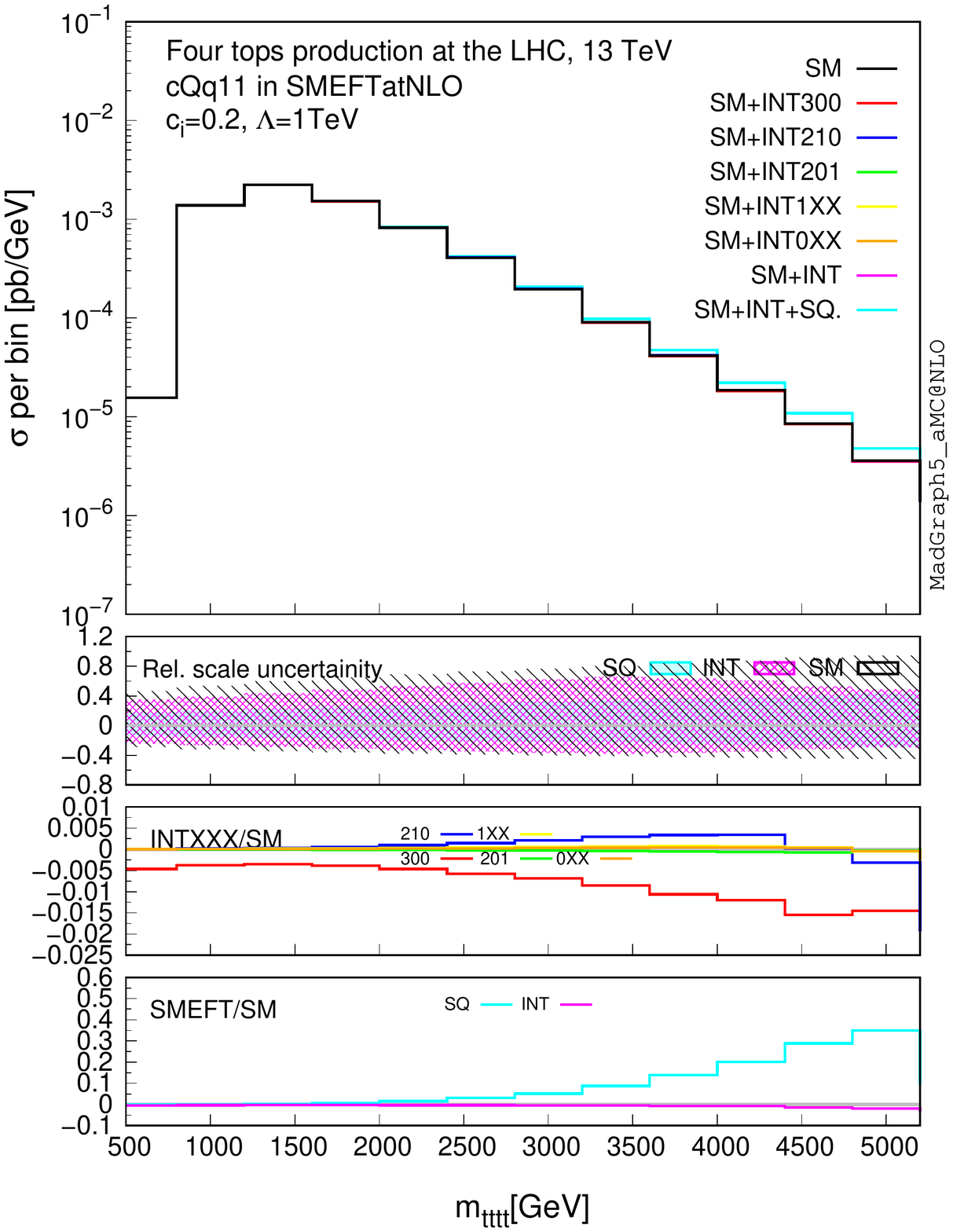}
    \includegraphics[trim=1.8cm 4.2cm 0.2cm 0.0cm, clip,width=.24\textwidth]{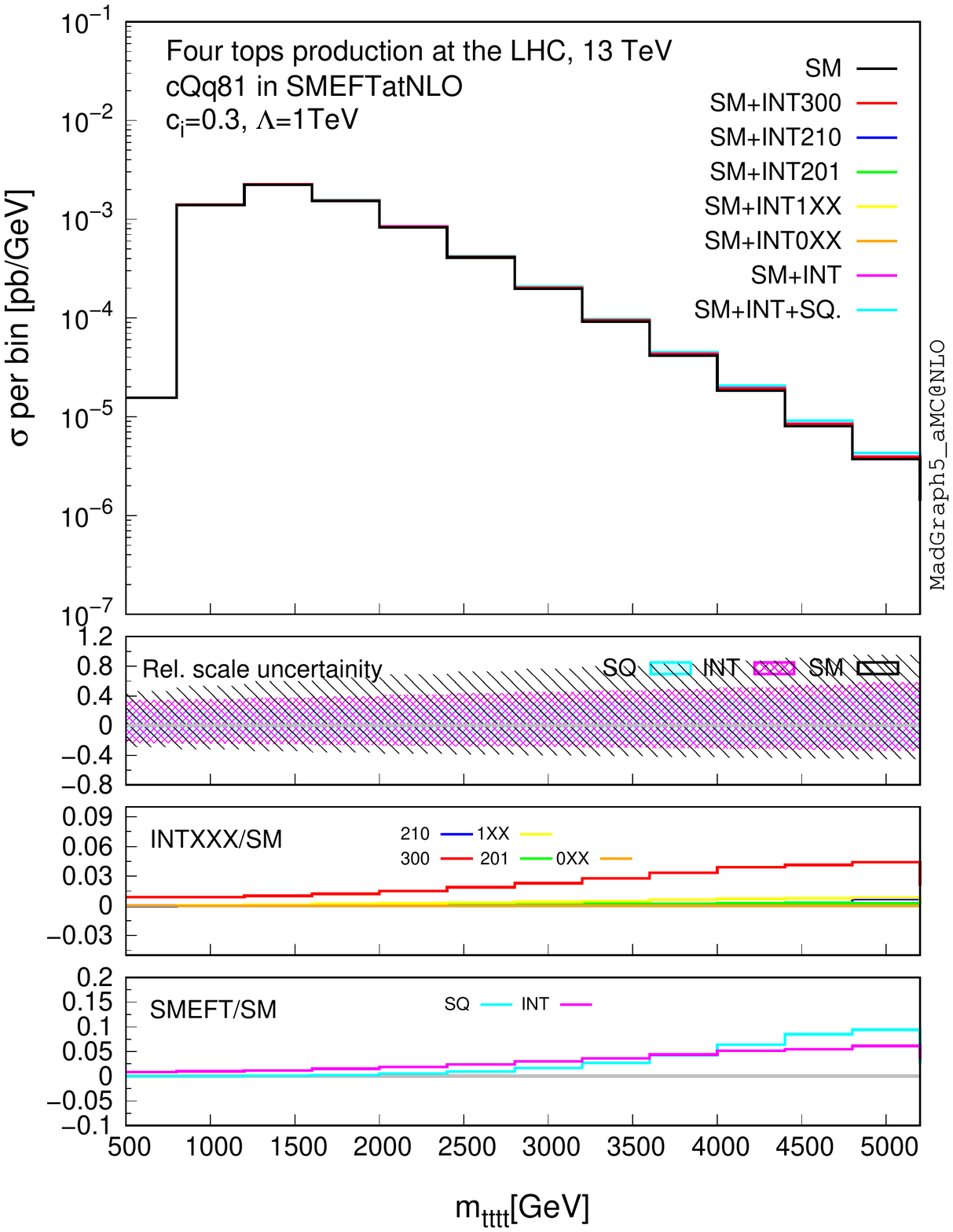}\\
    \includegraphics[trim=1.8cm 4.2cm 0.2cm 0.0cm, clip,width=.24\textwidth]{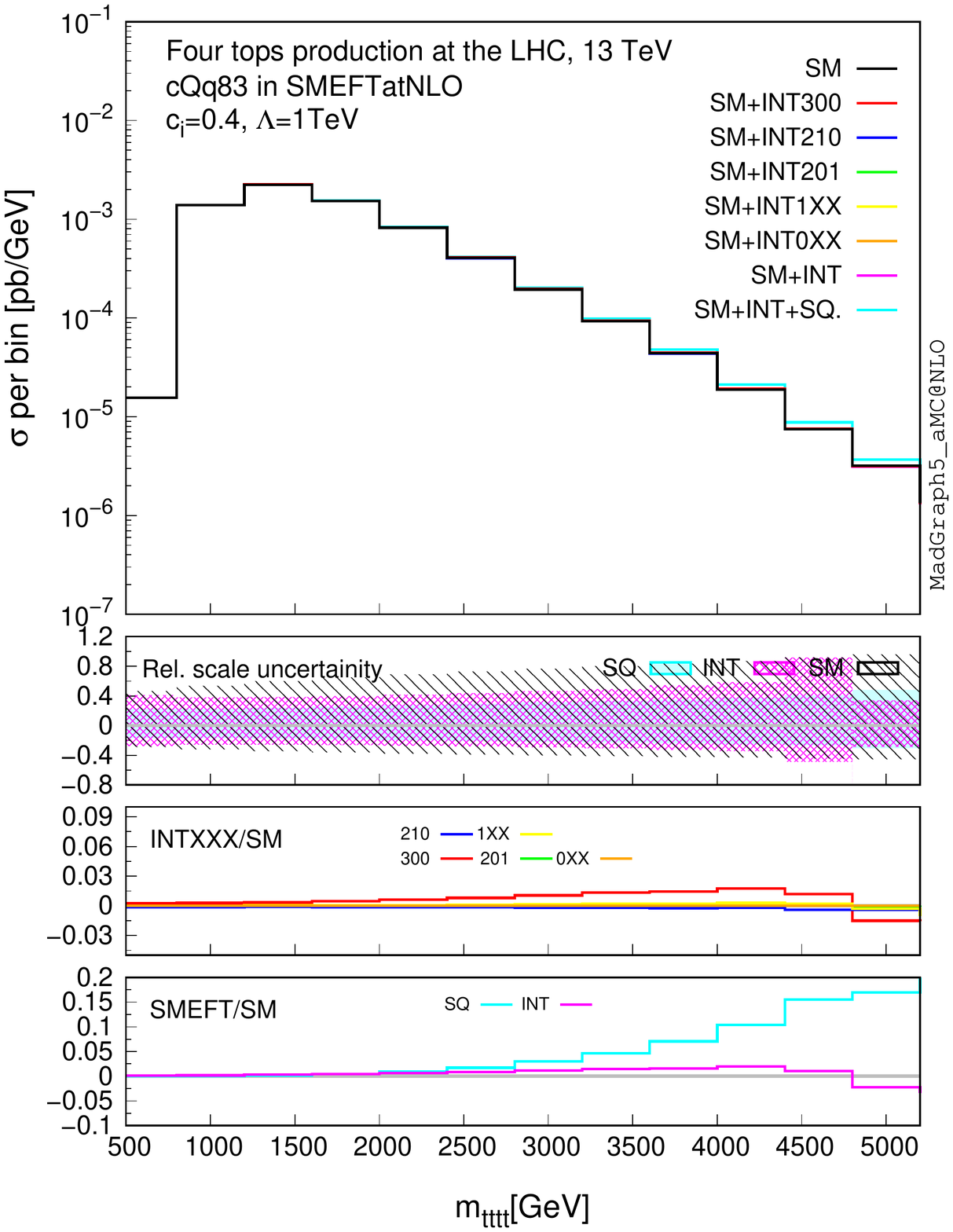}
    \includegraphics[trim=1.8cm 4.2cm 0.2cm 0.0cm, clip,width=.24\textwidth]{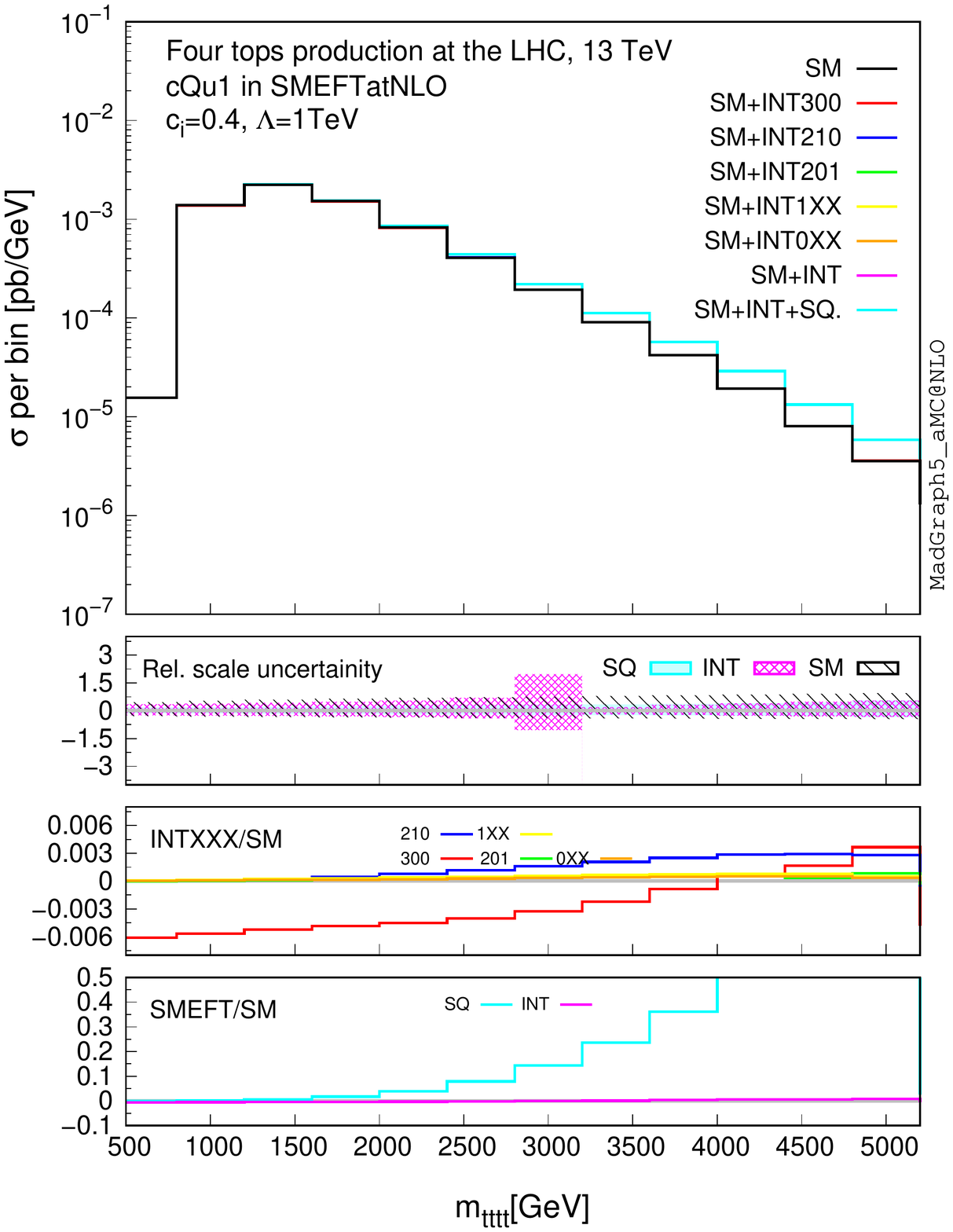}
    \includegraphics[trim=1.8cm 4.2cm 0.2cm 0.0cm, clip,width=.24\textwidth]{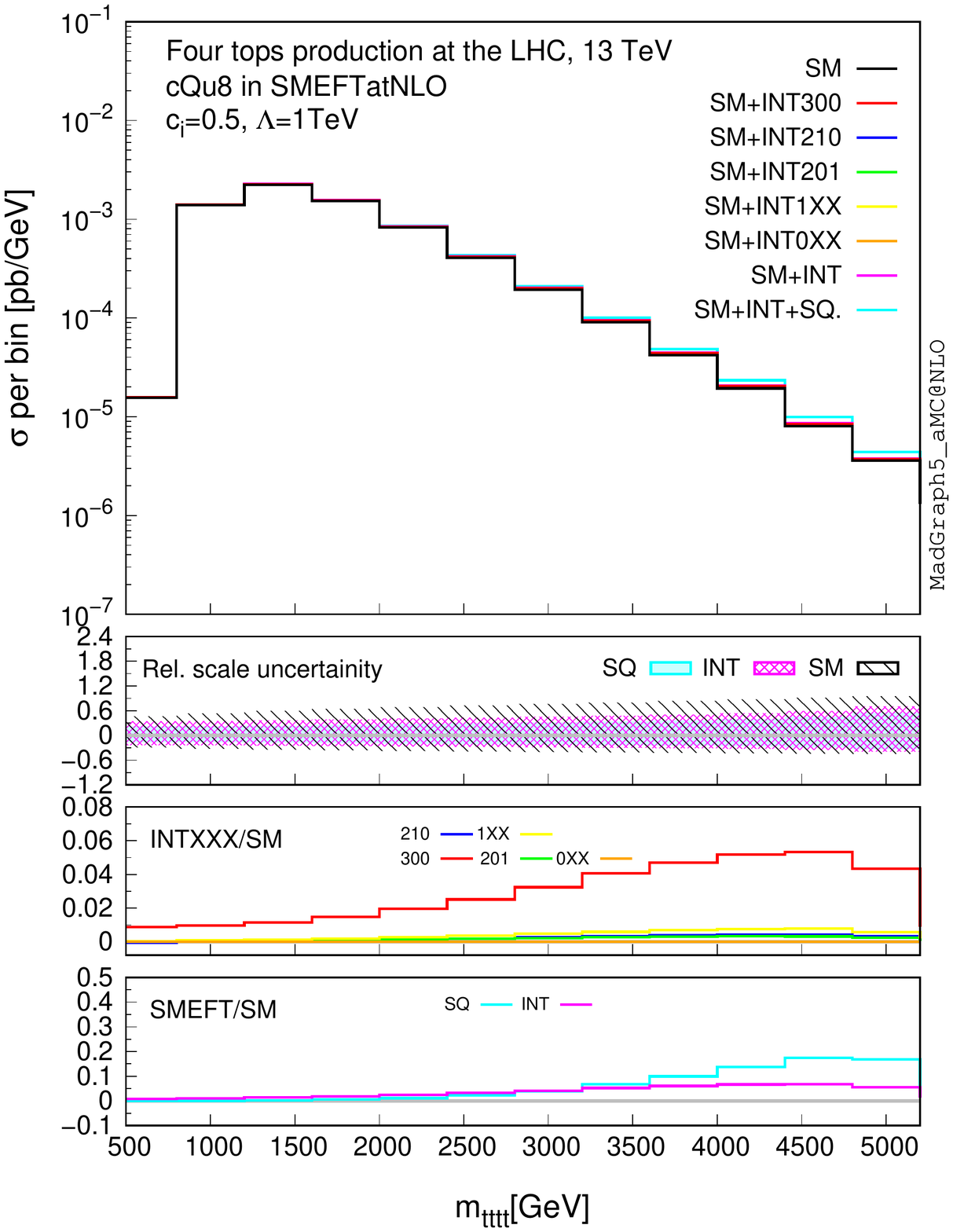}
    \includegraphics[trim=1.8cm 4.2cm 0.2cm 0.0cm, clip,width=.24\textwidth]{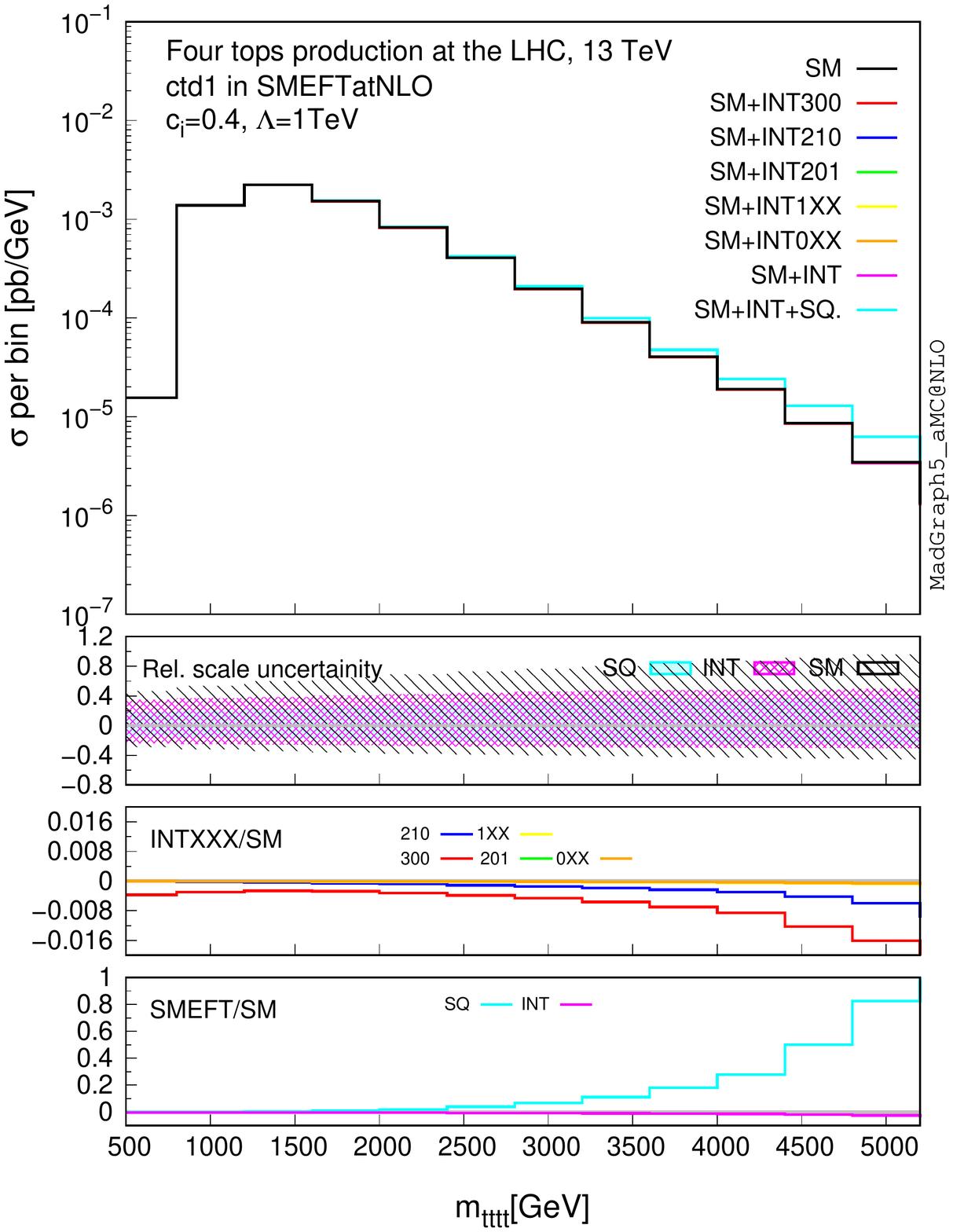}\\
    \includegraphics[trim=1.8cm 4.2cm 0.2cm 0.0cm, clip,width=.24\textwidth]{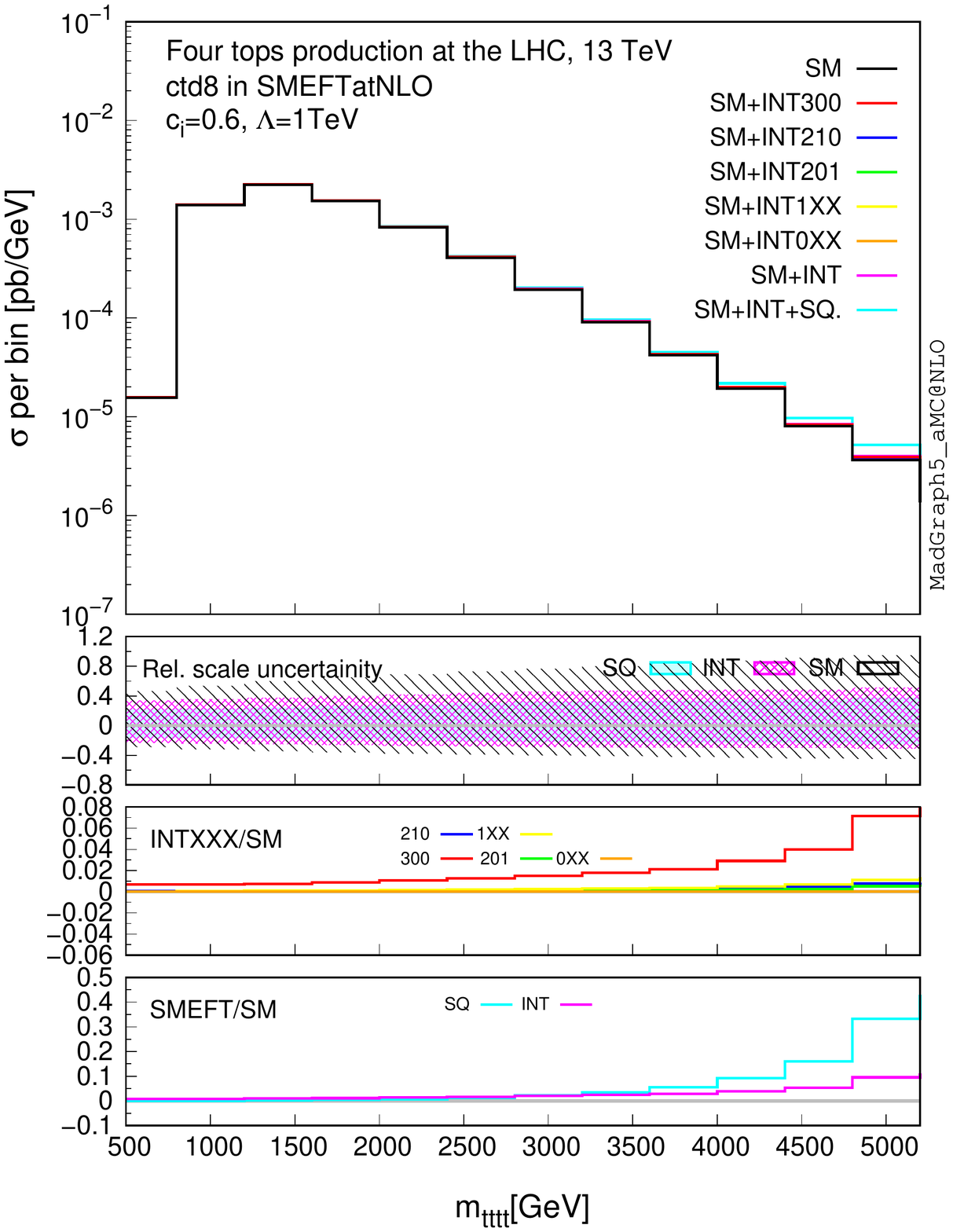}
    \includegraphics[trim=1.8cm 4.2cm 0.2cm 0.0cm, clip,width=.24\textwidth]{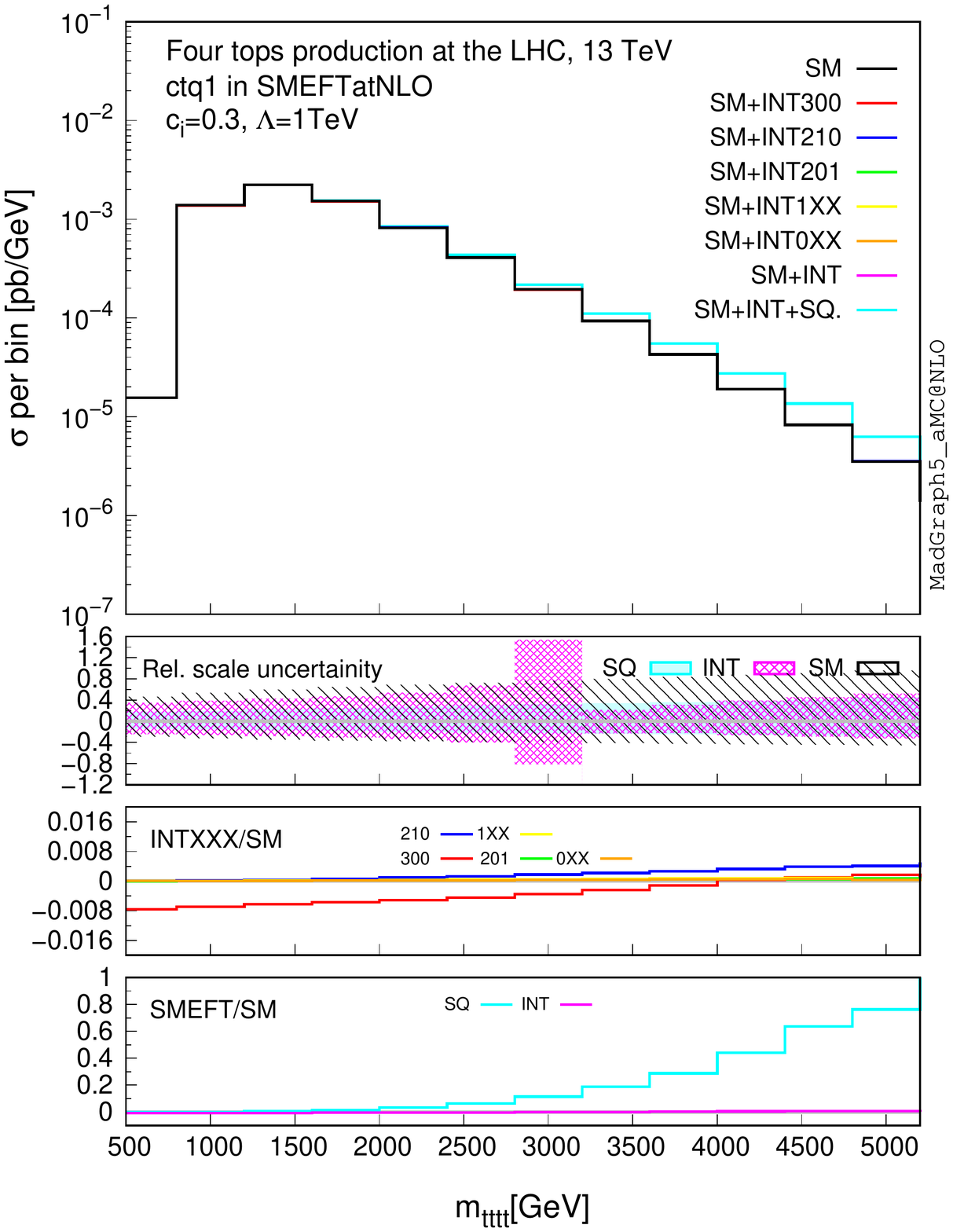}
    \includegraphics[trim=1.8cm 4.2cm 0.2cm 0.0cm, clip,width=.24\textwidth]{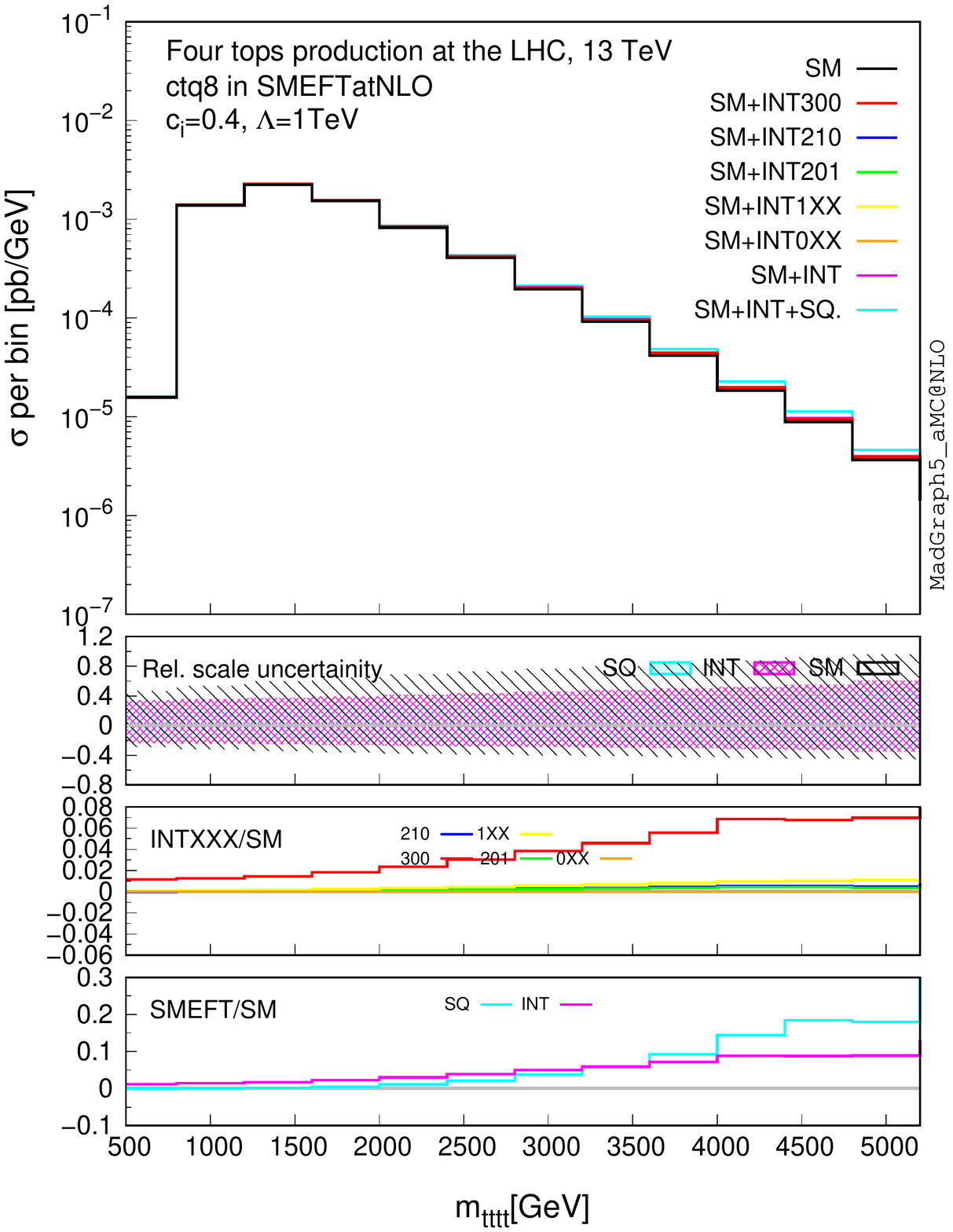}
    \includegraphics[trim=1.8cm 4.2cm 0.2cm 0.0cm, clip,width=.24\textwidth]{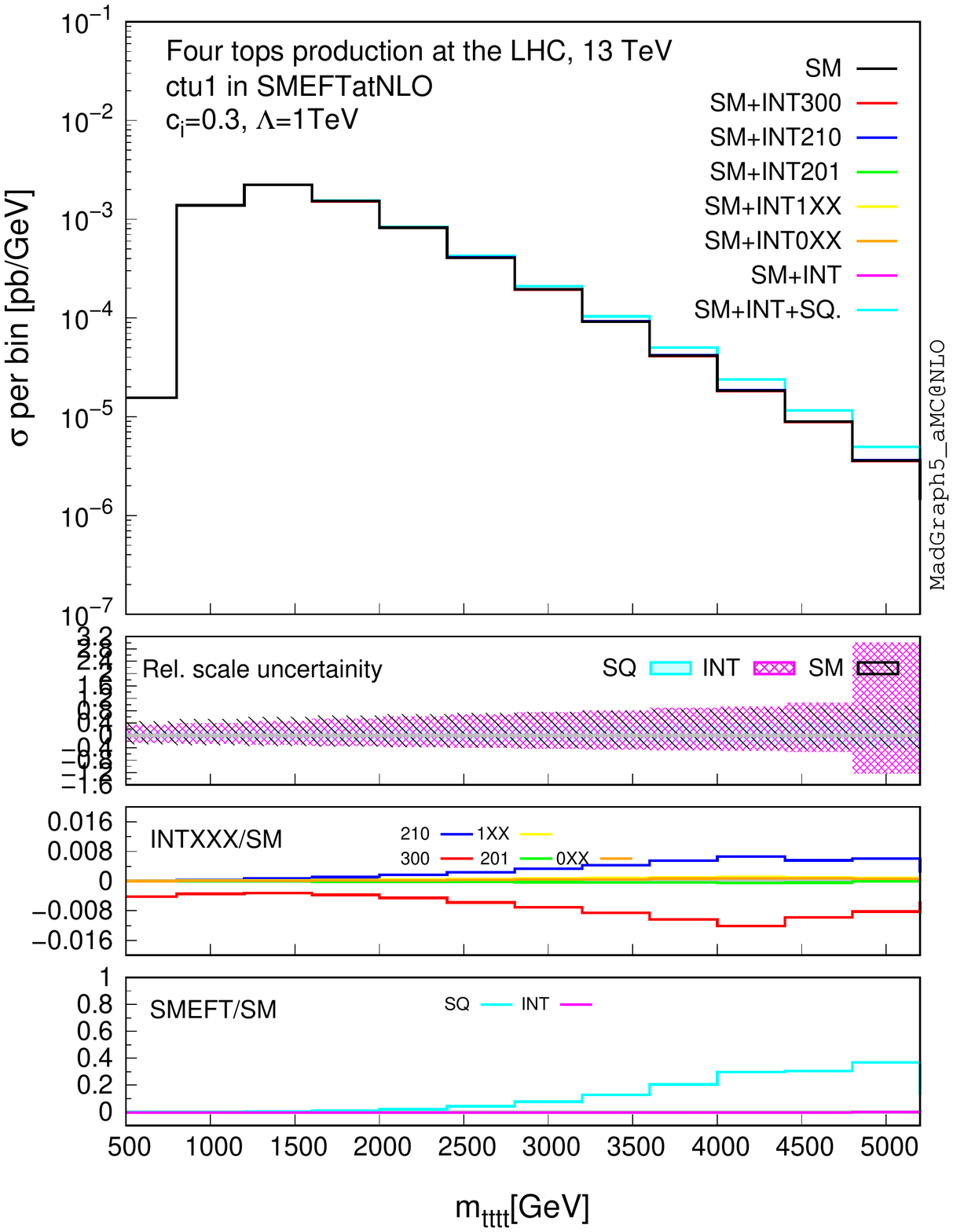}
    \includegraphics[trim=1.8cm 4.2cm 0.2cm 0.0cm, clip,width=.24\textwidth]{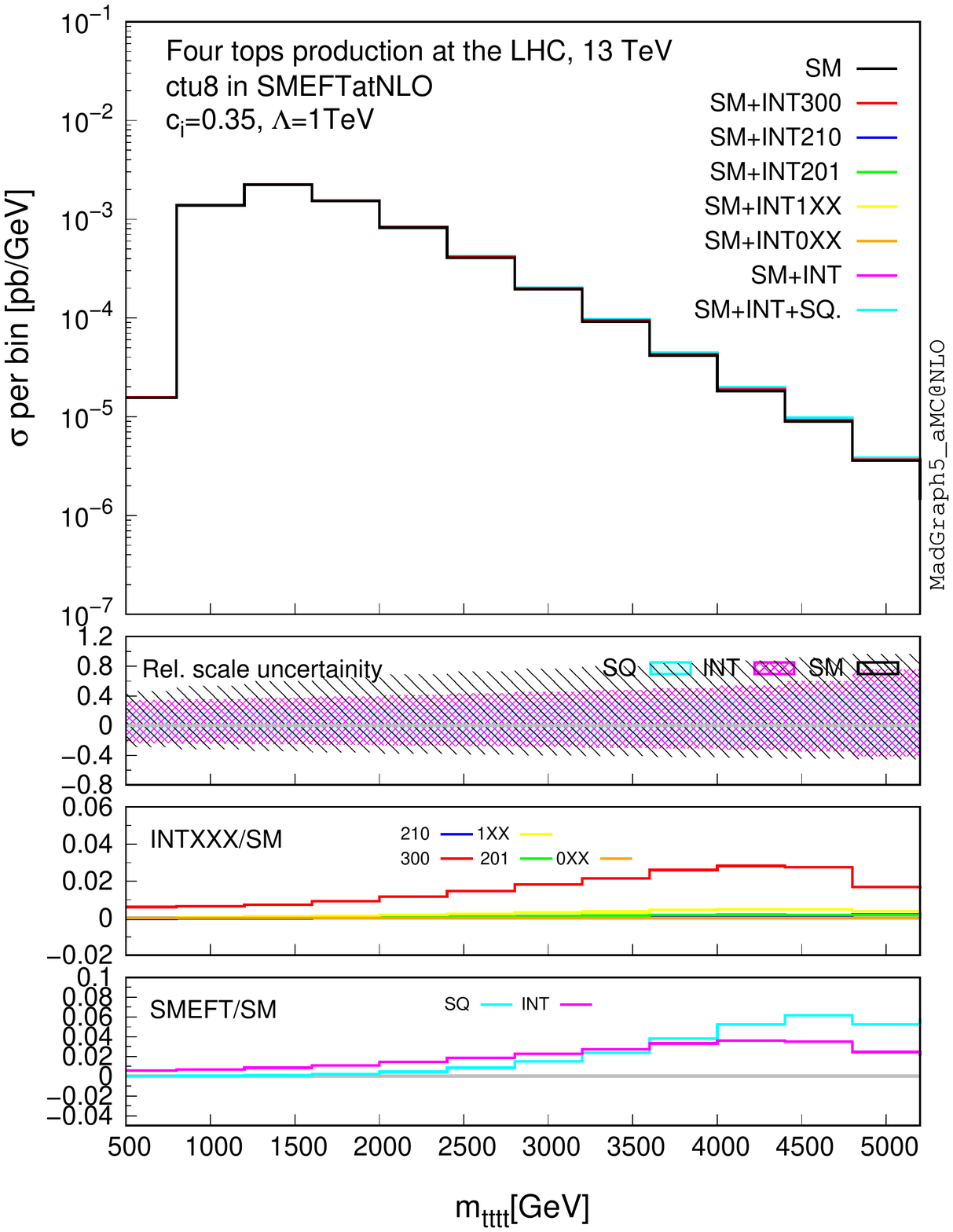}
    \caption{\label{fig:dim64f_rest_diff_13tev} Same as \cref{fig:dim64f_good_diff_13tev} but for the rest of the four-fermion operators.}
\end{figure}

\begin{figure}[h!]
    \centering
    \includegraphics[trim=1.8cm 4.2cm 0.2cm 0.0cm, clip,width=.24\textwidth]{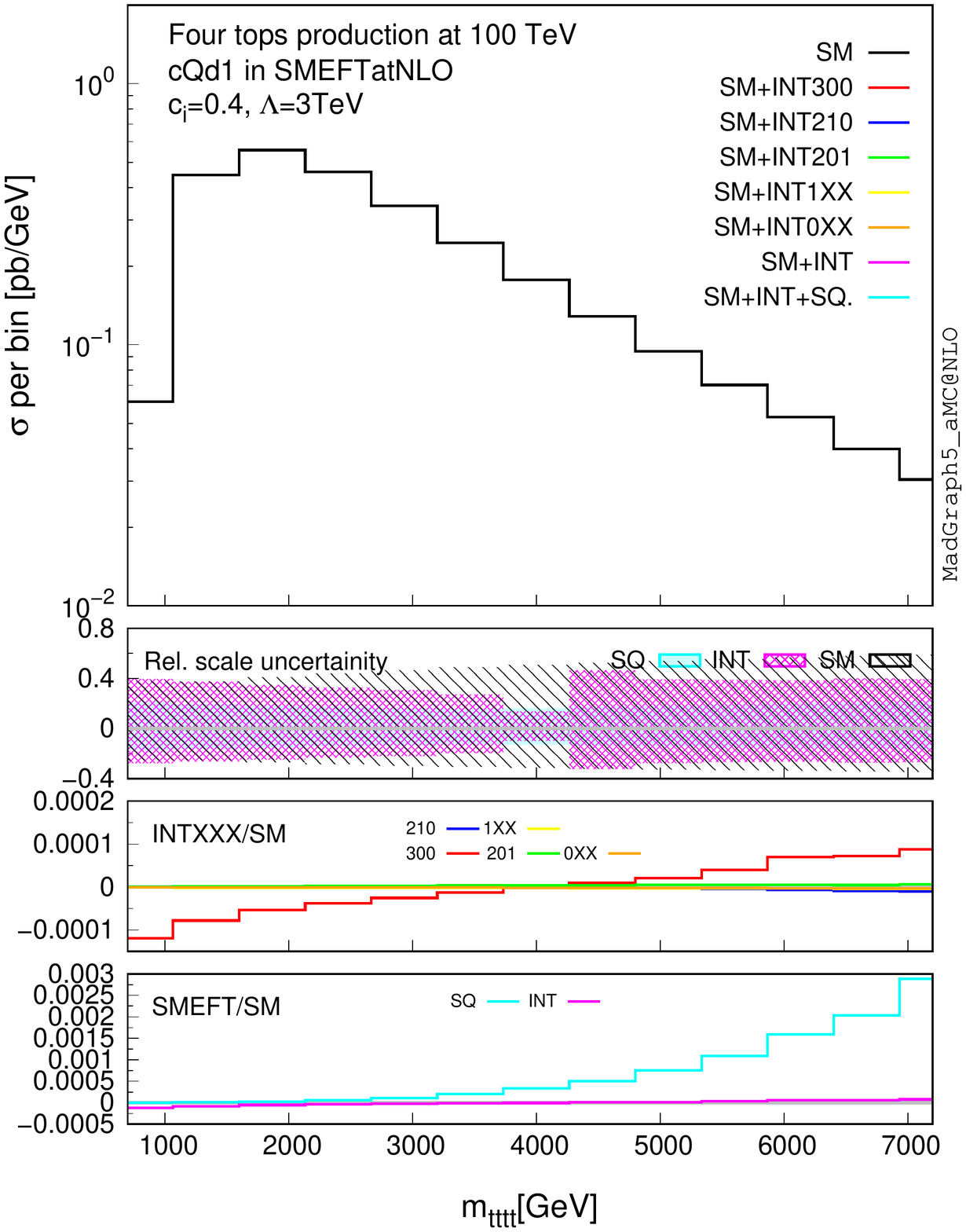}
    \includegraphics[trim=1.8cm 4.2cm 0.2cm 0.0cm, clip,width=.24\textwidth]{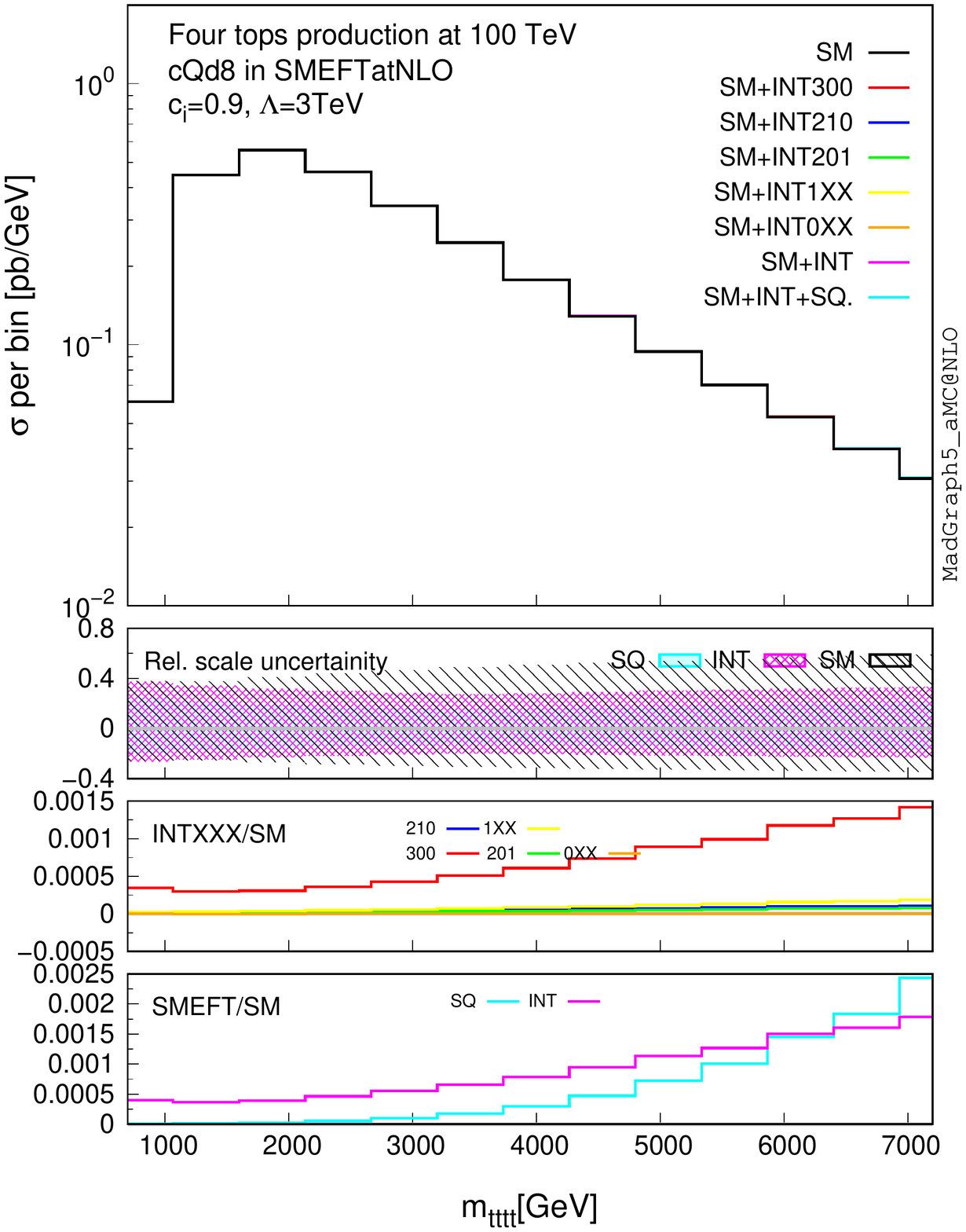}
    \includegraphics[trim=1.8cm 4.2cm 0.2cm 0.0cm, clip,width=.24\textwidth]{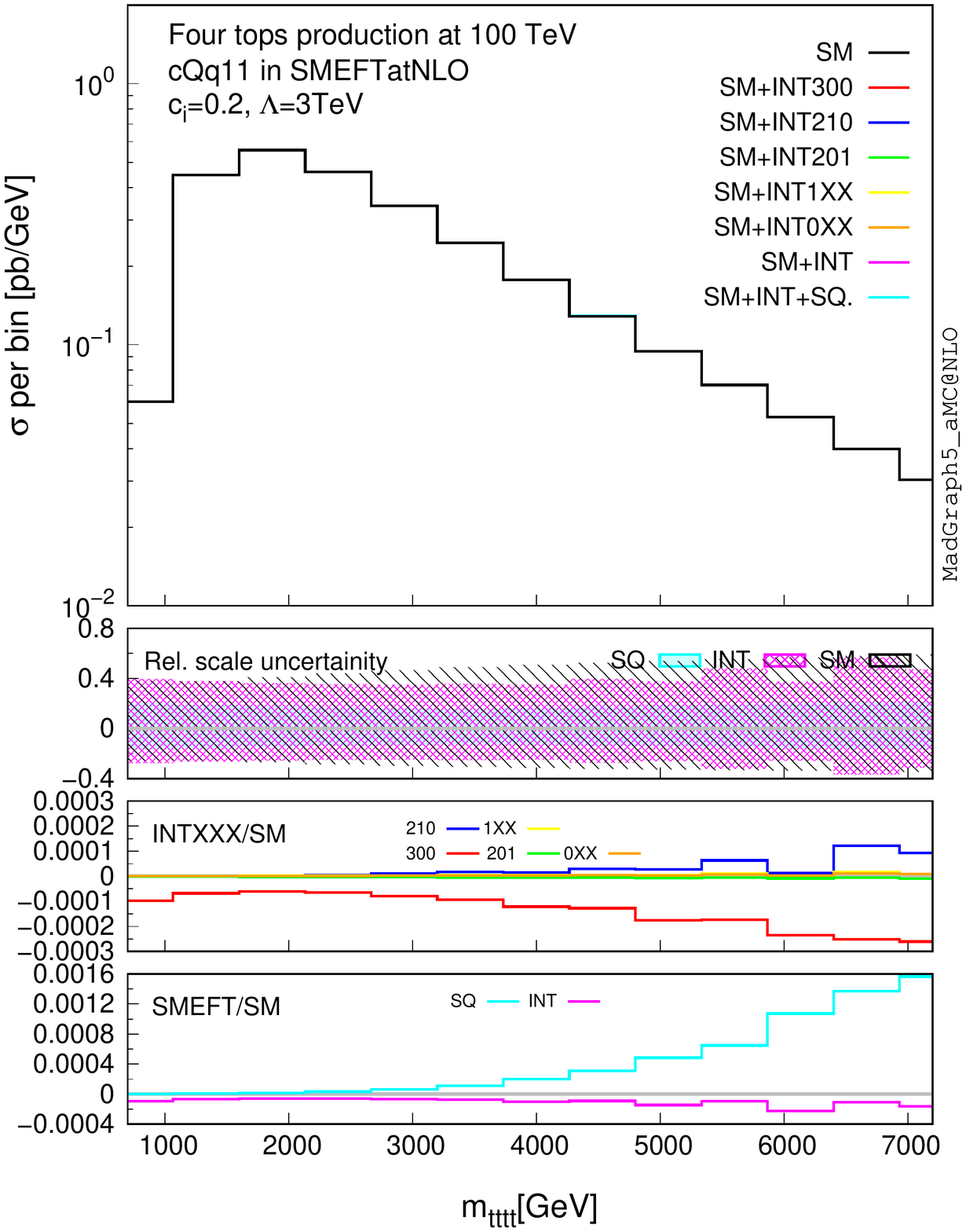}
    \includegraphics[trim=1.8cm 4.2cm 0.2cm 0.0cm, clip,width=.24\textwidth]{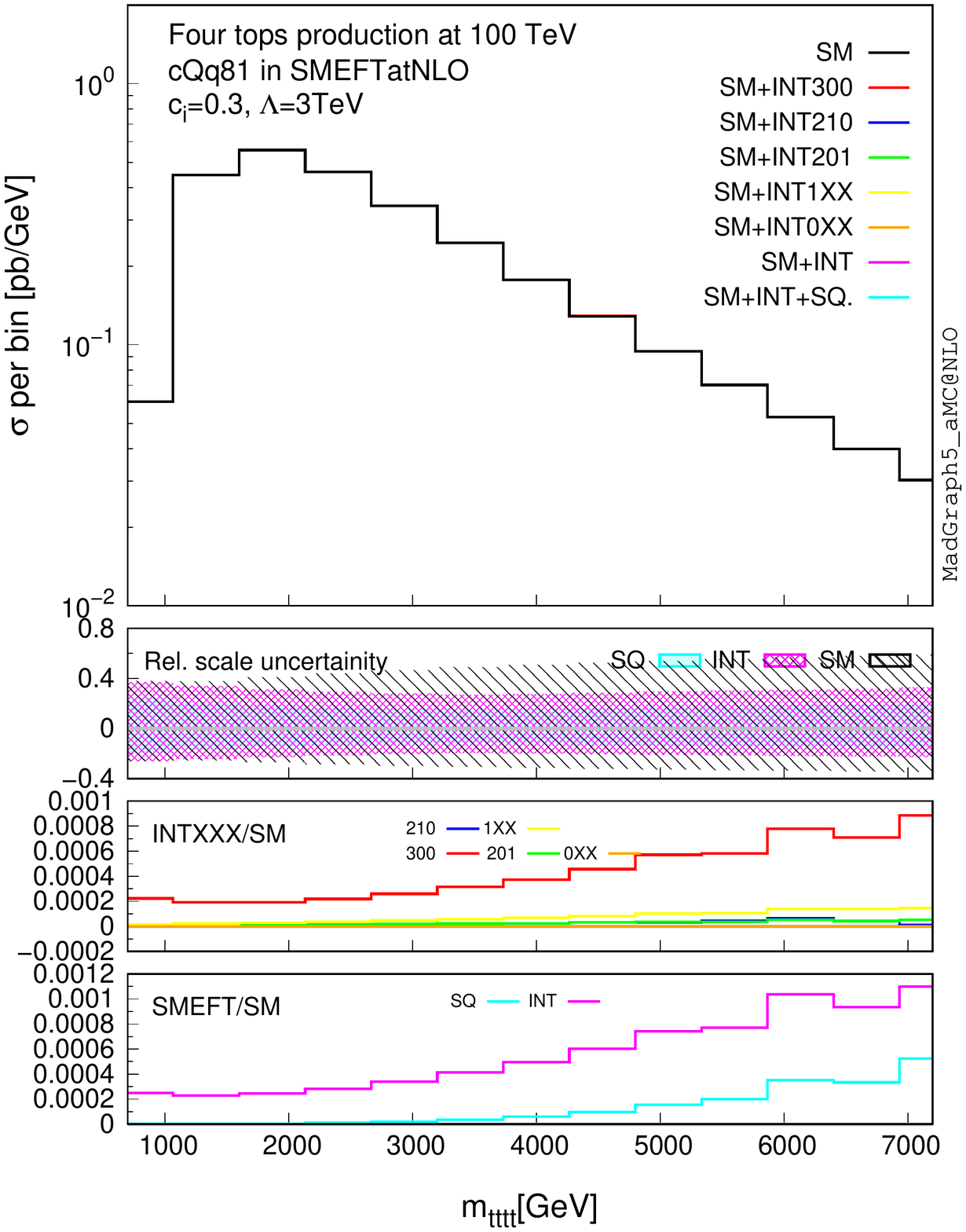}\\
    \includegraphics[trim=1.8cm 4.2cm 0.2cm 0.0cm, clip,width=.24\textwidth]{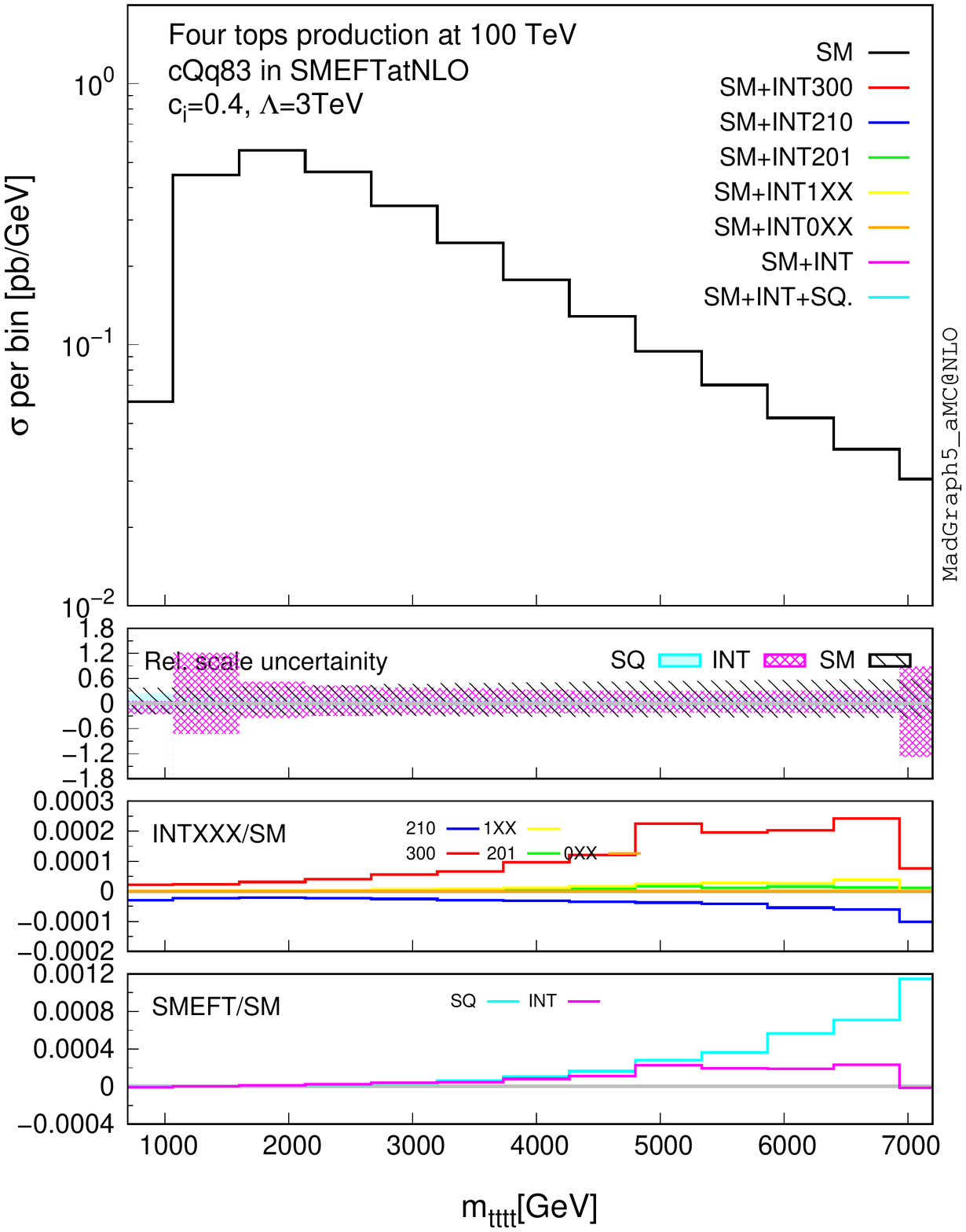}
    \includegraphics[trim=1.8cm 4.2cm 0.2cm 0.0cm, clip,width=.24\textwidth]{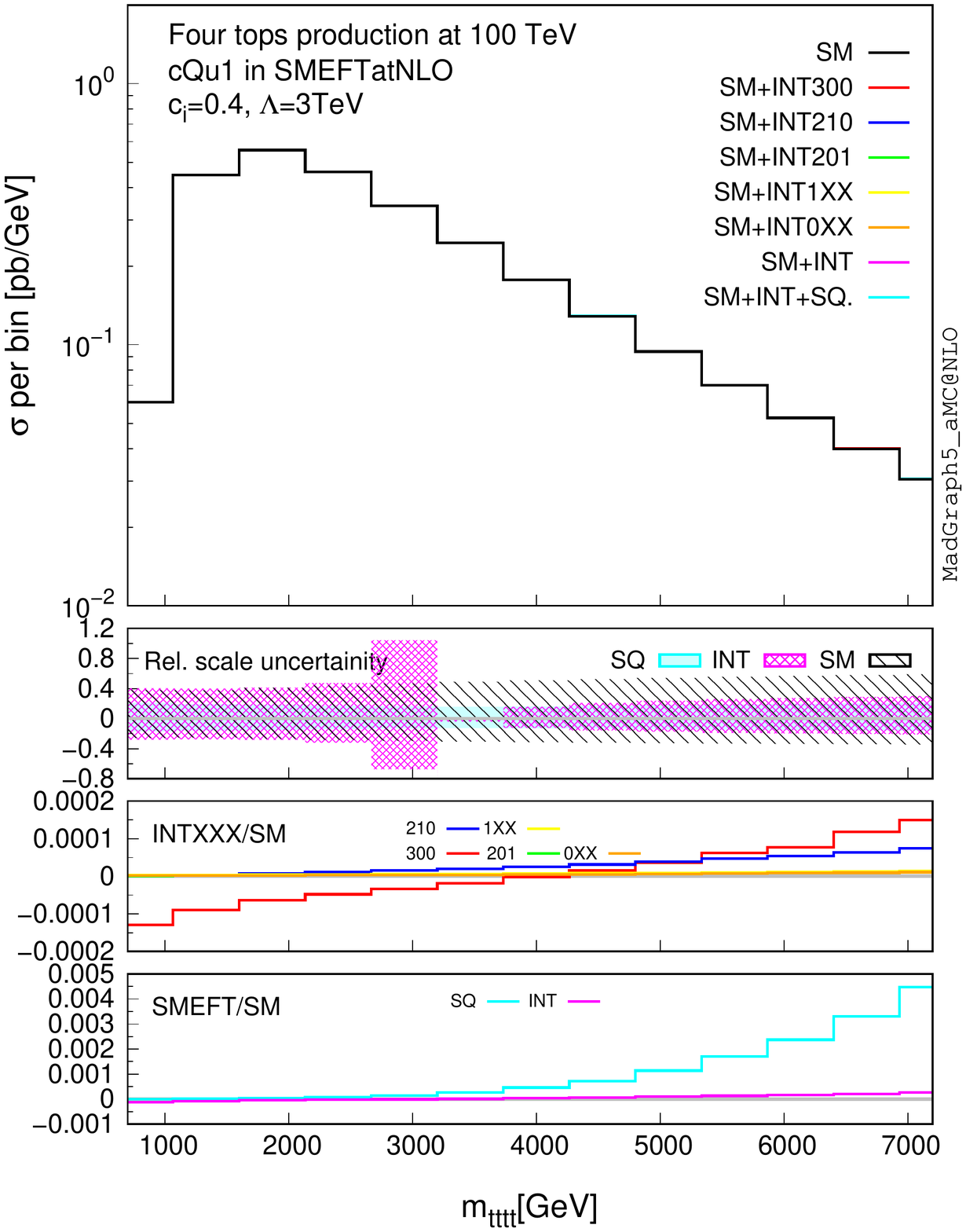}
    \includegraphics[trim=1.8cm 4.2cm 0.2cm 0.0cm, clip,width=.24\textwidth]{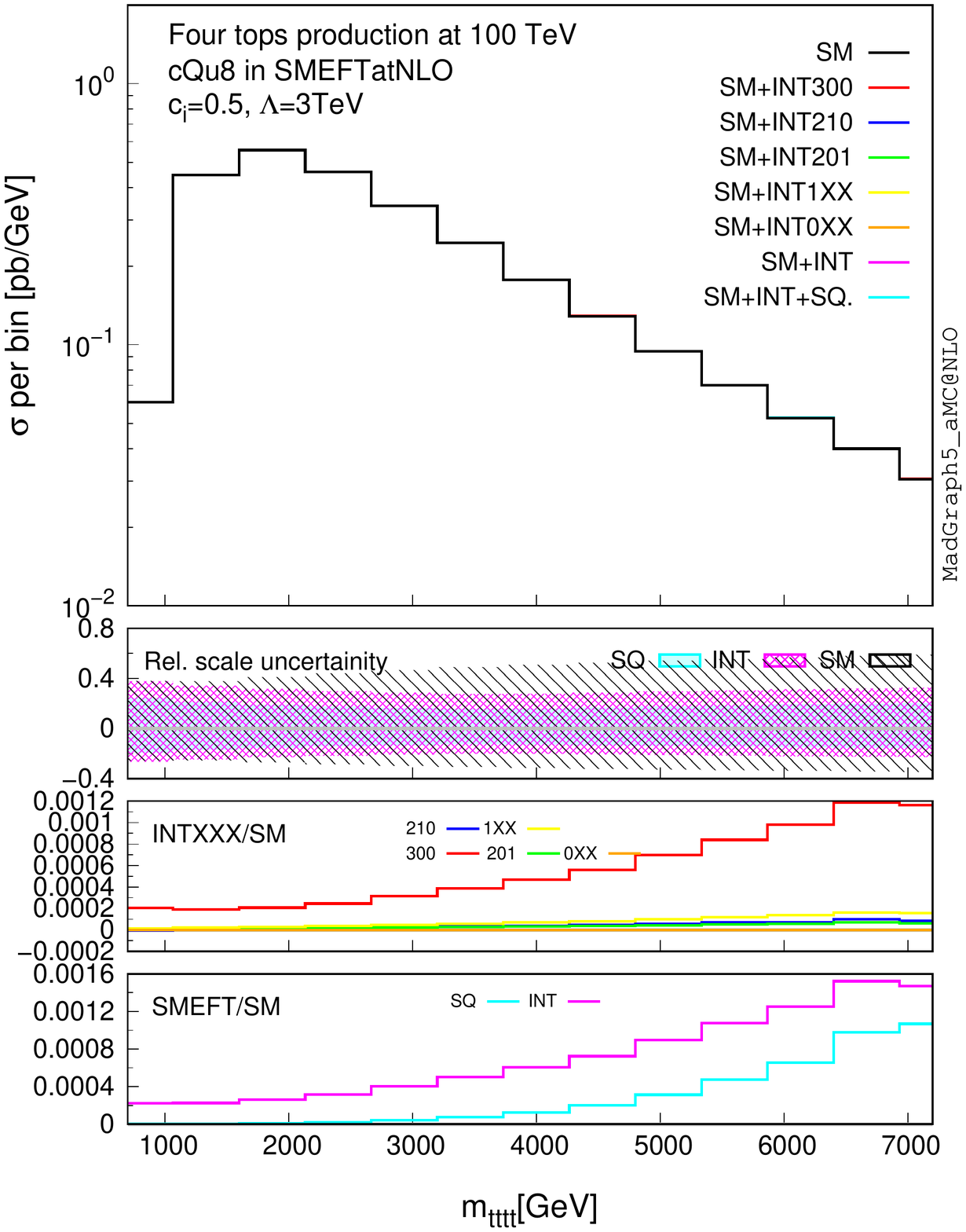}
    \includegraphics[trim=1.8cm 4.2cm 0.2cm 0.0cm, clip,width=.24\textwidth]{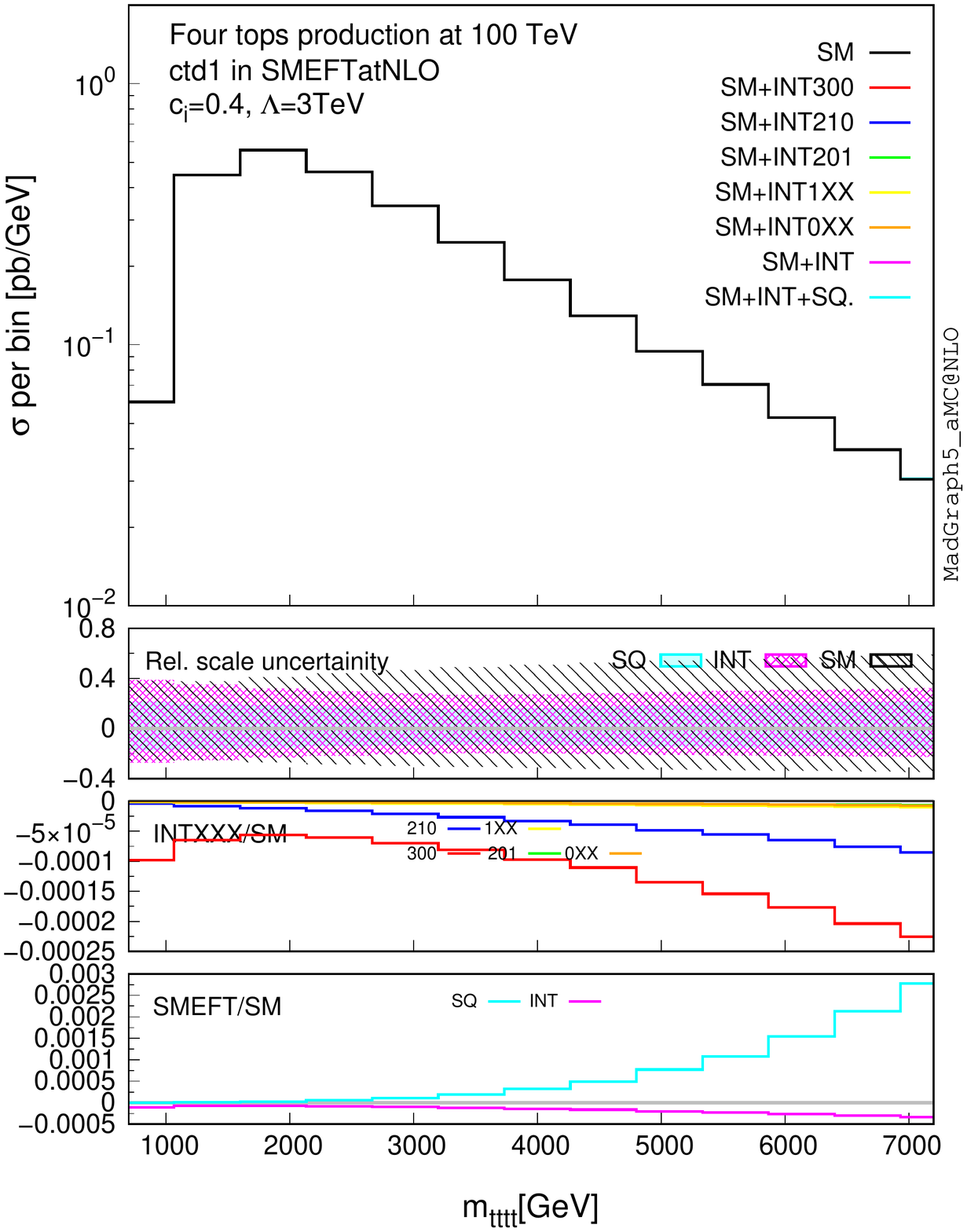}\\
    \includegraphics[trim=1.8cm 4.2cm 0.2cm 0.0cm, clip,width=.24\textwidth]{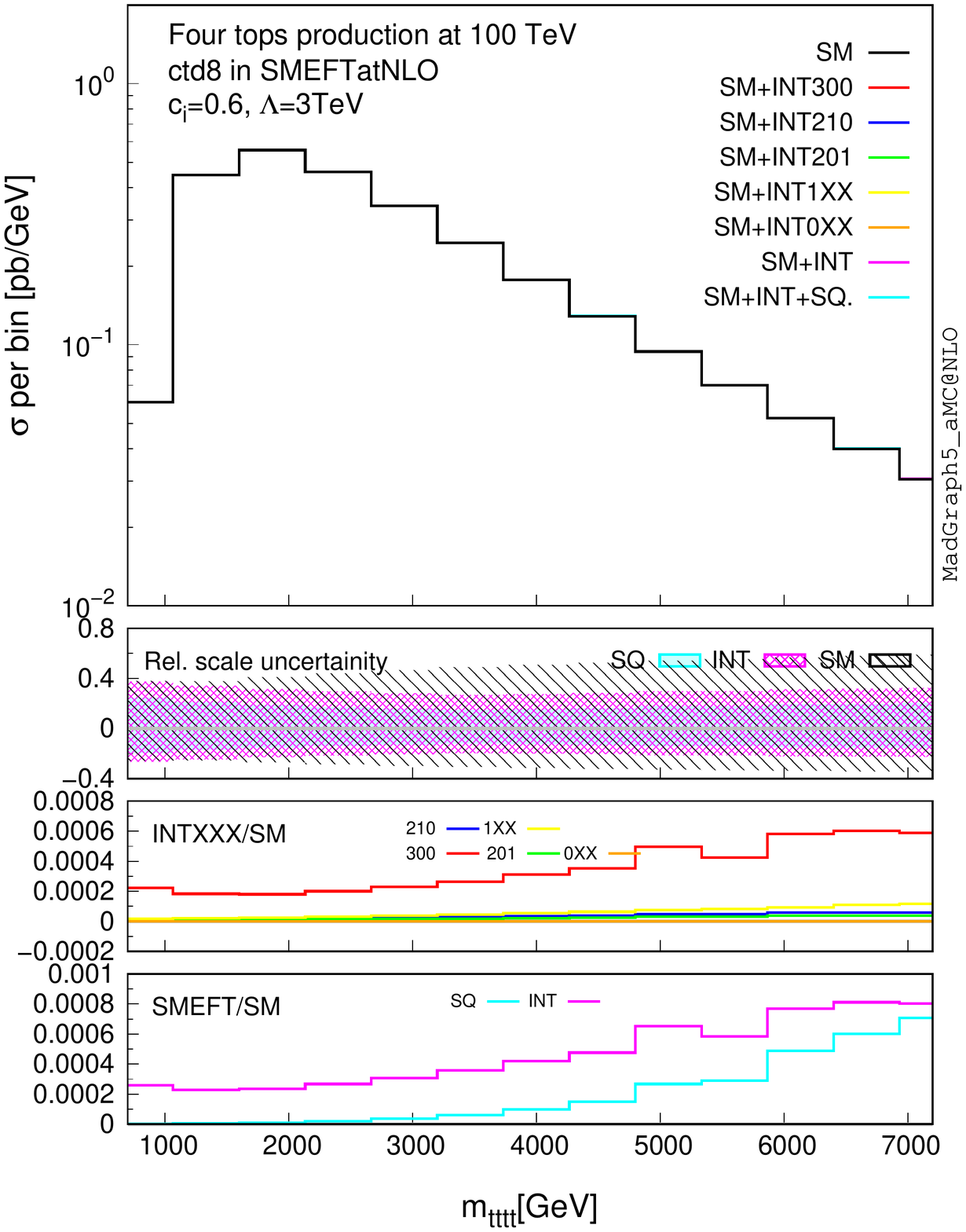}
    \includegraphics[trim=1.8cm 4.2cm 0.2cm 0.0cm, clip,width=.24\textwidth]{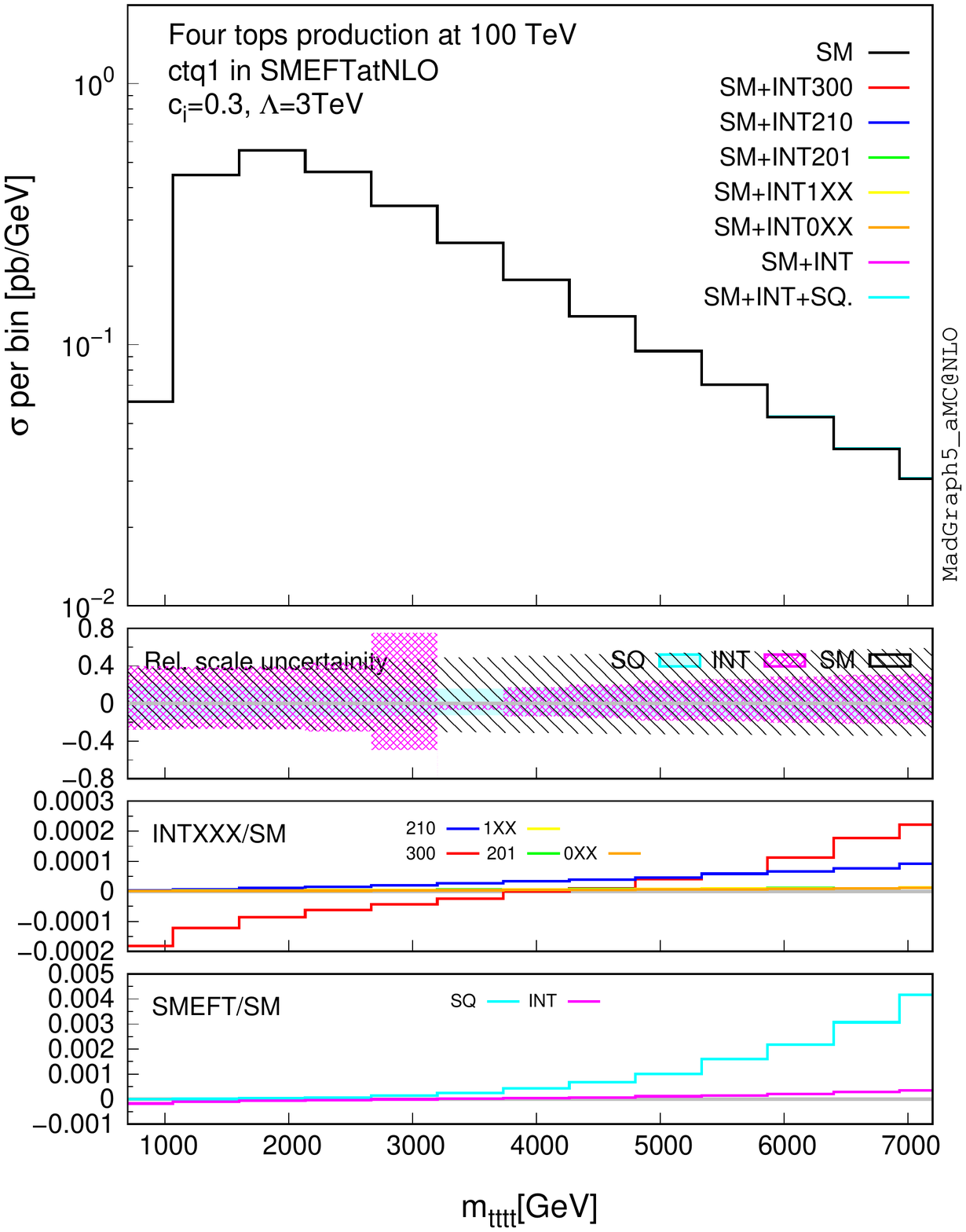}
    \includegraphics[trim=1.8cm 4.2cm 0.2cm 0.0cm, clip,width=.24\textwidth]{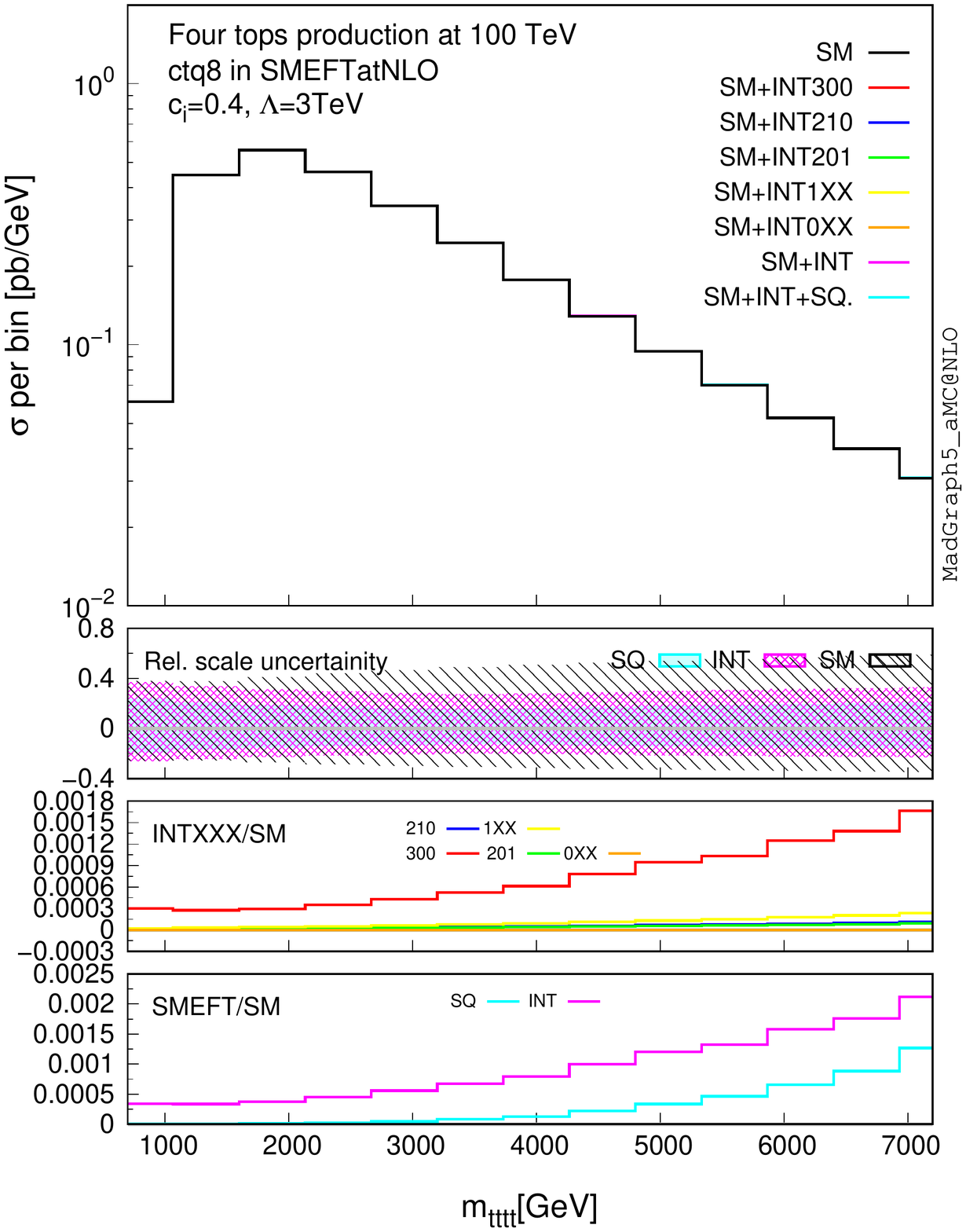}
    \includegraphics[trim=1.8cm 4.2cm 0.2cm 0.0cm, clip,width=.24\textwidth]{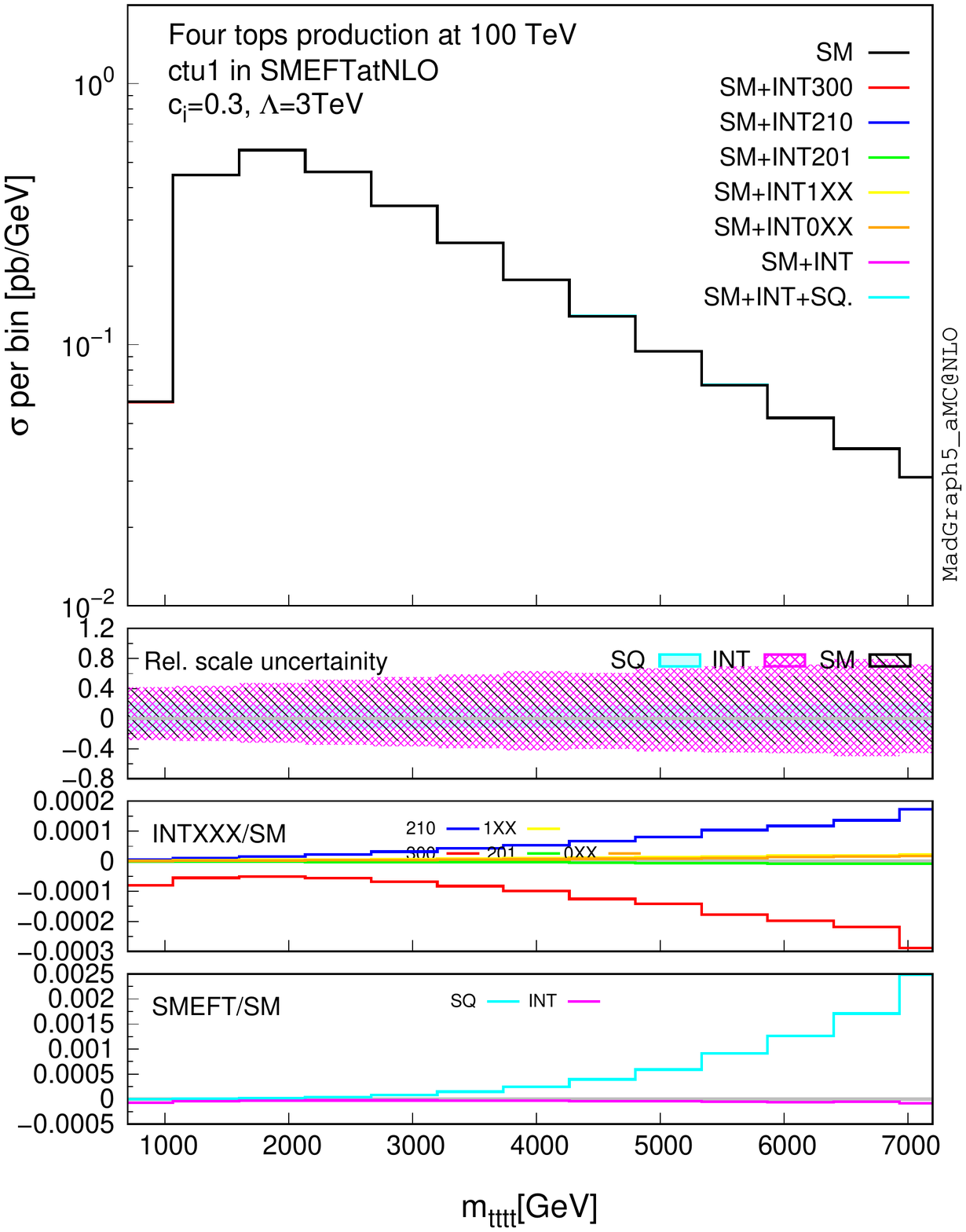}
    \includegraphics[trim=1.8cm 4.2cm 0.2cm 0.0cm, clip,width=.24\textwidth]{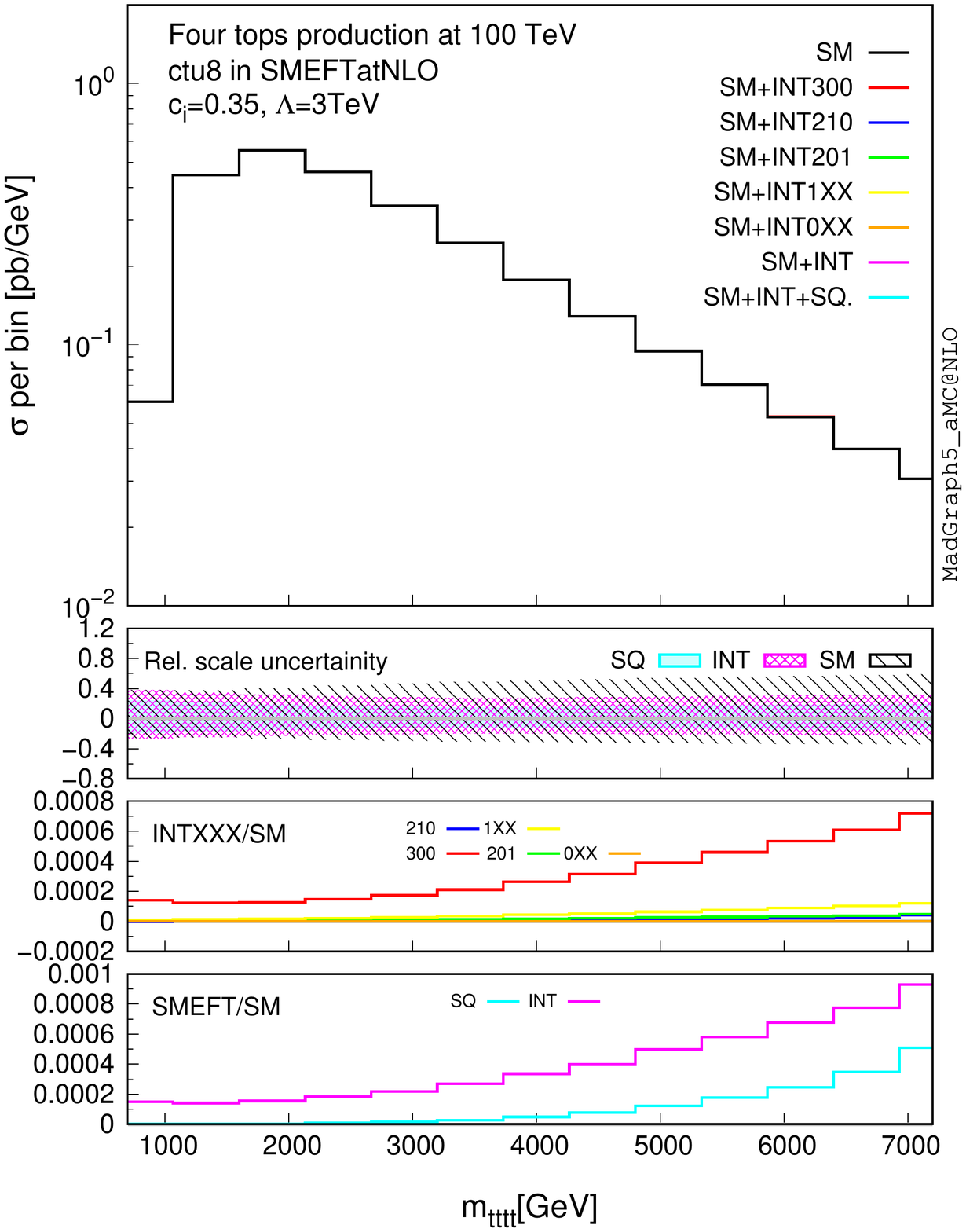}
    \caption{\label{fig:dim64f_rest_diff_100tev} Same as \cref{fig:dim64f_good_diff_100tev} but for the rest of the four-fermion operators.}
\end{figure}

\begin{figure}[h!]
    \centering
    \includegraphics[trim=1.8cm 4.2cm 0.2cm 0.0cm, clip,width=.24\textwidth]{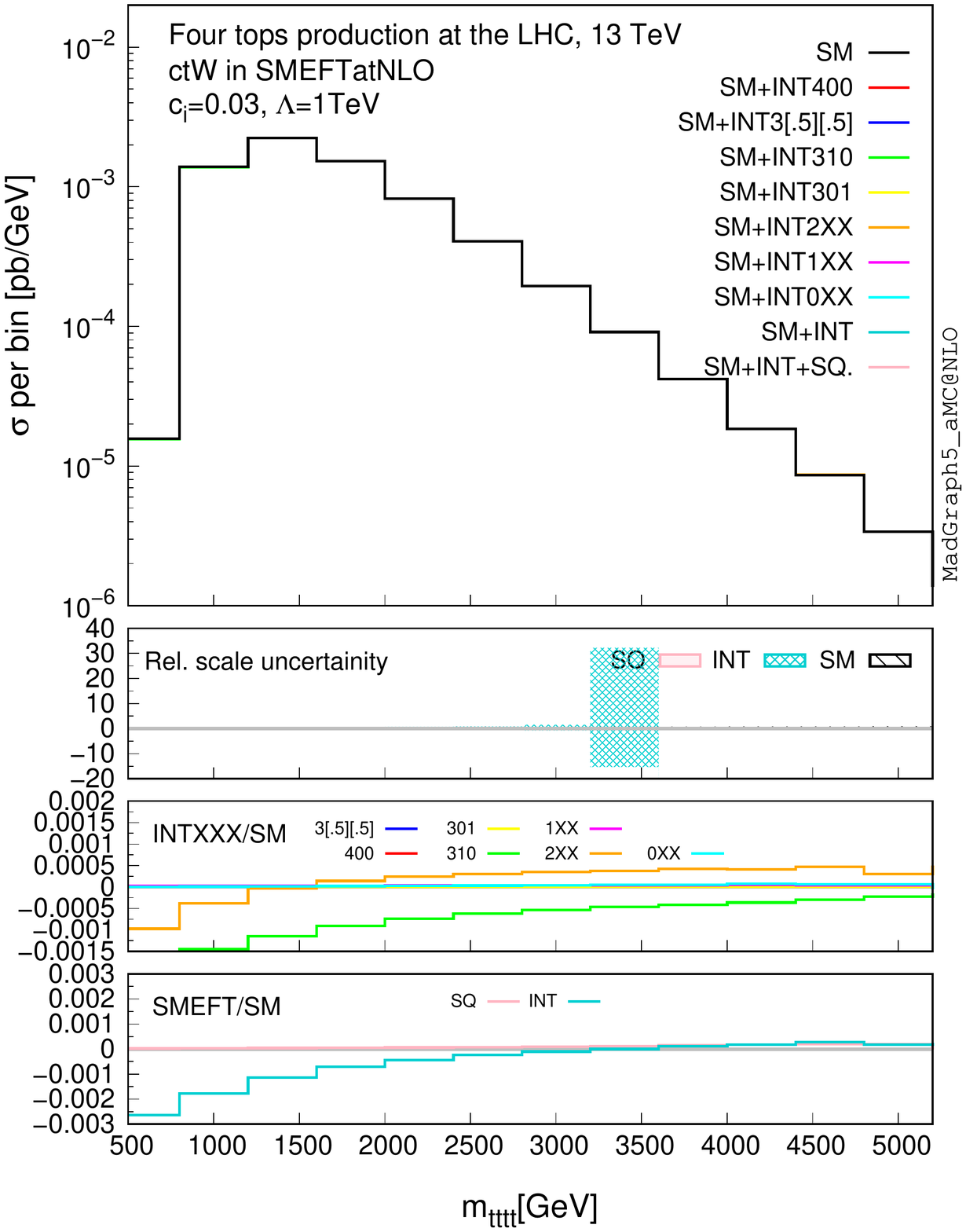}
    \includegraphics[trim=1.8cm 4.2cm 0.2cm 0.0cm, clip,width=.24\textwidth]{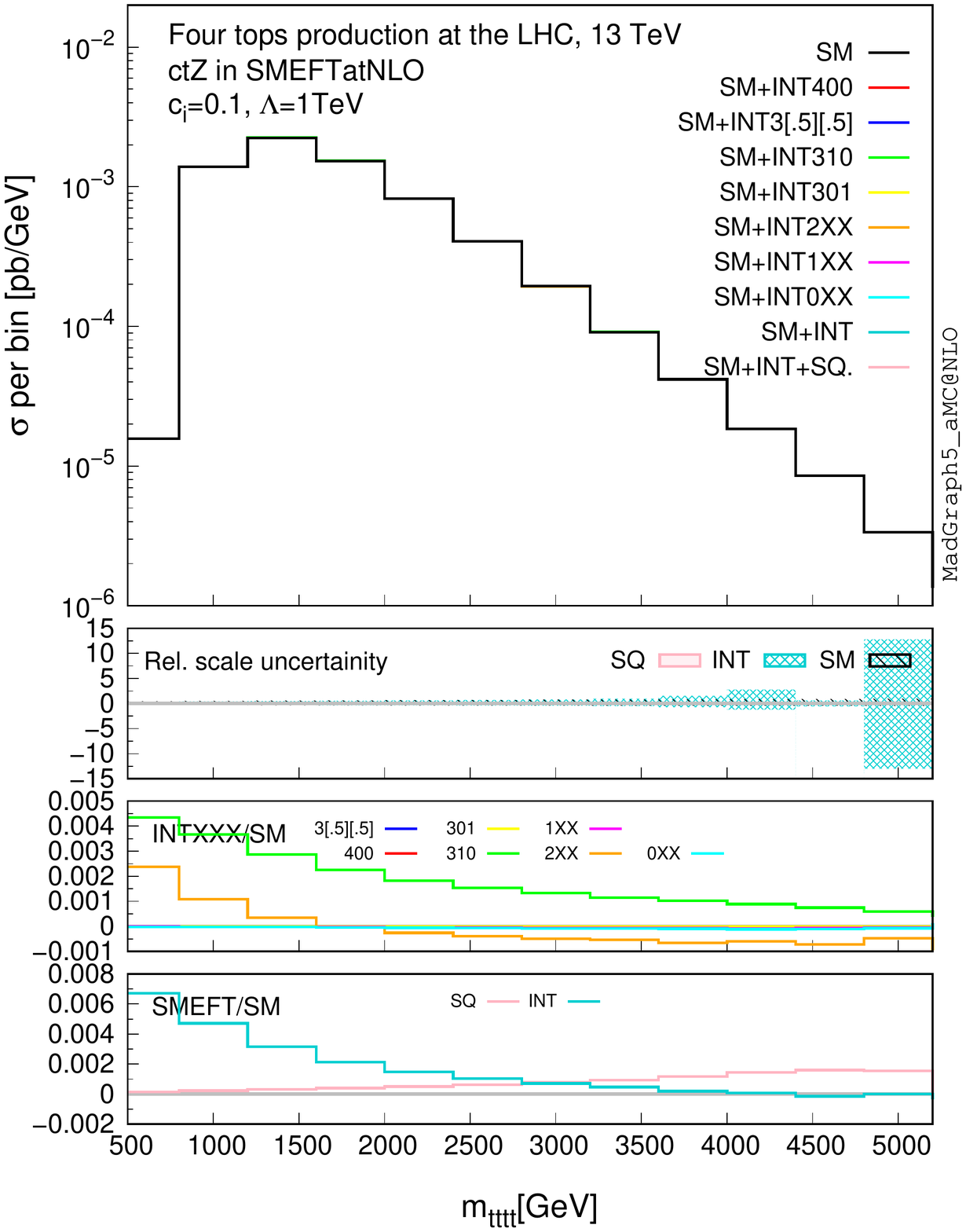}
    \includegraphics[trim=1.8cm 4.2cm 0.2cm 0.0cm, clip,width=.24\textwidth]{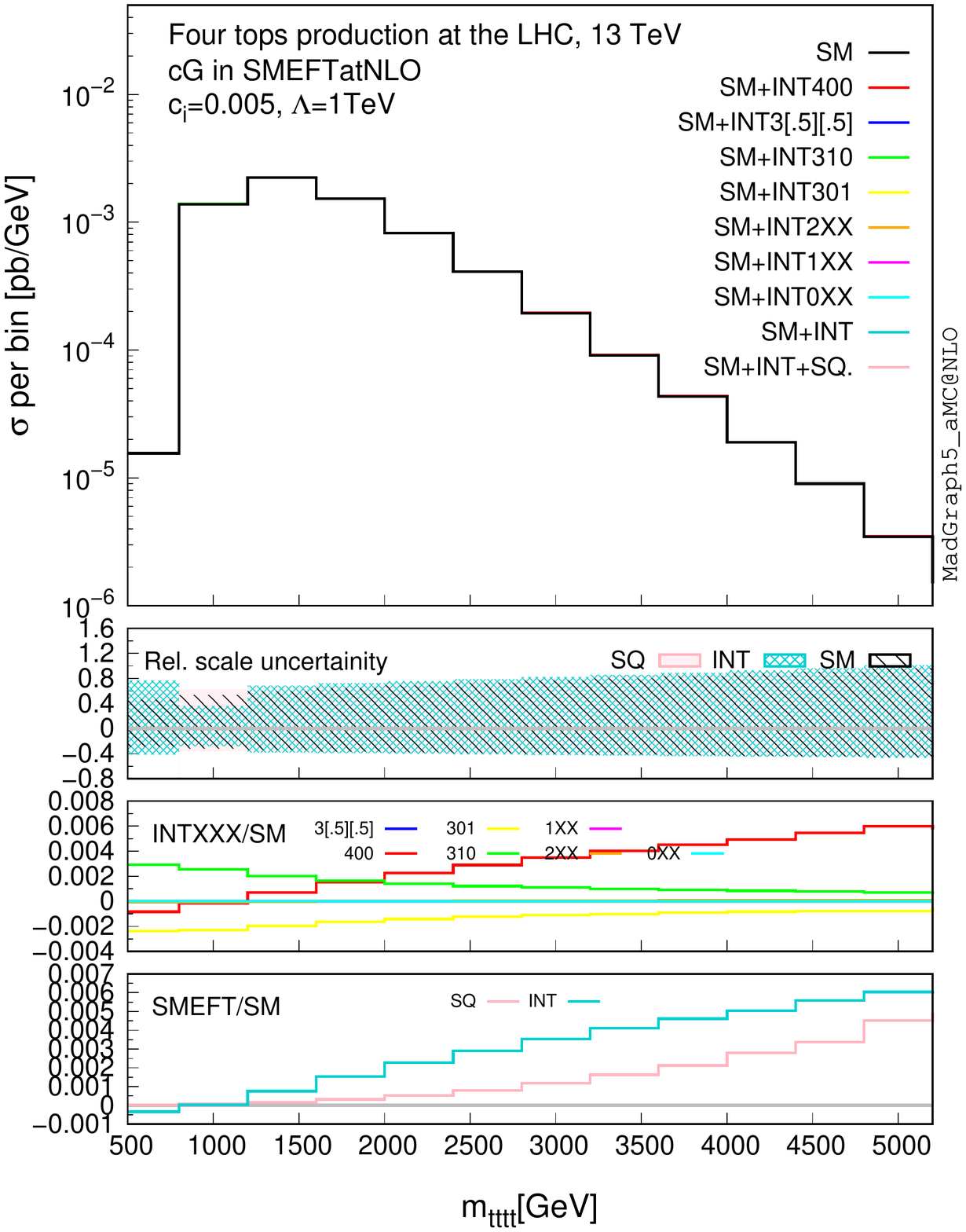}
    \caption{\label{fig:dim62f_dim60f_rest_diff_13tev} Same as \cref{fig:dim62f_dim60f_good_diff_13tev} but for the rest of the contributing two-fermion and purely-bosonic operators.}
\end{figure}

\begin{figure}[h!]
    \centering
    \includegraphics[trim=1.8cm 4.2cm 0.2cm 0.0cm, clip,width=.24\textwidth]{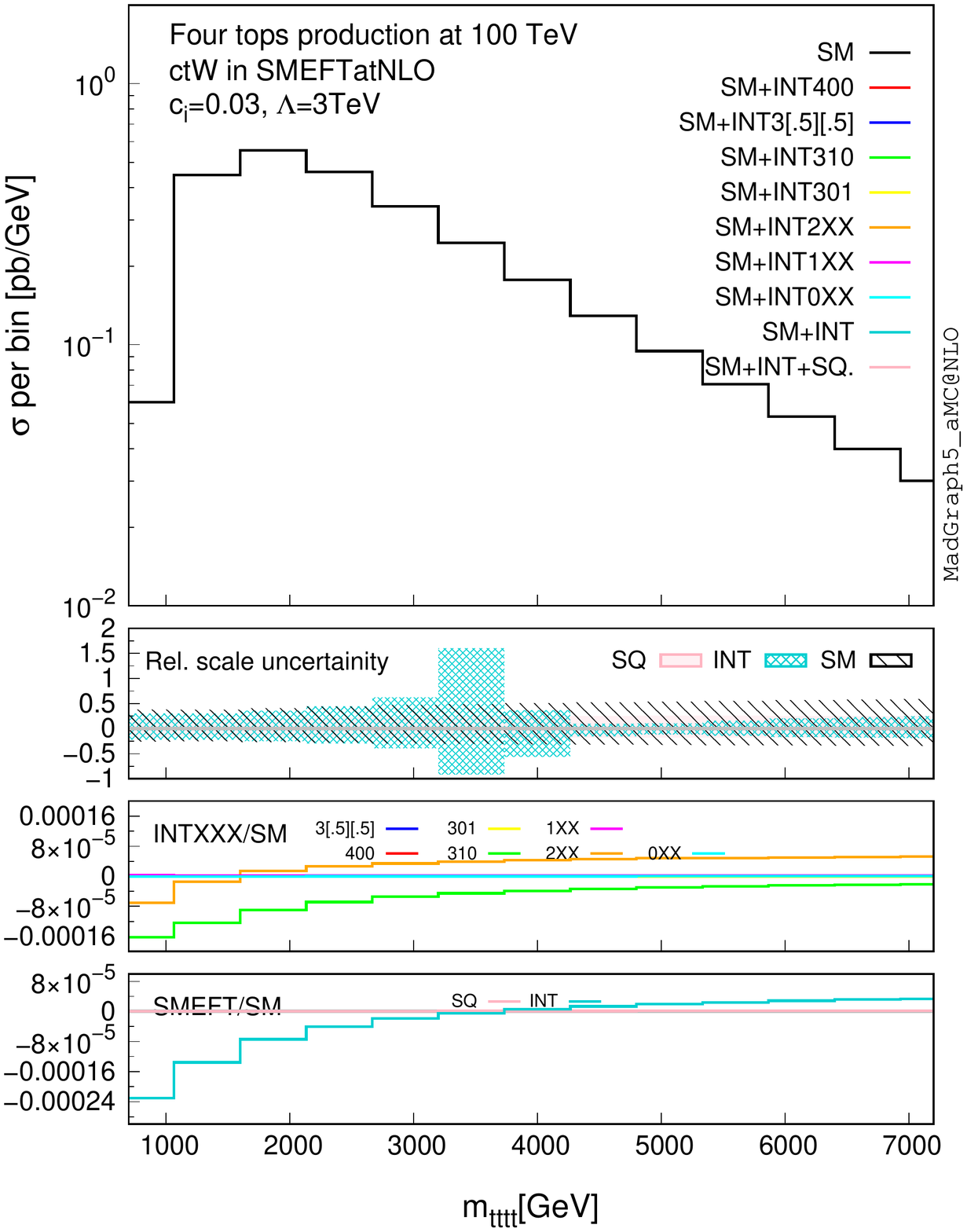}
    \includegraphics[trim=1.8cm 4.2cm 0.2cm 0.0cm, clip,width=.24\textwidth]{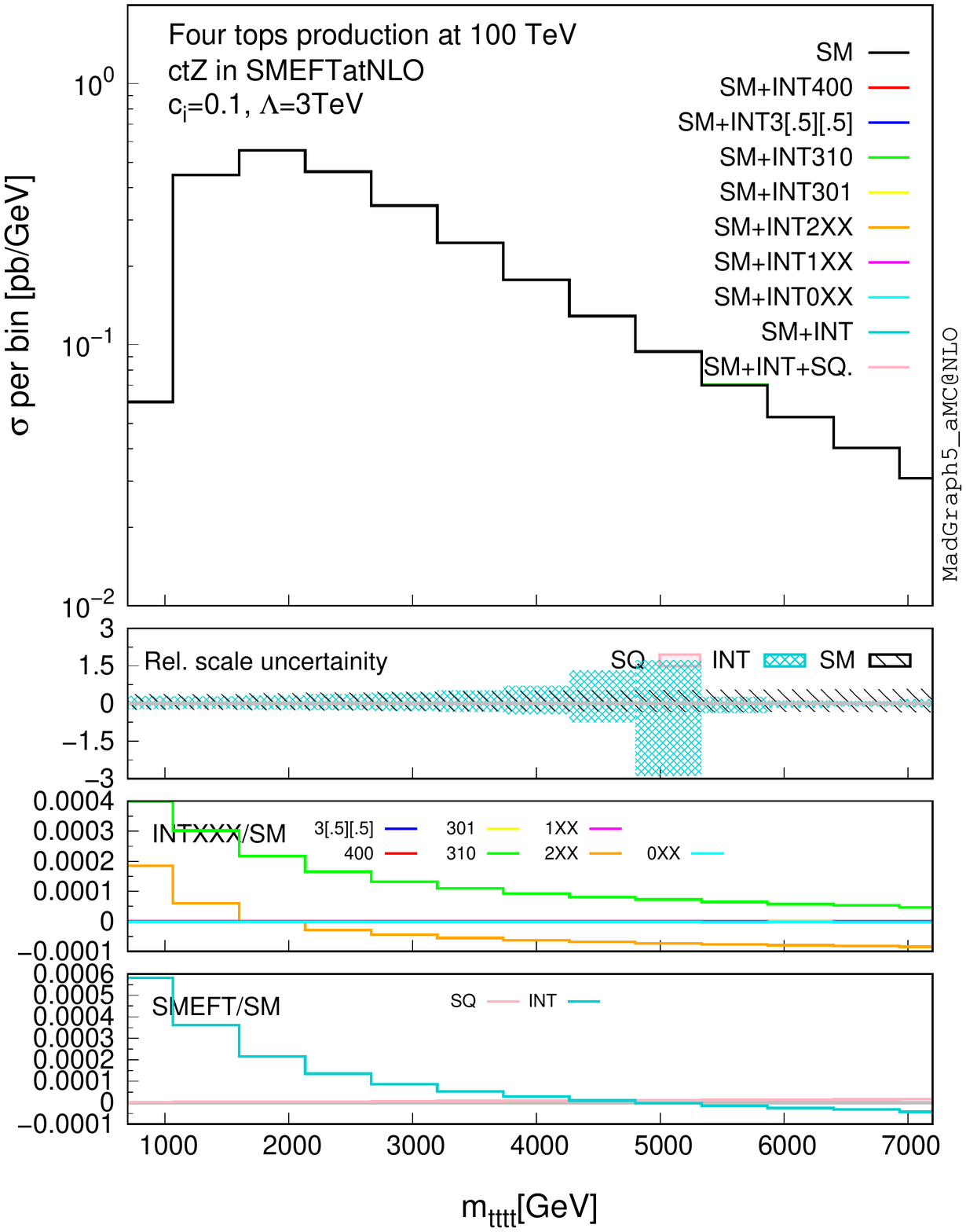}
    \includegraphics[trim=1.8cm 4.2cm 0.2cm 0.0cm, clip,width=.24\textwidth]{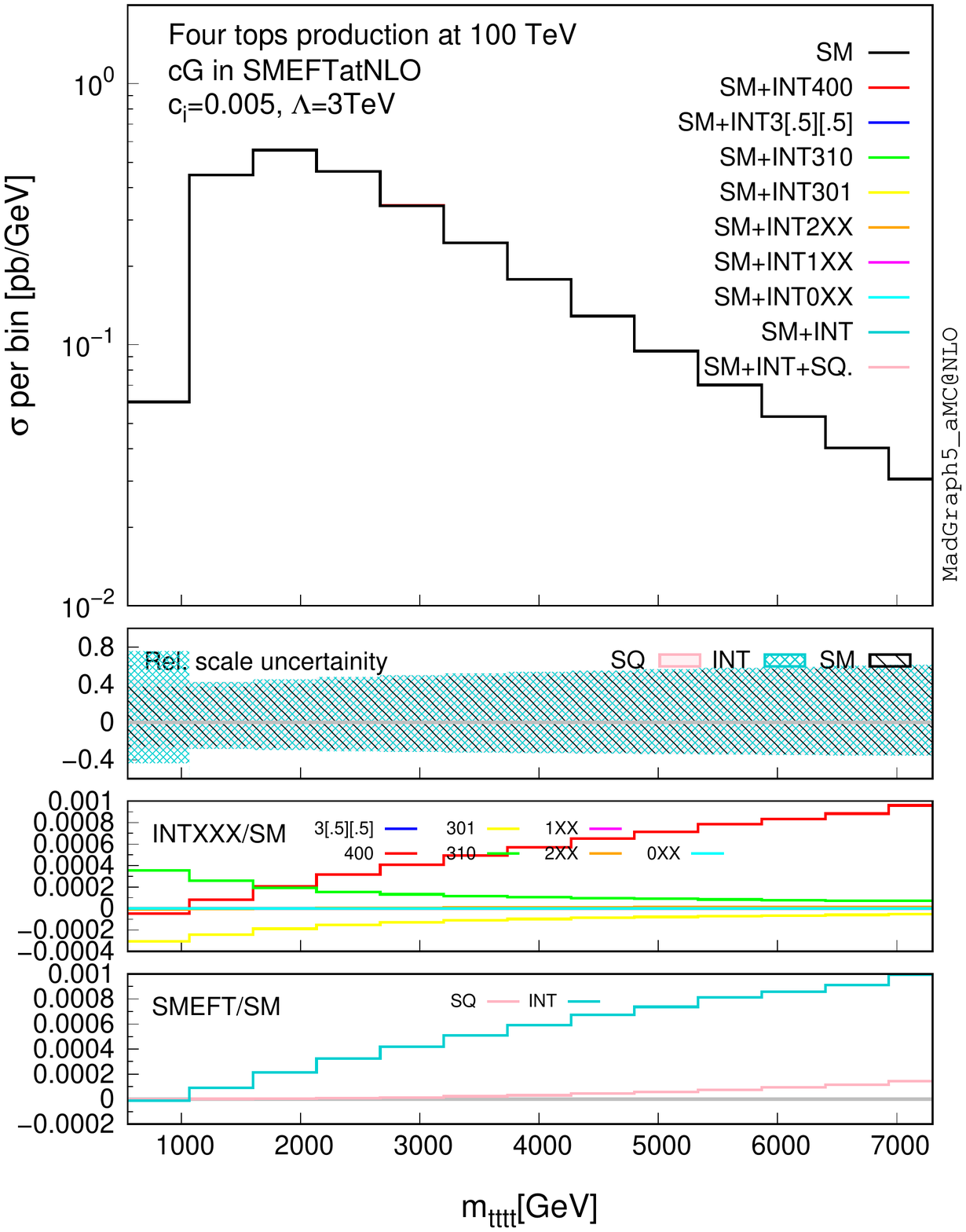}
    \caption{\label{fig:dim62f_dim60f_rest_diff_100tev} Same as \cref{fig:dim62f_dim60f_good_diff_100tev} but for the rest of the contributing two-fermion and purely-bosonic operators.}
\end{figure}

\Cref{tab:dim64f_4heavy_inclusive_within_scales} and \cref{tab:dim64f_2heavy_inclusive_within_scales} of this appendix present the LHC inclusive predictions for the 4-heavy and 2-heavy 2-light four-fermion operators within their scale uncertainties, respectively. 
Same is the case for \cref{tab:dim62f_dim60f_relevant_inclusive_within_scales} but for the set of contributing operators of \cref{eq:relevant_operators}.
Additional differential results at $\sqrt{s}=13$ TeV are presented in \cref{fig:dim64f_rest_diff_13tev} for the four-fermion operators, and in \cref{fig:dim62f_dim60f_rest_diff_13tev} for non-four-fermion ones.
Additional differential results for the FCC-hh are presented in \cref{fig:dim64f_rest_diff_100tev} for the four-fermion operators, and in \cref{fig:dim62f_dim60f_rest_diff_100tev} for non-four-fermion ones.
\begin{table}[h!]
\renewcommand{\arraystretch}{1.0}
\centering
\begin{tabular}{c|ccc}
\hline
$\mathcal{O}_{i}$ & $\mathscr{O}(\Lambda^{-2}):\sigma_{3}[\sigma_{2}]$[fb] & \multicolumn{1}{c}{$\sum\mathscr{O}(\Lambda^{-2})$[fb]} & $\sum\mathscr{O}(\Lambda^{-4})$[fb]
\\ \hline\hline
\QQOct    
&0.081 [-0.317]$^{+54\%}_{-32\%}$                           
&-0.235$^{+37\%}_{-25\%}$                             
&0.121$^{+45\%}_{-29\%}$                             

\\
\QtOct    
&0.273 [-0.577]$ ^{+54\%}_{-32\%}$                          
&-0.303$^{+29\%}_{-22\%}$                            
&0.354$^{+45\%}_{-29\%}$                            

\\ \hline 
\QQSi    
&0.242 [-0.948]$ ^{+54\%}_{-33\%}$                           
&-0.706$^{+37\%}_{-25\%}$                             
&1.086$(1) ^{+46\%}_{-29\%}$                             

\\ 
\QtSi    
&-0.005 [0.725]$ ^{+67\%}_{-61\%}$                           
&0.720$^{+41\%}_{-27\%}$                            
&1.471$(2) ^{+46\%}_{-29\%}$                            
                                       
\\ 
\ttSi    
&0.485 [-1.670]$ ^{+54\%}_{-33\%}$                         
&-1.185$(1) ^{+36\%}_{-24\%}$                            
&4.339$(2) ^{+46\%}_{-29\%}$                            
                                          
\\ \hline
\end{tabular}

\caption{\label{tab:dim64f_4heavy_inclusive_within_scales} Inclusive predictions within relative scale uncertainties for 4-heavy operators (scales are given on $\sigma_{3}$ in the first column).
$\sum\mathscr{O}(\Lambda^{-2})$ and $\sum\mathscr{O}(\Lambda^{-4})$ indicate total linear interference and total quadratic contributions, respectively.}
\end{table}
\begin{table}[h!]
\renewcommand{\arraystretch}{1.0}
\centering
\begin{tabular}{c|ccc}
\hline
$\mathcal{O}_{i}$ & $\mathscr{O}(\Lambda^{-2}):\sigma_{3}[\sigma_{2}]$[fb] & \multicolumn{1}{c}{$\sum\mathscr{O}(\Lambda^{-2})$[fb]} & $\sum\mathscr{O}(\Lambda^{-4})$[fb]
\\ \hline\hline
\OctTr    
&0.077 [-0.02]$^{+42\%}_{-27\%}$                           
&0.070$^{+41\%}_{-27\%}$                             
&0.274$(1)^{+29\%}_{-21\%}$                             

\\
\OctSi    
&0.278 [0.023]$ ^{+43\%}_{-28\%}$                          
&0.339$^{+40\%}_{-26\%}$                            
&0.275$(1)^{+30\%}_{-21\%}$                            

\\
\QuOct   
&0.202 [0.022]$ ^{+43\%}_{-28\%}$                           
&0.249$^{+40\%}_{-26\%}$                             
&0.211$(1)^{+30\%}_{-21\%}$                             

\\ 
\tqOct    
&0.315 [0.036]$ ^{+43\%}_{-28\%}$                           
&0.391$^{+40\%}_{-26\%}$                            
&0.335$(1)^{+30\%}_{-21\%}$                            
                                       
\\ 
\QdOct    
&0.115 [0.016]$ ^{+44\%}_{-28\%}$                         
&0.144$^{+40\%}_{-26\%}$                            
&0.129$(1)^{+31\%}_{-21\%}$                            
                                          
\\ 
\tuOct   
&0.178 [0.011]$ ^{+43\%}_{-28\%}$                         
&0.212$^{+40\%}_{-26\%}$                            
&0.167$(1)^{+30\%}_{-21\%}$                            
                                          
\\ 
\tdOct   
&0.101 [0.015]$ ^{+44\%}_{-28\%}$                         
&0.129$^{+40\%}_{-26\%}$                            
&0.103$(1)^{+30\%}_{-21\%}$                            
                                          
\\
\hline
\TriSi   
&-0.038 [0.079]$ ^{+41\%}_{-27\%}$                         
&0.071$^{+20\%}_{-16\%}$                            
&1.841$(4)^{+30\%}_{-21\%}$                            
                                          
\\ 
\SiSi  
&-0.140 [0.016]$ ^{+43\%}_{-28\%}$                         
&-0.113$^{+47\%}_{-30\%}$                            
&1.839$(4)^{+30\%}_{-21\%}$                            
                                          
\\ 
\QuSi    
&-0.083 [0.010]$ ^{+41\%}_{-27\%}$                         
&-0.066$^{+45\%}_{-29\%}$                            
&1.137$(1)^{+30\%}_{-21\%}$                            
                                          
\\ 
\tqSi    
&-0.131 [0.017]$ ^{+41\%}_{-27\%}$                         
&-0.106$^{+44\%}_{-29\%}$                            
&1.799$(1)^{+30\%}_{-21\%}$                            
                                          
\\ 
\QdSi   
&-0.048 [0.002]$ ^{+42\%}_{-27\%}$                         
&-0.049$^{+41\%}_{-27\%}$                            
&0.695$(1)^{+31\%}_{-21\%}$                            
                                          
\\ 
\tuSi   
&-0.089 [0.022]$ ^{+42\%}_{-27\%}$                         
&-0.056$^{+52\%}_{-32\%}$                            
&1.110$(4)^{+30\%}_{-21\%}$                            
                                          
\\ 
\tdSi   
&-0.051 [-0.011]$ ^{+43\%}_{-28\%}$                         
&-0.065$^{+40\%}_{-26\%}$                            
&0.684$(1) ^{+31\%}_{-21\%}$                            
                                          
\\ \hline
\end{tabular}
\caption{\label{tab:dim64f_2heavy_inclusive_within_scales} Same as \cref{tab:dim64f_4heavy_inclusive_within_scales} but for the 2-heavy 2-light operators.}
\end{table}
\begin{table}[h!]
\renewcommand{\arraystretch}{1.0}
\centering
\begin{tabular}{c|ccc}
\hline
$\mathcal{O}_{i}$ & $\mathscr{O}(\Lambda^{-2}):\sigma_{3}[\sigma_{2}]$[fb] & \multicolumn{1}{c}{$\sum\mathscr{O}(\Lambda^{-2})$[fb]} & $\sum\mathscr{O}(\Lambda^{-4})$[fb]
\\ \hline\hline
\ctW   
&$\times$ [-0.233]                           
&-0.220$^{+53\%}_{-32\%}$                             
&0.373$^{+37\%}_{-24\%}$                             

\\
\ctZ    
&$\times$ [0.176]                          
&0.187$^{+50\%}_{-31\%}$                            
&0.264$^{+37\%}_{-24\%}$                            

\\
\ctG   
&3.642$(1)$ [0.024]$^{+68\%}_{-38\%}$                           
&2.861$(1)^{*}$$^{+75\%}_{-40\%}$                             
&4.244$(2)^{+53\%}_{-32\%}$                             

\\ 
\ctp    
&$\times$ [0.072]                           
&-0.074$^{+26\%}_{-20\%}$                            
&0.012$^{+40\%}_{-26\%}$                            
                                       
\\ 
\hline
\cpQM    
&$\times$ [0.123]                         
&-0.302*$^{+35\%}_{-24\%}$                            
&0.030$^{+39\%}_{-26\%}$                            
                                          
\\ 
\cpt   
&$\times$ [-0.114]                         
&0.307*$^{+35\%}_{-24\%}$                            
&0.030$^{+40\%}_{-26\%}$                            
                                          
\\ 
\hline
\cG   
&1.633$(2)$ [0.113]$^{+75\%}_{-40\%}$                         
&1.715$(2)^{+75\%}_{-40\%}$                            
&94.5$(33)^{+75\%}_{-39\%}$                            
                                          
\\ 
\cpG   
&$\times$ [-0.107]                         
&-0.480$^{*}$$^{+41\%}_{-27\%}$                            
&2.229$(1)^{+28\%}_{-20\%}$                            
                                          
\\ \hline
\end{tabular}
\caption{\label{tab:dim62f_dim60f_relevant_inclusive_within_scales} Same as \cref{tab:dim64f_4heavy_inclusive_within_scales} but for the contributing operators.
The $\times$ denotes zero cross-section.
The asterisk indicates the operator receives non-negligible contributions at $\alpha_{s}$-orders lower than $\sigma_{2}$.
}
\end{table}


\clearpage
\bibliographystyle{JHEP}
\bibliography{main}
\end{document}